\documentclass[11pt,a4paper]{article}
\usepackage{jheppub}
\usepackage[T1]{fontenc}
\usepackage{amssymb}
\usepackage{mathtools}
\usepackage{stackengine}
\usepackage{scalerel}

\usepackage{hyperref}
\usepackage{epsfig}
\usepackage{amsmath,latexsym,amssymb}
\usepackage{graphicx}
\usepackage{braket}
\usepackage{pdfsync}

\usepackage{framed}
\usepackage{mathtools}

\usepackage{jheppub}
\usepackage[T1]{fontenc}
\usepackage{amssymb}
\usepackage{mathtools}
\usepackage{amsmath}
\usepackage{bm}
\usepackage{stackengine}

\usepackage{scalerel}
\usepackage{eufrak}
\usepackage{framed}

\usepackage{mathtools}

\usepackage{axodraw4j}
\usepackage{color}
\usepackage{pstricks}

\usepackage{amsmath}
\usepackage{bbold}
\usepackage{bm}
\usepackage{latexsym}
\usepackage{braket}
\usepackage{slashed}
\usepackage{graphicx,booktabs,multirow}
\usepackage{eufrak}
\numberwithin{equation}{section}

\usepackage{color}
\usepackage{pstricks}
\usepackage{amssymb}

\makeatletter
\newcommand{\doublewidetilde}[1]{{%
  \mathpalette\double@widetilde{#1}%
}}
\newcommand{\double@widetilde}[2]{%
  \sbox\z@{$\m@th#1\widetilde{#2}$}%
  \ht\z@=.9\ht\z@
  \widetilde{\box\z@}%
}
\makeatother

\usepackage{hyperref}
\usepackage{slashed}
\usepackage{epsfig}
\usepackage{amsmath,latexsym,amssymb}
\usepackage{graphicx}
\usepackage[latin1]{inputenc}
\usepackage{braket}
\usepackage{pdfsync}

\usepackage{framed}

\usepackage{mathtools}

\def\be{\begin{equation}}
\def\ee{\end{equation}}
\def\ba{\begin{eqnarray}}
\def\ea{\end{eqnarray}}

\def\braket#1{\mathinner{\langle{#1}\rangle}}

\newcommand{\bz}{\bar{z}}
\newcommand{\bw}{\bar{w}}
\newcommand{\bh}{\bar{h}}

\newcommand{\zbar}{\bar{z}}

\newcommand{\ha}{{1\over 2}}

\def\d{\delta}
\def\D{\Delta}

\def\p{\pi}

\def\om{\omega}

\def\cM{{\cal M}}

\def\cO{{\cal O}}

\def\cA{{\cal A}}

\newcommand{\comment}[1]{}

\def\fc#1#2{{\frac{#1}{#2}}}
\newcommand{\req}[1]{(\ref{#1})}

\def\h{\fc{1}{2}}

\def\p{\partial}

\newcommand{\eea}{\end{eqnarray}}
\def\lf{\left}
\def\ri{\right}

\def\ra{{\rightarrow}}

\author{
Angelos Fotopoulos${}^{1,2}$, Stephan Stieberger${}^{3}$,
Tomasz R.\ Taylor${}^{1}$,\, Bin Zhu${}^1$\\[0.5cm]
 $^1${\it Department of Physics \\
  Northeastern University, Boston, MA 02115, USA}\\[0.2cm]
 $^2${\it Department of Sciences\\
Wentworth Institute of Technology, Boston, MA 02115, USA} \\[.2 cm]
$^3${\it Max--Planck--Institut f\"{u}r Physik,	Werner--Heisenberg--Institut, \\80805 M\"unchen, Germany}}

\title{\boldmath Extended Super BMS Algebra of Celestial CFT \unboldmath}

\abstract{We study two-dimensional celestial conformal field theory describing four-dimensional ${\cal N}{=}1$ supergravity/Yang-Mills systems and show that the underlying symmetry is a supersymmetric generalization of BMS symmetry. We construct fermionic conformal primary wave functions and show how they are related via supersymmetry to their bosonic partners. We use soft and collinear theorems of supersymmetric Einstein-Yang-Mills theory to derive the OPEs of the operators associated to massless particles. The bosonic and fermionic soft theorems are shown to form a sequence under supersymmetric Ward identities. In analogy with the energy momentum tensor, the supercurrents are shadow transforms of soft gravitino operators and generate an infinite-dimensional supersymmetry algebra. The algebra of $\mathfrak{sbms}_4$ generators agrees with the expectations based on earlier work on the asymptotic symmetry group of supergravity. We also show that the supertranslation operator can be written as a product of holomorphic and anti-holomorphic supercurrents.}


\bigskip

\begin{document}
\maketitle

\section{Introduction}
Asymptotically flat space-times are of particular interest in physics.  The null infinity of four-dimensional asymptotically flat spacetimes is the product of conformal two-sphere (\textit{celestial sphere}) ${\cal C S}^2$ with a null line. In the 1960s, Bondi, van der Burg, Metzner and Sachs discovered an infinite-dimensional symmetry group,  the BMS group \cite{Bondi:1962px,Sachs:1962wk} that relates physically inequivalent, asymptotically flat, solutions of general relativity. The BMS group is an extension of the Poincar\'e group, where  the usual global translations become local symmetries -- supertranslations that depend on the  ${\cal C S}^2$ coordinates.    
In that sense the asymptotic symmetry group is the semi-direct product of the Lorentz $SL(2,\mathbb{C})$ and supertranslations groups.
In 2010, Barnich and Troessaert discovered that the BMS group can be further extended to the group of all {\em local\/} conformal transformations (superrotations), with the global subgroup of  $SL(2,\mathbb{C})$  Lorentz  transformations \cite{Barnich2009}. The algebra of the extended BMS group is infinite-dimensional, consisting of superrotations  and supertranslations.

It is also well known since the 1960s, that the soft theorem of Weinberg \cite{Weinberg1965}  imposes stringent constraints on scattering amplitudes, when a massless particle becomes soft. These constraints determine the universal properties of amplitudes in the infra-red. More recently in \cite{Cachazo1404},  the soft theorems were extended beyond the leading order  and it turns out that the universal structure of scattering processes persists at the  subleading and sub-subleading orders. In 2014, Strominger \cite{Strominger:2013jfa} showed that the extended BMS symmetry is also a symmetry of the scattering S-matrix. The Ward identities for the asymptotic symmetries \cite{Strominger1401} are actually equivalent to the soft theorems of Weinberg \cite{Weinberg1965} and Cachazo-Strominger \cite{Cachazo1404}.

One of the main motivations for further exploring the connection between soft theorems and asymptotic symmetries is the endeavour to construct a holographic description of quantum gravity in four-dimensional flat space-time  \cite{Strominger:2017zoo}. The emergence of the conformal (Virasoro) superrotations subgroup in the BMS group is a very strong indication.
Indeed, a Mellin transform over energies recasts the scattering amplitudes into conformal correlation functions on the celestial sphere~\cite{Pasterski1705, Strominger1706}.  A particular kind of two-dimensional celestial conformal field theory (CCFT) could then describe  four-dimensional dynamics. This putative CCFT is expected to provide a holographic description of four-dimensional physics.

The states of CCFT are labeled by their scaling dimension $\Delta$ and spin helicity $\ell$, which can be obtained from the conformal weights $(h,\bh)$ via $\Delta = h+ \bar{h},\, \ell=h-\bar{h}$. It turns out that states with particular values of the scaling dimension, the  conformally soft ones -- $\D\to1$ for gauge and $\D\to 1,0$ for gravity --  are identified as the generators of the BMS symmetries of CCFT and their correlators lead to Ward identities \cite{Strominger1810,Fan1903,Strominger1904,Volovich1904,Puhm1905,Mason1905,Guevara1906,Foto1906,Foto1912}. Conformally soft modes of gauge bosons are conserved currents of the CCFT, generate a Ka\v{c}-Moody symmetry and correspond to asymptotically large gauge transformations in 4d. Similarly, conformally soft gravitons give the Virasoro and supertranslation generators of the CCFT and represent asymptotically large diffeomorphisms in 4d.
On the other hand, collinear limits of 4d scattering amplitudes are particularly interesting from the CCFT point of view. They probe OPEs of CCFT operators, because identical momentum directions correspond to the operator insertion points coinciding on ${\cal C S}^2$. In \cite{Fan1903,Foto1912} we investigated
the structure of OPEs of the operators generating BMS transformations and established a connection between these OPEs and the extended BMS algebra $\mathfrak{bms}_4$.

As discussed above, the BMS algebra is an infinite-dimensional generalization of the  global bosonic algebra of the generators of the Poincar\'e group.  A natural question arises, whether such an infinite-dimensional generalization exists for supersymmetry. In the early days of BMS, such generalization was discussed in \cite{Awada1985}. In the last few years, recent developments have revived the interest in  supersymmetric generalizations of the BMS group and its corresponding algebra $\mathfrak{sbms}$. In particular for 3d space-times, a connection to the BTZ blackhole and asymptotically $AdS_3$ spacetimes can be established for asymptotically flat space-times. In three spacetime dimensions, the  BMS group of symmetries, for asymptotically flat space-times, is related to the asymptotic  symmetries  of  $AdS_3$ spaces in the limit where  the  $AdS$  radius  is  sent  to  infinity  \cite{Barnich1204}. In a similar context, an appropriate infinite radius limit of a BTZ black hole corresponds to flat space cosmologies \cite{Barnich1208,Bagchi1208}. Supersymmetric generalizations of $\mathfrak{bms}_3$ appeared in \cite{Barnich1407,Lodato1610,Fuentealba1706}.  Nevertheless our interest will be the asymptotically flat 4d space-times, which are relevant to the attempt to formulate a precise flat space holography in 4d with a CCFT.

 In the present work we discuss  $\mathfrak{sbms}_4$, the ${\cal N}=1$ supersymmetric extension of the $\mathfrak{bms}_4$ algebra.  In Section \ref{sec:prelim} we lay down the main formulas we use for conformal primary operators and their relation to the Mellin transforms of 4d scattering amplitudes. In section \ref{sec:super} we  construct supermultiplets of conformal primary wave functions for bosonic and fermionic states and demonstrate how they are related via supersymmetry transformations. We use the quantum mode expansion of these fields to identify the operators they correspond  to in the supersymmetric CCFT (SCCFT). In section \ref{sec:opegravitino} we use the collinear singularities of amplitudes in supersymmetric EYM theory to extract the OPEs of the SCCFT operators.

 As in our previous study of the bosonic CCFT \cite{Foto1912}, in sections \ref{sec:gaugino} and \ref{sec:softgravitino}, we shift gears towards the conformally soft theorems for fermionic states. The fermionic soft theorems of \cite{Strominger1511,Lysov1512,Avery1512}  are discussed from the point of view of the SCCFT. We use Mellin transform to recast the 4d supersymmetric Ward identities \cite{Grisaru:1976vm,Grisaru:1977px,Parke:1985pn} in CCFT language.  We find an interesting sequence of conformal soft limits and supersymmetry transformations which relate the subleading gauge theorem to  the soft gaugino theorem and subsequently the leading soft gauge theorem. The same picture arises in the case of the conformally soft gravitons and gravitinos. Supersymmetric Ward identities, leading gauge and gravity soft theorems are known to be exact, therefore this sequence supports earlier arguments \cite{Bern1406} that the BMS Ward identities are not anomalous. We discuss the importance of the shadow of the gravitino and its OPEs with primaries. In the bosonic case the stress energy tensor corresponds to the shadow of a spin two, dimension zero soft graviton operator \cite{Strominger1609,Sundrum1609,Foto1906}. In close analogy, we show that the shadows of the spin $\ell=\pm {3\over 2}$, soft gravitinos with dimension $\D={1\over 2}$  correspond to the supercurrents and generate supersymmetries of $\mathfrak{sbms}_4$. These generators are infinite-dimensional extensions of ${\cal N}=1$ supersymmetry generators.

 In section \ref{sec:sbms}, in close analogy with the bosonic case \cite{Foto1912}, we use soft theorems to derive the $\mathfrak{sbms}_4$ algebra.
 While the bosonic BMS group has a subgroup which is a product of holomorphic and antiholomorphic conformal (Virasoro) groups, this is not so for the super BMS group.  At the level of the algebra, the two supersymmetry generators do not anti-commute therefore $\mathfrak{sbms}_4$ does not split into superconformal algebras. Already in Ref.\cite{Awada1985}, by using the symmetries of supergravity in asymptotically flat space-times, it was  conjectured that the infinite-dimensional extension of supersymmetry in $\mathfrak{sbms}_4$ appears as a ``square root'' of supertranslations. Indeed we find that the supertranslation $ {\cal P}$ operator of \cite{Barnich1703}, which encodes all supertranslation modes, can be written as a composite operator of two supercharges,\footnote{This construction resembles the Sugawara construction of the energy momentum tensor for affine Lie algebras. See \cite{Fan2005} for a recent discussion, where it was shown that the energy momentum tensor and supertranslation current \cite{Strominger1810} can be constructed as composite operators of conformally soft gluon states.} therefore confirming these expectations. Finally, in section \ref{sec:modexp} we derive the mode expansions of the $\mathfrak{sbms}_4$ generators and derive their algebra. As in the bosonic case, the realization in terms of modes requires  studying the action of commutators of generators on the primaries. Appendix \ref{App:col} has a detailed analysis of the collinear limits of gaugino and gravitino states in scattering amplitudes as well as technical details of the 4d soft limits of scattering amplitudes and their Mellin transforms. In Appendix \ref{App:GluonSoft} we give a derivation, based on the method of \cite{Guevara1906}, of the leading and subleading conformal soft theorems for gluons.  We use this form of the theorems  in section \ref{sec:gaugino}.

\section{Notation}\label{sec:prelim}
The connection between light-like four-momenta $p^\mu$ of massless particles and points $z\in {\cal C S}^2$ relies on the following  parametrization:
\be\label{momparam}
p^\mu= \omega q^\mu, \qquad q^\mu={1\over 2} (1+|z|^2, -z-\bz, -i(z-\bz), 1-|z|^2)\ ,
\ee
where $\omega$ is the light-cone energy and $q^\mu$ is a null vector -- the direction along which the massless state propagates,  determined by $z$.
The basis of wave functions required for transforming scattering amplitudes into CCFT correlators consist of conformal wave packets characterized by $z$, dimension $\D$ and helicity $\ell$ or equivalently by $z$ and the conformal weights $h=(\D+\ell)/2, ~\bar h=(\D-\ell)/2$.

We will be using conventions and spinor helicity notation of Ref.\cite{tt}.
The four-dimensional momentum vector can be written as
\begin{eqnarray}
p_{\alpha\dot{\alpha}} =p_\mu(\sigma^\mu)_{\alpha\dot{\alpha}}=
\omega\left(\begin{array}{cc}
|z|^2 & \,z\,\, \\
\bar{z} & 1
\end{array} \!\right)=\omega q_{\alpha\dot{\alpha}}\ ,\\[1mm]
\bar{p}^{\,\dot{\alpha}\alpha}\; =p_\mu(\bar\sigma^\mu)^{\,\dot{\alpha}\alpha}=
\omega\left(\begin{array}{cc}
 1& -z \\
-\bar{z} & ~~|z|^2
\end{array} \right) = \omega \bar{q}^{\,\dot{\alpha}\alpha}\ ,
\end{eqnarray}
where $\sigma^{\mu} = (\mathbb{1},\vec{\sigma})$ and $\bar{\sigma}^{\mu} = (\mathbb{1},-\vec{\sigma})$. The spinor helicity variables are
\begin{eqnarray}
|p\rangle_{\alpha} =
\sqrt{\omega}\left(\begin{matrix}
  z \\
  1
\end{matrix} \right)=\sqrt{\omega} |q\rangle_\alpha\qquad
\,[ p|_{\dot{\alpha}} =
\sqrt{\omega}\left(\begin{matrix}
  \bar{z} \\
  1
\end{matrix} \right)=\sqrt{\omega} [ q|_{\dot{\alpha}}\\
\langle p|^{\alpha} =
\sqrt{\omega}\left(\begin{matrix}
  1\\
  -z
\end{matrix} \right)= \sqrt{\omega} \langle q|^\alpha \qquad
 |p]^{\dot{\alpha}} =
\sqrt{\omega}\left(\begin{matrix}
  1 \\
  -\bar{z}
\end{matrix} \right)= \sqrt{\omega} | q]^{\dot{\alpha}}\ .\nonumber
\end{eqnarray}
The invariant spinor products are defined as
\begin{equation}
\langle 12\rangle=\langle p_1|^\alpha |p_2\rangle_\alpha=\sqrt{\omega_1\omega_2}(z_2-z_1)\ ,\qquad [12]=
[ p_1|_{\dot{\alpha}} |p_2]^{\dot{\alpha}}=\sqrt{\omega_1\omega_2}(\bz_1-\bz_2)\ ,
\end{equation}
so that
\be \langle 12\rangle[21]=2p_1p_2=\omega_1\omega_2|z_{12}|^2\ ,\qquad z_{12}\equiv z_1-z_2\ .\ee
Note that under $SL(2,C)$ Lorentz transformations, $|q\rangle$ and $[q|$ transform in the fundamental and anti-fundamental representations with additional chiral weight factors of $(-1/2,0)$ and $(0,-1/2)$, respectively.

\section{Supermultiplets}\label{sec:super}
We begin by constructing spin $1\over 2$ conformal wave packets, by following the same route as for the scalar fields. We will set our discussion in the framework of supersymmetry. The massless scalar conformal wave packets of weight $\Delta$ are obtained by Mellin transforms of plane wave functions:
\begin{equation}\label{swv}
\varphi_{\Delta}^{\pm}(X^{\mu},z,\bz)= \int_0^{\infty}d\omega \, \omega^{\Delta-1}e^{\pm i\omega q\cdot X-\epsilon \omega}= \frac{(\mp i)^{\Delta}\Gamma(\Delta)}{(-q\cdot X\mp i\epsilon)^\Delta}\ .
\end{equation}
They are normalizable with respect to the Klein-Gordon norm only if $\mathrm{Re}(\Delta)=1$ \cite{Pasterski1705}. Then
\begin{eqnarray}
(\varphi^{\pm}_{\Delta_1},\varphi^{\pm}_{\Delta_2}) &=& -i\int d^3X[\varphi_{\Delta_1}^{\pm}(X)\partial_{X^0}(\varphi^{\pm}_{\Delta_2}(X))^* -\partial_{X^0}\varphi_{\Delta_1}^{\pm}(X)(\varphi^{\pm}_{\Delta_2}(X))^*  ] \nonumber\\
 &=&  \pm 8\pi^{4}\delta(\lambda_1-\lambda_2)\delta^{(2)}(z_1-z_2)\ ,
\end{eqnarray}
where $\lambda\equiv\mathrm{Im}(\Delta)$, i.e.\ $\Delta=1+i
\lambda$. In order to construct conformal wave packets of fermions, we take the Mellin transforms of plane waves of helicity spinors. For helicity $\ell=-1/2$,
\begin{eqnarray}
\psi^{\pm}_{\Delta,\,\alpha}(X,z,\bar{z}) &&\!\!\!\!\!= \int_0^{\infty} d\omega|p\rangle_{\alpha} \omega^{\Delta-1}e^{\pm i\omega q\cdot X-\epsilon\omega}\nonumber=| q\rangle_{\alpha}
\int d\omega \omega^{\Delta +\frac{1}{2}-1}e^{\pm i\omega q\cdot X-\epsilon\omega}\nonumber\\
&=&
| q\rangle_{\alpha}
\varphi_{\Delta+\frac{1}{2}}^{\pm}(X,z,\bz)\label{fwv}.
\end{eqnarray}
Note that the conformal wave packets $\psi^{\pm}_{\Delta,\,\alpha}(X,z,\bar{z})$ have chiral weights $(h,\bar h)=(\Delta/2-1/4,\Delta/2+1/4)$.
The spinor wave functions (\ref{fwv}) satisfy Weyl equation
\begin{eqnarray}\label{eq:weylcon}
(\bar{\sigma}^{\mu})^{\dot{\alpha}\alpha }\partial_{\mu}\psi^{\pm}_{\Delta,\dot{\alpha}}(X,z,\bar{z})
\sim (q\cdot\bar{\sigma})^{ \dot{\alpha}\alpha} | q\rangle_{\alpha} = 0\ .
\end{eqnarray}
{}For helicity $\ell = +\frac{1}{2}$, the wave functions
\begin{equation}\label{fwv1}
\bar\chi^{\pm,\,\dot{\alpha}}_{\Delta}(X,z,\bar{z})_{l = +\frac{1}{2}} =|q]^{\dot{\alpha}}\varphi_{\Delta+1/2}^{\pm}(X,z,\bz)
\end{equation}
have chiral weights $(\Delta/2+1/4,\Delta/2-1/4)$.
We can also construct Dirac spinors
\begin{eqnarray}
\Psi^\pm_{\Delta,\ell = -\frac{1}{2}}(X,z,\bar{z})
=\left(\begin{matrix}
  \psi_{\Delta,\alpha}^\pm \\
  0
\end{matrix} \right)\ , \qquad
\Psi^\pm_{\Delta,\ell = +\frac{1}{2}}(X,z,\bar{z})
=\left(\begin{matrix}
  0 \\
  \bar\chi^{\pm\dot{\alpha}}_{\Delta}
\end{matrix} \right).
\end{eqnarray}
Fermionic wave functions will be normalized with respect to the Dirac inner product
\begin{eqnarray}
(\Psi_{\Delta_1,\ell}^\pm,\Psi_{\Delta_2,\ell'}^\pm) &=&\frac{1}{2}\int\! d^3X[ \overline{\Psi}_{\Delta_2,\ell'}(X,z_2,\bar{z}_2)\gamma^0\nonumber
\Psi_{\Delta_1,\ell}(X,z_1,\bar{z}_1)\\ &&\qquad\qquad+~
\overline{\Psi}_{\Delta_1,\ell}(X,z_1,\bar{z}_1)\gamma^0
\Psi_{\Delta_2,\ell'}(X,z_2,\bar{z}_2)]\\[1mm] &=&\delta_{\ell,\ell'}
(\Psi_{\Delta_1,\ell}^\pm,\Psi_{\Delta_2,\ell}^\pm)\ .\nonumber
\label{prod1/2}
\end{eqnarray}
{}For $\ell=+1/2$,
\begin{eqnarray}(\Psi_{\Delta_1,\frac{1}{2}}^\pm, \nonumber
\Psi_{\Delta_2,\frac{1}{2}}^\pm)&=&
\langle q_1|^\alpha\sigma^0_{\alpha\dot{\alpha}}|q_2]^{\dot{\alpha}}
\int d^3X\varphi_{\Delta_2+\frac{1}{2}}^{\pm*}(X)\varphi^{\pm}_{\Delta_1+
\frac{1}{2}}(X)+(1\leftrightarrow 2)\\
&=&\langle q_1|^\alpha\sigma^0_{\alpha\dot{\alpha}}|q_2]^{\dot{\alpha}}\int d\omega_1 d\omega_2 (2\pi)^3 \delta^{(3)}(\omega_1 q_1^i-\omega_2 q_2^i)\omega_1^{\Delta_1-\frac{1}{2}}\omega_2^{\Delta_2^*-\frac{1}{2}}
+(1\leftrightarrow 2)\nonumber\\
 &=&2i\int d^3X[\varphi_{\Delta_1}^{\pm}(X)\partial_{X^0}(\varphi^{\pm}_{\Delta_2}(X))^* -\partial_{X^0}\varphi_{\Delta_1}^{\pm}(X)(\varphi^{\pm}_{\Delta_2}(X))^*  ] ,
\end{eqnarray}
where we used $q_1^0\delta^{(3)}(\omega_1 q_1^i - \omega_2 q_2^i) = \frac{1}{4\omega_1^2} \delta(\omega_1-\omega_2)\delta^{(2)}(z_1-z_2)$.
As in the scalar case, this inner product is well-defined only if $\mathrm{Re}(\Delta)=1$, i.e.\ $\Delta=1+i\lambda$. Then
\begin{eqnarray}
(\Psi_{\Delta_1,\ell}^\pm,\Psi_{\Delta_2,\ell'}^\pm)=
\pm 8\pi^4\delta(\lambda_1-\lambda_2)\delta^{(2)}(z_1-z_2)
\delta_{\ell\ell'}.
\end{eqnarray}
From now on, we will often use $\lambda$ instead of $\Delta$ to label conformal dimensions.

We can write the mode expansion of the quantum field operators in the following way:
\begin{eqnarray}
\varphi(X) &=& \int d^2z d\lambda \,\Big[a_{\Delta+}(z)\varphi^-_{\Delta^*}(X,z) + \label{phiexpand} a^\dagger_{\Delta-}(z)\varphi^+_\Delta(X,z)
\Big],\\
\psi_\alpha(X) &=& \int d^2z d\lambda \,\Big[b_{\Delta+}(z)\psi^-_{\Delta^*,\alpha}(X,z) + b^\dagger_{\Delta-}(z)\psi^+_{\Delta,\alpha}(X,z)
\Big].
\label{psiexpand}
\end{eqnarray}
The scalar annihilation operator $a_{\Delta+}(z)$ has conformal weights $(1-\Delta^*/2,1-\Delta^*/2)=(\Delta/2,\Delta/2)$, that is dimension $\Delta=1+i\lambda$ and helicity $\ell=0$. The fermionic annihilation operator $b_{\Delta+}(z)$ has conformal weights $(1-\Delta^*/2+1/4,1-\Delta^*/2-1/4)=(\Delta/2+1/4,\Delta/2-1/4)$,
that is dimension $\Delta=1+i\lambda$ and helicity $\ell=\frac{1}{2}$.

The canonical commutation relations for Klein-Gordon and anti-commutation relations for Weyl fields imply
\begin{eqnarray}
\, [a_{\lambda\pm}(z), a_{\lambda' \pm}^{\dagger}(z')] &=& 8\pi^4\delta(\lambda-\lambda')\delta^{(2)}(z-z'), \\[1mm]
\, \{b_{\lambda \pm}(z), b_{\lambda'\pm}^{\dagger}(z')\}  &=&8\pi^4\delta(\lambda-\lambda')\delta^{(2)}(z-z').
\end{eqnarray}
It is easy to see that these operators are related by Mellin transformations to the standard creation and annihilation operators in momentum space:
\begin{eqnarray}\label{amel}
a_{\Delta \pm}(z)=  \int_0^{\infty} d\omega\, \omega^{\Delta-1}\, a_{\pm}(\vec{p})\ ,\qquad b_{\Delta \pm}(z)=  \int_0^{\infty} d\omega\, \omega^{\Delta-1}\, b_{\pm}(\vec{p})\ .
\end{eqnarray}
The amplitudes describing the scattering of plane waves with definite momentum are given by vacuum expectation values of {\em in} creation and {\em out} annihilation operators. Their Mellin transforms evaluate vacuum expectation values of the conformal creation and annihilation operators. Hence celestial holography amounts to the mapping
\begin{equation}
a_{\Delta \pm},b_{\Delta \pm}\mapsto {\cal O}_{\Delta,\ell}(z,\zbar)\, ,\qquad
|0\rangle_{D=4}\mapsto|0\rangle_{{\cal CS}_2}\ .\label{corrcs}
\end{equation}

The simplest way of constructing supersymmetry generators on ${\cal CS}_2$ is to analyze supersymmetry transformations of a chiral multiplet consisting of a complex scalar and a chiral fermion.  For free (on-shell) fields, the supersymmetry transformations read:
\begin{eqnarray}
&&\delta_{\eta,\bar\eta}\varphi =  [\langle\eta Q\rangle+[\bar\eta \bar Q],\,\varphi]  = \sqrt 2\eta\psi \nonumber \\[1mm] &&
\delta_{\eta,\bar\eta}\psi =  [\langle\eta Q\rangle+[\bar\eta \bar Q],\,\psi] =
i\sqrt 2\sigma^\mu\bar\eta \partial_\mu\varphi\ .
\label{susyvarphi}
\end{eqnarray}
In order to simplify further discussion, we will absorb $\sqrt 2$ into the definition of $\eta$.
Eqs.(\ref{phiexpand}) (\ref{psiexpand}) and (\ref{susyvarphi}) imply the following supersymmetry transformations properties of conformal annihilation operators:
\begin{eqnarray}
&&[\langle\eta Q\rangle,a_{\Delta+}]  = \langle\eta q\rangle b_{(\Delta+\frac{1}{2})+} \ , \qquad [[\bar\eta \bar Q],a_{\Delta+}]=0,\nonumber \\[1mm]
&&[\langle\eta Q\rangle,b_{\Delta+}]  = 0 \ , \qquad [[\bar\eta \bar Q],b_{\Delta+}]= [\bar\eta \bar q]a_{(\Delta+\frac{1}{2})+}\ .
\label{susyvara}
\end{eqnarray}
Acting on the bosonic annihilation operator, the supersymmetry generator $Q$ increases both the dimension and helicity by 1/2, which is equivalent to shifting the conformal weight $h\to h+\frac{1}{2}$. Acting on the fermionic annihilation operator, the supersymmetry generator $\bar Q$ increases the dimension by 1/2 but decreases helicity by $-1/2$, which is equivalent to shifting the conformal weight $\bar h\to \bar h+\frac{1}{2}$.
These operators can be mapped to ${\cal CS}_2$ according to Eq.(\ref{corrcs}):
\begin{equation}a_{\Delta+}\to {\cal O}_{h,\bar h}(z,\zbar)\ ,\quad b_{\Delta+}
\to {\cal O}_{h+\frac{1}{4},\bar h-\frac{1}{4}}(z,\zbar)\ ,\quad\makebox{with}~~ (h,\bar h)=\Big(\frac{1+i\lambda}{2},\frac{1+i\lambda}{2}\Big)\ .
\end{equation}
We can also introduce a two-dimensional superfield depending on $z,\bar z$ and one Grassmann variable $\theta$:
\begin{equation}{\cal O}_\Delta(z,\bz,\theta)={\cal O}_{h,\bar h}(z,\bz)+{\cal O}_{h+\frac{1}{4},\bar h-\frac{1}{4}}(z,\bz)\theta\qquad (\Delta=1+i\lambda).\label{superf}\end{equation}
In ${\cal CS}_2$ superspace, supersymmetry is generated by
\begin{equation} Q_\alpha={\partial\over\partial\theta}|q\rangle_\alpha e^{\partial_h+\partial_{\bar h}\over 4}\ ,\qquad {\bar Q}^{\dot{\alpha}}=\theta|q]^{\dot{\alpha}} e^{\partial_{h}+\partial_{\bar h}\over 4}\ ,
\end{equation}
which satisfy the usual four-dimensional supersymmetry algebra
\begin{equation}
\{Q_\alpha,\bar Q_{\dot{\alpha}}\}=\sigma_{\alpha\dot{\alpha}}^\mu P_\mu \ ,~\makebox{with}~~ P_\mu =q_\mu
e^{\partial_h+{\partial_{\bar h}}\over 2}\ .
\end{equation}
Note that  the above supersymmetry transformations
rely on the relation between the bosonic and fermionic wave functions of Eqs.(\ref{swv}) and (\ref{fwv}):
\begin{equation}
\psi^{\pm}_{\Delta,\,\alpha}(X,z,\bar{z}) =
| q\rangle_{\alpha}e^{\partial_h+\partial_{\bar h}\over 4}\varphi^{\pm}_{\Delta}(X,z,\bar{z})\ .
\end{equation}

{}For spin 1,  Maxwell's equations are solved by
\be
v_{\D , \ell}^{\mu\pm} (X, z, \bz)=\epsilon^\mu_\ell(q,r) \varphi^\pm_\Delta
 (X, z, \bz)\label{vgauge},
\ee
with the polarization vectors
\be \epsilon^\mu_{\ell=+1}(q,r)=\frac{\langle r|\sigma^\mu| q]}{\sqrt{2}\langle rq\rangle} \ ,\qquad
\epsilon^\mu_{\ell=-1}(q,r)=\frac{[ r|\bar\sigma^\mu| q\rangle}{\sqrt{2}[qr]}\ ,
\ee
where $r$ is an arbitrary reference spinor. The gauge fixing constraint is $r^\mu v_\mu=0$, with  $r^\mu=\langle r|\sigma^\mu| r]$. Note that $r$ can be also identified with a point on ${\cal CS}_2$. The conformal mode expansion of the gauge field is
\begin{eqnarray}
v^\mu(X) &=& \sum_{\ell=\pm 1}\int d^2z d\lambda \,\Big[a_{\Delta,\ell}(z)
v_{\Delta^* ,- \ell}^{\mu-} (X, z, \bz)
+ \label{vexpand} a^\dagger_{\Delta,-\ell}(z)v_{\Delta , \ell}^{\mu+} (X, z, \bz)
\Big].\end{eqnarray}
The vector supermultiplet consists of the gauge field $v^\mu$ and gaugino $\chi_\alpha$, with the conformal mode expansions given in Eq.(\ref{psiexpand}):
\begin{eqnarray}
\chi_\alpha(X) &=& \int d^2z d\lambda \,\Big[b_{\Delta+}(z)\psi^-_{\Delta^*,\alpha}(X,z) + b^\dagger_{\Delta-}(z)\psi^+_{\Delta,\alpha}(X,z)\label{chiexp1}
\Big],\\
\bar\chi^{\dot\alpha}(X) &=& \int d^2z d\lambda \,\Big[b_{\Delta-}(z)\bar\psi^{-,\dot\alpha}_{\Delta^*}(X,z) + b^\dagger_{\Delta+}(z)\bar\psi^{+,\dot\alpha}_{\Delta}(X,z)
\Big].
\label{chiexp2}\end{eqnarray}
Under supersymmetry transformations,
\begin{eqnarray}&&
\delta_{\eta,\bar\eta} v_\mu=-i\bar\chi\bar\sigma^\mu\eta +i\bar\eta\bar\sigma^\mu\chi \ , \\[1mm]
&&
\delta_{\eta,\bar\eta} \chi_\alpha=(\sigma^{\mu\nu})_{\alpha}^{\beta}\eta_\beta(\partial_\mu v_\nu-
\partial_\nu v_\mu)\label{atran1} \ , \\[1mm]
&&
\delta_{\eta,\bar\eta} \bar\chi^{\dot\alpha}=(\bar\sigma^{\mu\nu})^{\dot\alpha}_{\dot\beta}
\bar\eta^{\dot\beta}
(\partial_\mu v_\nu-
\partial_\nu v_\mu)\label{atran2}.
\end{eqnarray}

The gauge boson and gaugino wave functions are related in the following way:
\begin{eqnarray}\psi^{\pm}_{\Delta}(X,z, \bz)&=&\frac{1}{\sqrt 2}e^{\partial_h+\partial_{\bar h}\over 4}v_{\D , \ell=-1}^{\mu\pm} (X, z, \bz)\sigma_\mu|q] \ ,\nonumber\\[1mm]
\bar\psi^{\pm}_{\Delta}(X,z, \bz)&=&\frac{1}{\sqrt 2}e^{\partial_h+\partial_{\bar h}\over 4}v_{\D , \ell=+1}^{\mu\pm} (X, z, \bz)\bar\sigma_\mu|q\rangle \ . \label{psirel}
\end{eqnarray}
With the wave functions normalized as in Eqs.(\ref{fwv},\ref{fwv1}) and (\ref{vgauge}), the conformal annihilation operators are direct Mellin transforms of the momentum space operators, c.f.\ Eq.(\ref{amel}).

The supersymmetry transformations (\ref{atran1},\ref{atran2}) combined with the relations (\ref{psirel}) lead to the following supersymmetry transformations of the annihilation operators:
\begin{eqnarray}
&&[\langle\eta Q\rangle,a_{\Delta,-1}]  = \langle\eta q\rangle b_{(\Delta+\frac{1}{2})-} \ , \qquad [[\bar\eta \bar Q],a_{\Delta,+1}]=[\bar\eta \bar q]b_{(\Delta+\frac{1}{2})+} \ ,\nonumber \\[1mm]
&&[\langle\eta Q\rangle,b_{\Delta+}]  = \langle\eta q\rangle a_{(\Delta+\frac{1}{2})+1}  \ , ~\qquad [[\bar\eta \bar Q],b_{\Delta-}]= [\bar\eta \bar q]a_{(\Delta+\frac{1}{2})-1} \ ,
\label{susyvec}
\end{eqnarray}
with all other commutators vanishing. As in the chiral multiplet, $Q$ increases $h$ by 1/2 while $\bar Q$ increases $\bar h$ by the same amount.
Here again, one could organize the supermultiplet into $\theta$-dependent  ${\cal CS}_2$ superfields similar to Eq.(\ref{superf}).

The gravitational multiplet is very similar to the gauge multiplet. It consists of spin 2 graviton $h^{\mu\nu}$ and spin 3/2 gravitino $\psi^\mu_\alpha$. Their conformal wave functions are given by
\begin{eqnarray}
h_{\D , \ell=\pm 2}^{\mu\nu\pm} (X, z, \bz)&=&\epsilon^\mu_{\ell=\pm 1}(q,r)v_{\D , \ell=\pm 1}^{\nu\pm} (X, z, \bz) \ ,\nonumber\\[1mm]
 \psi^{\mu\pm}_{\Delta,\ell=-3/2}(X,z, \bz)&=&\epsilon^\mu_{\ell=- 1}(q,r)\psi^{\pm}_{\Delta}(X,z, \bz) \ ,\\[1mm]
\bar\psi^{\mu\pm}_{\Delta,\ell=+3/2}(X,z, \bz)&=&\epsilon^\mu_{\ell=+ 1}(q,r)\bar\psi^{\pm}_{\Delta}(X,z, \bz)
 \nonumber
\end{eqnarray}
and the mode expansions have the same form as Eqs.(\ref{vexpand},\ref{chiexp1},\ref{chiexp2}).
Furthermore, the supersymmetry transformations are
\begin{eqnarray}
&&[\langle\eta Q\rangle,a_{\Delta,-2}]  = \langle\eta q\rangle b_{(\Delta+\frac{1}{2})-3/2} \ , \qquad [[\bar\eta \bar Q],a_{\Delta,+2}]=[\bar\eta \bar q]b_{(\Delta+\frac{1}{2})+3/2} \ ,\nonumber \\[1mm]
&&[\langle\eta Q\rangle,b_{\Delta,+3/2}]  = \langle\eta q\rangle a_{(\Delta+\frac{1}{2}),+2}  \ , ~\qquad [[\bar\eta \bar Q],b_{\Delta,-3/2}]= [\bar\eta \bar q]a_{(\Delta+\frac{1}{2}),-2}\ .
\label{susygra}
\end{eqnarray}

To summarize, dimension $\Delta$ conformal wave packets of fermions and bosons are properly normalized  only if $\mathrm{Re}(\Delta) = 1$, i.e.\ $\Delta=1+i\lambda$ with real $\lambda$. Through the conformal mode expansions of quantum fields, these packets are associated with conformal annihilation (and creation) operators. These are in turn  related by Mellin transformations to the usual momentum space operators. Hence the Mellin transforms of scattering amplitudes evaluate the vacuum expectation values of the respective chains  of {\em in} and {\em out} conformal creation and annihilation operators. According to the rules  of ${\cal CS}_2$ holography, they are mapped into conformal correlators of the corresponding primary field operators:
\begin{eqnarray}&&\hskip -5mm
\Big(  \prod_{n=1}^N\int d \omega_n  \ \omega_n^{\D_n-1} \Big)  \d^{(4)}\big(\sum_{n=1}^N  \epsilon_n\om_n q_n\big)\nonumber
 \cM_{\ell_1\dots \ell_N}(\omega_n, z_n, \bz_n)=\\[1mm]&&~~~~=~\langle 0|
a_{\lambda_1}^{out}\cdots b_{\lambda_k}^{out}\cdots a_{\lambda_l}^{\dagger in}\cdots b_{\lambda_N}^{\dagger in}|0\rangle~\mapsto~
\Big\langle \prod_{n=1}^N\cO_{\D_n,\ell_n}(z_n,\zbar_n)\Big\rangle \ ,
\end{eqnarray}
where $\epsilon_n=+1$ or $-1$ for {\em out} or {\em in}, respectively.

The supersymmetry generators $Q$ and $\bar Q$ raise the conformal weights of the primary fields by 1/2. They act in the following way:
\begin{eqnarray}
[\langle\eta Q\rangle,\cO_{\Delta,\ell^c}]  &=& \langle\eta q\rangle \cO_{(\Delta+\frac{1}{2}),\ell} \label{qsusy} \ , \\[1mm]
[[\bar\eta\bar Q],\cO_{\Delta,\ell}]  &=& [\bar\eta q]\label{qbarsusy} \cO_{(\Delta+\frac{1}{2}),\ell^c}\ ,
\end{eqnarray}
with $\ell$ and its complement set of $\ell^c=\ell-\frac{1}{2}$ restricted by the content of supermultiplets:
\be
\begin{array}{l|l|l|}
\makebox{\bf supermultiplet}&~~~~\boldsymbol{\ell}&
~~~~\boldsymbol{\ell^c}\\ \hline
\makebox{chiral} &~\,0 ,+\frac{1}{2}\, &\,-\frac{1}{2}\, ,0 \\ \hline
\makebox{gauge} & -\frac{1}{2}\, ,{+}1&-1\, ,+\frac{1}{2} \\ \hline
\makebox{gravitational} &-\frac{3}{2}~, {+}2& -2\, , +\frac{3}{2}~  \\ \hline
\end{array}\label{lsusy}
\ee
We will be always pairing $\ell^c$ with $\ell=\ell^c+\ha$.
\section{OPEs of fermion fields in supersymmetric EYM theory}\label{sec:opegravitino}
In this section, we discuss OPEs involving the fermions of SEYM: gauginos and gravitinos.
We follow the same path as in \cite{Foto1912} and extract OPEs from the collinear singularities of scattering amplitudes  involving particles in parallel momentum configurations ($z_{12},\bz_{12}\to 0)$.  The relevant collinear limits are derived in Appendix A. Here, we limit ourselves to collecting the results.

We begin with gauginos. The OPE of two gauginos with identical helicities is regular. Two gauginos of opposite helicities can fuse into a gauge boson or, if they carry opposite gauge charges, they can also fuse into a graviton via gravitational interactions.\footnote{We are setting the gauge coupling $g=1$ and the gravitational coupling $\kappa=2$. The CCFT operators associated to particles will be normalized here as the creation/annihilation operators, that is in a different way than in Ref.\cite{Fan1903}.}
In momentum space, the collinear limit of the respective Feynman matrix element reads
\ba
\mathcal{M}(1^{a,-\frac{1}{2}},2^{b,+\frac{1}{2}},\cdots) &=& \sum_c f^{abc}\Bigg[\frac{1}{\bar{z}_{12}}\frac{\omega_2^{\frac{1}{2}}}{\omega_1^{\frac{1}{2}}\omega_P}\mathcal{M}(P^{c,+1},\cdots) + \frac{1}{z_{12}}\frac{\omega_1^{\frac{1}{2}}}{\omega_2^{\frac{1}{2}}\omega_P}\mathcal{M}(P^{c,-1},\cdots) \Bigg]\nonumber\\
&-& \delta^{ab} \Bigg[ \frac{z_{12}}{\bar{z}_{12}}\frac{\omega_2^{\frac{3}{2}}\omega_1^{\frac{1}{2}}}{\omega_P^2}
\mathcal{M}(P^{+2},\cdots) + \frac{\bar{z}_{12}}{z_{12}}\frac{\omega_2^{\frac{1}{2}}\omega_1^{\frac{3}{2}}}{\omega_P^2}
\mathcal{M}(P^{-2},\cdots)\Bigg] +{\rm regular} ,\nonumber\ea
where $a,b,c$ denote gauge charges. The corresponding OPE is:
\begin{eqnarray}
\mathcal{O}^{a}_{\Delta_1,-\frac{1}{2}}(z_1,\bar{z}_1)\mathcal{O}^{b}_{\Delta_2,+\frac{1}{2}}(z_2,\bar{z}_2) &=& \sum_c f^{abc}\Bigg[\frac{1}{\bar{z}_{12}}B\left( \Delta_1-\frac{1}{2},\Delta_2+\frac{1}{2}\right)\mathcal{O}^c_{\Delta_1+\Delta_2-1,+1}(z_2,\bar{z}_2) \nonumber\\
&+&\frac{1}{z_{12}} B\left( \Delta_1+\frac{1}{2},\Delta_2-\frac{1}{2}\right)\mathcal{O}^c_{\Delta_1+\Delta_2-1,-1}(z_2,\bar{z}_2) \Bigg] \nonumber\\
&-& \delta^{ab} \Bigg[ \frac{z_{12}}{\bar{z}_{12}} B\left( \Delta_1+\frac{1}{2},\Delta_2+\frac{3}{2}\right)\mathcal{O}_{\Delta_1+\Delta_2,+2}(z_2\bar{z}_2) \\
&+& \frac{\bar{z}_{12}}{z_{12}}B\left( \Delta_1+\frac{3}{2},\Delta_2+\frac{1}{2}\right)\mathcal{O}_{\Delta_1+\Delta_2,-2}(z_2\bar{z}_2)\Bigg] +{\rm regular}.\nonumber
\end{eqnarray}
{}Similar singularities appear in the gaugino -- gauge boson channels:
\ba
\mathcal{M}(1^{a,+\frac{1}{2}},2^{b,+1},\cdots) &=& \sum_c f^{abc}\frac{1}{z_{ 12}}\frac{\omega_P^{1/2}}{\omega_1^{1/2}\omega_2}\mathcal{M}(P^{c,+\frac{1}{2}},\cdots) 
+ \delta^{ab}\frac{\omega_1^{1/2}}{\omega_P^{1/2}}\mathcal{M}(P^{+\frac{3}{2}},\cdots) \ ,\\
\mathcal{M}(1^{a,+\frac{1}{2}},2^{b,-1},\cdots) &=&  \sum_c f^{abc}\frac{1}{\bar{z}_{12}}\frac{\omega_1^{1/2}}{\omega_2\omega_P^{1/2}}
\mathcal{M}(P^{c,+\frac{1}{2}},\cdots) 
+\delta^{ab}\frac{z_{12}}{\bar{z}_{12}}\frac{\omega_1^{3/2}}{\omega_P^{3/2}}
\mathcal{M}(P^{+\frac{3}{2}},\cdots)\ .\nonumber
\end{eqnarray}
After performing the Mellin transforms, we obtain
\begin{eqnarray}
\mathcal{O}^a_{\Delta_1,+\frac{1}{2}}(z_1,\bar{z}_1)\mathcal{O}^b_{\Delta_2,+1}(z_2,\bar{z}_2) &=&\sum_c f^{abc}\frac{1}{z_{ 12}}B\Big( \Delta_1-\frac{1}{2},\Delta_2-1\Big)\mathcal{O}^c_{\Delta_1+\Delta_2-1,+\frac{1}{2}}(z_2,\bar{z}_2)\nonumber\\
&+&\delta^{ab} B\Big( \Delta_1+\frac{1}{2},\Delta_2\Big)\mathcal{O}_{\Delta_1 +\Delta_2 ,+\frac{3}{2}}(z_2,\bar{z}_2)\ , \\[1mm]
\mathcal{O}^a_{\Delta_1,+\frac{1}{2}}(z_1,\bar{z}_1)\mathcal{O}^b_{\Delta_2,-1}(z_2,\bar{z}_2) &=& \sum_c f^{abc}\frac{1}{\bar{z}_{12}} B\Big(\Delta_1+\frac{1}{2},\Delta_2-1 \Big)\mathcal{O}^c_{\Delta_1+\Delta_2-1,+\frac{1}{2}}(z_2,\bar{z}_2)\nonumber\\
&+&\delta^{ab} \frac{z_{12}}{\bar{z}_{12}}B\Big( \Delta_1+\frac{3}{2},\Delta_2\Big)\mathcal{O}_{\Delta_1+\Delta_2,+\frac{3}{2}}(z_2,\bar{z}_2) \ .
\end{eqnarray}
Finally, in the gaugino -- graviton channel,
\begin{equation}
\mathcal{M}(1^{-2},2^{a,+\frac{1}{2}},\cdots) =  \frac{z_{12}}{\bar{z}_{12}}\frac{\omega_2^{3/2}}{\omega_1
\omega_P^{1/2}}\mathcal{M}(P^{a,+\frac{1}{2}},
\cdots) \ ,
\end{equation}
\begin{equation}
\mathcal{M}(1^{+2},2^{a,+\frac{1}{2}},\cdots) = \frac{\bar{z}_{12}}{{z}_{12}} \frac{\omega_2^{1/2}\omega_P^{1/2}}{\omega_1}\mathcal{M}(P^{a,+\frac{1}{2}},\cdots)\ ,
\end{equation}
which yields
\begin{equation}
\mathcal{O}_{\Delta_1,-2}(z_1,\bar{z}_1)\mathcal{O}^a_{\Delta_2,+\frac{1}{2}}(z_2,\bar{z}_2) =  \frac{z_{12}}{\bar{z}_{12}}B\Big(\Delta_1-1,\Delta_2+\frac{3}{2}\Big)
\mathcal{O}^a_{\Delta_1+\Delta_2,+\frac{1}{2}}(z_2,\bar{z}_2)+{\rm regular} \ ,
\end{equation}
\begin{equation}
\mathcal{O}_{\Delta_1,+2}(z_1,\bar{z}_1)\mathcal{O}^a_{\Delta_2,+\frac{1}{2}}(z_2,\bar{z}_2) =\frac{\bar{z}_{12}}{{z}_{12}}B\Big( \Delta_1-1,\Delta_2+\frac{1}{2}\Big)\mathcal{O}^a_{\Delta_1+\Delta_2,+\frac{1}{2}}(z_2,\bar{z}_2)+{\rm regular} \ .
\end{equation}

Next, we consider the OPEs involving the gravitino field.
The OPE of two gravitinos with identical helicities is regular.  For two gravitinos with opposite helicities,
\begin{equation}
\mathcal{M}(1^{-\frac{3}{2}},2^{+\frac{3}{2}},\cdots)= \frac{z_{12}}{\bar{z}_{12}}\frac{\omega_2^{\frac{5}{2}}}{\omega_1^{\frac{1}{2}}\omega_P^2}
\mathcal{M}(P^{+2},\cdots)+\frac{\bar{z}_{12}}{z_{12}}\frac{\omega_1^{\frac{5}{2}}}{
\omega_2^{\frac{1}{2}}\omega_P^2}\mathcal{M}(P^{-2},\cdots) \ .
\end{equation}
After Mellin transformation, we obtain
\begin{eqnarray}
\mathcal{O}_{\Delta_1,\,-\frac{3}{2}}(z_1,\bar{z}_1)\mathcal{O}_{\Delta_2,\,+\frac{3}{2}}(z_2,\bar{z}_2) &=& \frac{z_{12}}{\bar{z}_{12}}B\Big(\Delta_1-\frac{1}{2},\Delta_2+\frac{5}{2}\Big)\mathcal{O}_{\Delta_1+\Delta_2,\,+2}(z_2,\bar{z}_2)\nonumber\\
&+&\frac{\bar{z}_{12}}{z_{12}}B\Big( \Delta_1+\frac{5}{2},\Delta_2-\frac{1}{2}\Big)\mathcal{O}_{\Delta_1+\Delta_2,\,-2}(z_2,\bar{z}_2)\label{G-3/2G+3/2} \  .
\end{eqnarray}
The collinear limit of the gravitino and gaugino is singular only if the two particles carry opposite sign helicities:
\begin{equation}
\mathcal{M}(1^{+\frac{3}{2}},2^{a,-\frac{1}{2}},\cdots) = \frac{\bar{z}_{12}}{z_{12}}\frac{\omega_2^{3/2}}{\omega_P \omega_1^{1/2}}\mathcal{M}(P^{a,-1},\cdots)  ]\ ,
\end{equation}
which leads to
\begin{equation}
\mathcal{O}_{\Delta_1,+\frac{3}{2}}(z_1,\bar{z}_1)\mathcal{O}^a_{\Delta_2,-\frac{1}{2}}(z_2,\bar{z}_2) = \frac{\bar{z}_{12}}{z_{12}}B(\Delta_1-\frac{1}{2},\Delta_2+\frac{3}{2})
\mathcal{O}^a_{\Delta_1+\Delta_2,\,-1}(z_2,\bar{z}_2) ]\  .
\end{equation}
Similarly, in the gravitino-gauge boson channel only one helicity configuration is singular:
\be
\mathcal{M}(1^{+\frac{3}{2}},2^{a,+1},\cdots) = \frac{\bar{z}_{12}}{z_{12}}\left( \frac{\omega_P}{\omega_1}\right)^{\frac{1}{2}}\mathcal{M}(P^{a,\frac{1}{2}},\cdots)\ .
\ee
This leads to
\begin{eqnarray}
\mathcal{O}_{\Delta_1,+\frac{3}{2}}(z_1,\bar{z}_1) \mathcal{O}^a_{\Delta_2,+1}(z_2,\bar{z}_2)= \frac{\bar{z}_{12}}{z_{12}}B(\Delta_1-\frac{1}{2}, \D_2)\mathcal{O}^a_{\Delta_1+\Delta_2,\,+\frac{1}{2}}(z_2,\bar{z}_2) \ .
\end{eqnarray}
Finally, in the gravitino -- graviton channel
\begin{equation}
\mathcal{M}(1^{+\frac{3}{2}},2^{+2},\cdots) = \frac{\bar{z}_{12}}{z_{12} } \frac{\omega_P^{3/2}}{\omega_1^{1/2}\omega_2}\mathcal{M}(P^{+3/2},\cdots) \ ,
\end{equation}
\begin{equation}
\mathcal{M}(1^{+\frac{3}{2}},2^{-2},\cdots) =\frac{z_{12} }{\bar{z}_{12}} \frac{\omega_1^{5/2}}{\omega_P^{3/2}\omega_2}\mathcal{M}(P^{+3/2},\cdots) \ .
\end{equation}
At the end, we obtain
\begin{equation}
\mathcal{O}_{\Delta_1,+\frac{3}{2}}(z_1,\bar{z}_1)\mathcal{O}_{\Delta_2,+2}(z_2,\bar{z}_2) = B(\Delta_1-\frac{1}{2},\Delta_2-1)\frac{\bar{z}_{12}}{z_{12}}\mathcal{O}_{\Delta_1+
\Delta_2,\,+\frac{3}{2}}(z_2,\bar{z}_2)\ ,
\end{equation}
\begin{equation}
\mathcal{O}_{\Delta_1,\,+\frac{3}{2}}(z_1,\bar{z}_1)\mathcal{O}_{\Delta_2,\,-2}(z_2,\bar{z}_2) = B(\Delta_1+\frac{5}{2},\Delta_2-1)
\frac{z_{12}}{\bar{z}_{12}}
\mathcal{O}_{\Delta_1+\Delta_2,\,+\frac{3}{2}}(z_2,\bar{z}_2)\ .
\end{equation}

Another way of obtaining fermionic OPEs is by applying supersymmetry transformations to bosonic OPEs, along the lines of \cite{Foto1912}, where the symmetries implied by bosonic soft theorems were employed to determine OPEs. In the following sections, we will show how the relevant bosonic and fermionic soft theorems are related by supersymmetry.
\section{Soft gaugino}\label{sec:gaugino}
Yang-Mills amplitudes diverge when the energies of one or more gluons approach zero. The divergent terms are of order $\omega^{-1}$. After Mellin transformation leading to celestial amplitudes, these soft divergences appear as simple poles $(\Delta-1)^{-1}$ in the correlators of the primary field operators $\cO_{\Delta,\pm 1}$ associated to gauge bosons. In the ``soft conformal'' $\D\to 1$  limit, these operators can be identified with the holomorphic currents and the soft theorem takes the form of a Ward identity.  A very interesting feature of the $\omega\to 0$ limit is that the ``subleading'' {\em finite} terms of order $\omega^0$ are also universal: any amplitude with a soft gluon can be factorized into an amplitude without the gluon times a universal factor. In celestial amplitudes, this universal factor is encoded as the divergence of conformal correlators in the limit of $\Delta\to 0$.

The amplitudes involving gauginos are also divergent in the soft limit although in a milder way, as $\omega^{-1/2}$. This leads to single poles in the $\Delta\to 1/2$ limit. The amplitudes may also involve a number of external gravitons and gravitinos however, as explained in Appendix A, gravitational interactions are regular in the soft gaugino limit. In Appendix A, soft gaugino terms are extracted from Feynman diagrams.
The splitting factors yield soft divergences only for specific helicity configurations. For an outgoing gaugino with helicity $+1/2$, it needs to emerge from the SYM vertex together with another gaugino of helicity $-1/2$ or a gauge boson with helicity ${+}1$.

In the amplitudes involving gauginos and other fermions, attention needs to be paid to the ordering of operator insertions. For us, the relevant characteristics of the correlators are $\sigma_i$, defined as the number of fermions {\em preceding} particle $i$, i.e.\ the number of fermion operators inserted to the left of $\mathcal{O}_i$.
In the $\Delta\to 1/2$ limit,
the gaugino correlators diverge as
\ba &&\Big\langle \cO^a_{\Delta,+1/2}(z,\zbar) \cO^{a_1}_{\Delta_1,\ell_1}(z_1,\bz_1)\cO^{a_2}_{\Delta_2,\ell_2}
(z_2,\bz_2)
\cdots\cO^{a_N}_{\Delta_N,\ell_N}(z_N,\bz_N)\Big\rangle~ {\rightarrow}\label{sgth}\\ &&\to\frac{1}{\Delta-\frac{1}{2}}
\sum_{i=1}^N\sum_{b}\frac{(-1)^{\sigma_i}f^{aa_ib}}{z-z_i}\Big
\langle \cO^{a_1}_{\Delta_1,\ell_1}(z_1,\bz_1)\cdots
\cO^{b}_{(\Delta_i-1/2),(\ell_i-1/2)}
(z_i,\bz_i)
\cdots\cO^{a_N}_{\Delta_N,\ell_N}(z_N,\bz_N)\Big\rangle \ . \nonumber
\ea
The sum on the r.h.s.\ is restricted to the operators with $\ell_i\in\{-\frac{1}{2},{+}1\}$, i.e.\ to the  helicities of gauge supermultiplets labeled as $\ell$ in the first column of Eq.(\ref{lsusy}). There is a similar anti-holomorphic expression for gaugino helicity $-\ha$, with the helicities restricted to the complement set of $\{ {-}1,{+}\frac{1}{2}\}$ and $z\to\zbar$.

The soft gaugino limit of Eq.(\ref{sgth}) is related by supersymmetry to both leading $(\Delta\to 1)$ and subleading $(\Delta\to 0)$ soft limits of gauge bosons. In order to exhibit these relations, it is convenient to consider partial amplitudes associated to a single group factor $\mathrm{Tr}(T\,T^1\cdots T^N)$, with helicity ${+}\frac{1}{2}$ gaugino associated to the group generator $T$.  We first focus on the subset
\be\Big\langle(\Delta,{+}\frac{1}{2})(\Delta_1,\ell_1)\cdots(\Delta_k,\ell_k)
(\Delta_{k+1},\ell_{k+1}^c)\cdots (\Delta_N,\ell_N^c)\Big\rangle\ ,\ee
with $\ell_i,i=1,\dots,k$ in the set of $\ell$ in Eq.(\ref{lsusy}) and $\ell^c_i,i=k{+}1,\dots, N$ is in the complement set of $\ell^c$.
As in Eq.(\ref{sgth}), soft gaugino is inserted at point $z$ while the remaining particles at $z_i, i=1,\dots, N$.
According to Eq.(\ref{sgth}), in the $\Delta=\frac{1}{2}$ limit
\begin{eqnarray}\Big\langle&& \!\!\!\! \!\!\!\! \! (\Delta,{+}\frac{1}{2})(\Delta_1,\ell_1)\cdots
(\Delta_k,\ell_k)
(\Delta_{k+1},\ell_{k+1}^c)\cdots (\Delta_N,\ell_N^c)\Big\rangle\label{sgt4}\\[1mm]
&=& \frac{1}{(\Delta-\frac{1}{2})(z-z_1)}\Big\langle
(\Delta_1{-}\frac{1}{2},\ell_1^c)\cdots
(\Delta_k,\ell_k)
(\Delta_{k+1},\ell_{k+1}^c)\cdots (\Delta_N,\ell_N^c)\Big\rangle\ ,\nonumber
\end{eqnarray}
where $\ell^c_1=\ell_1-\frac{1}{2}$.

First, we will show that the $\Delta\to\frac{1}{2}$ soft gaugino limit leads to the well-known $\Delta= 1$ singularity of the gauge boson operator. To that end, we use the Ward identity implied by the supersymmetry transformation $\langle\eta Q\rangle$:
\begin{eqnarray}\langle\eta q\rangle\Big\langle&& \!\!\!\! \!\!\!\! \! (\Delta,{+}1)(\Delta_1,\ell_1)\cdots
(\Delta_k,\ell_k)
(\Delta_{k+1},\ell_{k+1}^c)\cdots (\Delta_N,\ell_N^c)\Big\rangle\\[1mm]
&=& -\sum_{i=k+1}^N(-1)^{\sigma_i}\langle \eta q_i\rangle\Big\langle
(\Delta-\frac{1}{2},{+}\frac{1}{2})(\Delta_1,\ell_1)\cdots
(\Delta_k,\ell_k)\cdots
(\Delta_{i}+\frac{1}{2},\ell_i)\cdots (\Delta_N,\ell_N^c)\Big\rangle\ ,
\nonumber
\end{eqnarray}
where $\ell_i=\ell_i^c+\ha$.
The limit of $\Delta=1$ corresponds to $\Delta-\ha=\ha$, therefore  we can use Eq.(\ref{sgth}) to transform the r.h.s.\ into
\ba &-&\!\!\sum_{i=k+1}^N\frac{(-1)^{\sigma_i}\langle \eta q_i\rangle}{(\Delta-1)(z-z_1)}\Big\langle
(\Delta_1-\ha,\ell_1^c)\cdots
(\Delta_k,\ell_k)\cdots
(\Delta_{i}+\frac{1}{2},\ell_i)\cdots \nonumber (\Delta_N,\ell_N^c)\Big\rangle\\[1mm]
&&-~\frac{\langle \eta q_N\rangle}{(\Delta-1)(z-z_N)}
\Big\langle
(\Delta_1,\ell_1)\cdots
(\Delta_k,\ell_k)
(\Delta_{k+1},\ell_{k+1}^c)\cdots (\Delta_N,\ell_N^c)\Big\rangle \ .
\ea
After using Ward identity again in the first term, we obtain
\begin{eqnarray}\langle\eta q\rangle\Big\langle&& \!\!\!\! \!\!\!\! \! (\Delta,{+}1)(\Delta_1,\ell_1)\cdots
(\Delta_k,\ell_k)
(\Delta_{k+1},\ell_{k+1}^c)\cdots (\Delta_N,\ell_N^c)\Big\rangle\\[1mm]
&=&\frac{1}{\Delta-1}\left(
\frac{\langle \eta q_1\rangle}{z-z_1}-\frac{\langle \eta q_N\rangle}{z-z_N}\right)\Big\langle (\Delta_1,\ell_1)\cdots
(\Delta_k,\ell_k)
(\Delta_{k+1},\ell_{k+1}^c)\cdots (\Delta_N,\ell_N^c)\Big\rangle \ .
\nonumber
\end{eqnarray}
We can choose $\eta$ such that $\langle\eta q\rangle=\langle\eta q_i\rangle=\eta_0$, where $\eta_0$ is a constant Grassmann number. This leads to the well known expression for the $\Delta\to 1$ limit of the gauge boson operator. The same argument can be repeated for all other helicity configurations.

Next, we will show that the $\Delta\to\frac{1}{2}$ soft gaugino limit can be obtained from the ``subleading'' $\Delta= 0$ singularity of the gauge boson operator. To that end, we use the Ward identity implied by the supersymmetry transformation $[\bar\eta \bar Q]$:
\begin{eqnarray}
[\bar\eta\bar q]\Big\langle&& \!\!\!\! \!\!\!\! \! (\Delta,{+}\ha)(\Delta_1,\ell_1)\cdots
(\Delta_k,\ell_k)
(\Delta_{k+1},\ell_{k+1}^c)\cdots (\Delta_N,\ell_N^c)\Big\rangle\label{sgt2}\\[1mm]
&=& -\sum_{i=1}^k(-1)^{\sigma_i}[\bar\eta \bar q_i]\Big\langle
(\Delta-\frac{1}{2},{+}1)(\Delta_1,\ell_1)\cdots(\Delta_{i}+\frac{1}{2},\ell_i^c)\cdots
(\Delta_k,\ell_k)\cdots
(\Delta_N,\ell_N^c)\Big\rangle\ ,
\nonumber
\end{eqnarray}
where $\ell_i^c=\ell_i-\ha$. The limit of $\Delta\to\ha$ corresponds to $\Delta-\ha=0$,
in which we can use the ``subleading'' gauge boson limit
\begin{eqnarray}\Big\langle&& \!\!\!\! \!\!\!\! \! (\Delta,{+}1)(\Delta_1,\ell_1)\cdots
(\Delta_k,\ell_k)\nonumber
\cdots (\Delta_N,\ell_N^c)\Big\rangle=\\[1mm]
&=&\frac{1}{\Delta}\left\{
\frac{1}{z-z_1}\Big[(\bz-\bz_1)\partial_{\bz_1}-2\bar h_1+1\Big]\langle(\Delta_1-1,\ell_1)\cdots
(\Delta_k,\ell_k)\cdots (\Delta_N,\ell_N^c)\Big\rangle\right.
\\[1mm]&&~~~~\left.
-
\frac{1}{z-z_n}\Big[(\bz-\bz_N)\partial_{\bz_N}-2\bar h_N+1
\Big]\Big\langle (\Delta_1,\ell_1)\cdots
(\Delta_k,\ell_k)\cdots (\Delta_N-1,\ell_N^c)\Big\rangle\right\} \ .
\nonumber
\end{eqnarray}
Once we insert this into Eq.(\ref{sgt2}) and choose $\bar\eta$ such that $[\bar\eta \bar q]=[\bar \eta \bar q_i]=\bar\eta_0$, where $\bar\eta_0$ is a constant Grassmann number, all terms on the r.h.s.\ will cancel as a consequence of supersymmetric Ward identity, except for the single term proportional to $\bar h_1$ which  is raised by $1/2$ inside the $i=1$ contribution. As a result, we obtain Eq.(\ref{sgt4}).

To summarize, supersymmetry implies the following sequence of soft limits:\begin{center}
  $(\Delta=0)$ subleading gauge boson $\stackrel{\bar Q}{\longrightarrow}$ $(\Delta=\ha)$ gaugino $\stackrel{Q}{\longrightarrow}$  $(\Delta=1)$ gauge boson.\end{center}
Since $\Delta=0$ follows from $\Delta=1$ by gauge invariance, supersymmetry and gauge invariance create a closed symmetry loop of soft limits.
Although the soft gaugino operator ${\cal O}^a_{\D=1/2,\ell=1/2}(z)$ is a $(1/2,0)$ holomorphic form, there is no global symmetry associated to the soft limit (\ref{sgth}) because its correlators are not sufficiently suppressed at infinity; it is easy to see that they fall off only as $1/z$.
In the next section we will show how supersymmetric Ward identities follow from the soft limit of the gravitino operator which has a faster suppression at infinity.

\section{Soft gravitino, supercurrent and SUSY Ward identities }\label{sec:softgravitino}

At low energies, spin 3/2 gravitinos behave in a similar way to gauginos. In the soft limit, their amplitudes diverge as $\omega^{-1/2}$. These divergences are extracted from Feynman diagrams
in Appendix A. As the four-momentum $p_s=\omega_sq_s\to 0$, the gravitino amplitudes behave as
\begin{eqnarray}
\mathcal{M}(p_s\, \ell_s \!\!&=&\!\! +\frac{3}{2}, p_1 \, \ell_1, \cdots ,p_N \, \ell_N)\nonumber\\
  &=&\sum_{i=1}^N  (-1)^{\sigma_i} \frac{\bar{z}_{si}}{z_{si}}\frac{z_{ri}}{z_{rs}}\Big(\frac{\omega_i}{\omega_s}\Big)^{
  \frac{1}{2}}\mathcal{M}(p_1\,\ell_1,\cdots, p_i\,\ell^c_i,\cdots,p_N,\,\ell_N)~+~\mathcal{O}(\omega_s^{\ha})\ ,\label{sgrav}
\end{eqnarray}
where the sum on the r.h.s.\ is restricted to particles with $\ell_i\in\{-\frac{3}{2},-\frac{1}{2},{+}1, {+}2\}$, i.e.\ to the  helicities labeled as $\ell$ in the first column of Eq.(\ref{lsusy}), while $\ell^c_i=\ell_i-\ha$ are in the complement set. The expression on the r.h.s.\ involves an arbitrary reference point $z_r$ however it does not depend on its choice. This can be shown by using supersymmetric Ward identities. After performing Mellin transformations, the leading soft singularity appears
in celestial amplitudes as the pole at dimension $\Delta=1/2$ of the gravitino operator. Near $\Delta=1/2$,
\begin{eqnarray}\Big\langle(\Delta,\!\!&{+}&\!\!\frac{3}{2})(\Delta_1,\ell_1)
\cdots(\Delta_k,\ell_k)
(\Delta_{k+1},\ell_{k+1}^c)\cdots (\Delta_N,\ell_N^c)\Big\rangle\label{half1}
  \\
&=& \frac{1}{\Delta-\ha} \sum_{i=1}^k(-1)^{\sigma_i}\frac{z_{ri}}{z_{rs}} \frac{\bar{z}_{si}}{z_{si}} \Big\langle
(\Delta_1,\ell_1)\cdots(\Delta_{i}+\frac{1}{2},\ell_i^c)\cdots
(\Delta_k,\ell_k)\cdots
(\Delta_N,\ell_N^c)\Big\rangle \ . \nonumber
\end{eqnarray}
As in Eq.(\ref{sgrav}), soft gravitino is inserted at point $z_s$ while the remaining particles at $z_i, i=1,\dots, N$. For opposite helicity,
\begin{eqnarray}\Big\langle(\Delta,\!\!&{-}&\!\!\frac{3}{2})(\Delta_1,\ell_1)
\cdots(\Delta_k,\ell_k)
(\Delta_{k+1},\ell_{k+1}^c)\cdots (\Delta_N,\ell_N^c)\Big\rangle\label{half2}
  \\
&=& \frac{1}{\Delta-\ha} \sum_{i=k+1}^N(-1)^{\sigma_i}\frac{\bar z_{ri}}{\bar z_{rs}} \frac{{z}_{si}}{\bar z_{si}} \Big\langle
(\Delta_1,\ell_1)          \cdots(\Delta_k,\ell_k)\cdots
(\Delta_{i}+\frac{1}{2},\ell_i)\cdots
\cdots
(\Delta_N,\ell_N^c)\Big\rangle \ . \nonumber
\end{eqnarray}

The subleading term of order $\mathcal{O}(\omega_s^{\ha})$ in Eq.(\ref{sgrav}) is also universal to all gravitino amplitudes. It is similar to subleading terms present in the graviton amplitudes and, as we will show below, related to them by supersymmetry. It can be extracted from the amplitudes by using local supersymmetry invariance, along the lines of \cite{Bern1406}. In celestial amplitudes, this subleading soft term is encoded in the residue of the pole at $\Delta=-\ha$. After performing Mellin transformations, one finds that near $\Delta=-\ha$,
\begin{eqnarray}\Big\langle(\Delta,\!\!&{+}&\!\!\frac{3}{2})(\Delta_1,\ell_1)
\cdots(\Delta_k,\ell_k)
(\Delta_{k+1},\ell_{k+1}^c)\cdots (\Delta_N,\ell_N^c)\Big\rangle\label{ssgravi}\\
&=& \frac{1}{\Delta+\ha} \sum_{i=1}^k(-1)^{\sigma_i} \frac{\bar{z}_{si}}{z_{si}}
\big(\bar{z}_{si}\partial_{\bar{z}_i}-2\bar{h}_i\big)
\Big\langle
(\Delta_1,\ell_1)\cdots(\Delta_{i}-\frac{1}{2},\ell_i^c)\cdots
(\Delta_N,\ell_N^c)\Big\rangle \ . \nonumber
\end{eqnarray}

The relations between soft gravitino and graviton limits are
very similar to the relations between gauginos and gauge bosons, as explained  in the previous section. The following sequence emerges as a consequence of supersymmetric Ward identities:\begin{center}
{  $(\Delta=-1)$ sub-subleading graviton $\stackrel{\bar Q}{\rightarrow}$ $(\Delta=-\ha)$ subleading gravitino ${\longrightarrow}$\\[1mm]  $\stackrel{Q}{\rightarrow}$  $(\Delta=0)$ subleading graviton}$\stackrel{\bar Q}{\rightarrow}$ $(\Delta=\ha)$  leading gravitino $\stackrel{Q}{\rightarrow}$  $(\Delta=1)$ leading graviton.
\end{center}
All these links can be demonstrated in a similar way, therefore we will limit ourselves to the proof of the first one which is implied by $\bar Q$ supersymmetry.
To that end, we use the following Ward identity:
\begin{eqnarray}
[\bar\eta\bar q]\Big\langle&& \!\!\!\! \!\!\!\! \! (\Delta,{+}\frac{3}{2})(\Delta_1,\ell_1)\cdots
(\Delta_k,\ell_k)
(\Delta_{k+1},\ell_{k+1}^c)\cdots (\Delta_N,\ell_N^c)\Big\rangle\label{sgravi2}\\[1mm]
&=& -\sum_{i=1}^k(-1)^{\sigma_i}[\bar\eta \bar q_i]\Big\langle
(\Delta-\frac{1}{2},{+}2)(\Delta_1,\ell_1)\cdots(\Delta_{i}+\frac{1}{2},\ell_i^c)\cdots
(\Delta_k,\ell_k)\cdots
(\Delta_N,\ell_N^c)\Big\rangle\ ,
\nonumber
\end{eqnarray}
where $\ell_i^c=\ell_i-\ha$. The limit of $\Delta\to -\ha$ corresponds to $\Delta-\ha=-1$,
in which we can use the sub-subleading graviton limit \cite{Guevara1906,Strominger1910}:
\begin{eqnarray}\Big\langle&& \!\!\!\!(\Delta,{+}2)(\Delta_1,\ell_1)\cdots
(\Delta_k,\ell_k)\nonumber
\cdots (\Delta_N,\ell_N^c)\Big\rangle=\frac{1}{2(\Delta+1)}\sum_{j=1}^N \frac{\bar{z}_{sj}}{z_{sj}}\Big[2\bar{h}_j(2\bar{h}_j-1)
\\[1mm]
&&
-~4\bar{h}_j\bar{z}_{sj}\partial_{\bar{z}_j} +\bar{z}_{sj}^2\partial^2_{\bar{z}_j}\Big] \Big\langle (\Delta_1,\ell_1)\cdots
(\Delta_j-1,\ell_j)
\cdots (\Delta_N,\ell_N^c)\Big\rangle
\end{eqnarray}
Once we insert this into Eq.(\ref{sgravi2}) and choose $\bar\eta$ such that $[\bar\eta \bar q]=[\bar \eta \bar q_i]=\bar\eta_0$, where $\bar\eta_0$ is a constant Grassmann number, most of terms on the r.h.s.\ will cancel as a consequence of supersymmetric Ward identities, except for the terms involving $\bar h_j$ which are raised by $1/2$ inside the $i=j=1,2,\dots, k$ contributions. As a result, we obtain Eq.(\ref{ssgravi}).

The formulas describing soft limits of operator insertions are exact statements about the properties of CCFT and
are expected to reflect the underlying symmetries. In \cite{Foto1906} the energy-momentum tensor was constructed by performing the shadow transformations on $\Delta=0$ graviton operator. Here, we are interested in the supercurrent that generates supersymmetry transformations and supersymmetric Ward identities. The holomorphic and anti-holomorphic supercurrents can be constructed from the gravitino operator by taking the limits of following shadow transforms:
\ba
S(z)&=& \lim_{\Delta\rightarrow \frac{1}{2}}\frac{\Delta-\ha}{\pi}\int d^2 z' \frac{1}{(z-z')^3}\mathcal{O}_{\Delta,-\frac{3}{2}}(z',\bar{z}')\ ,\label{G-3/2def}\\[1mm]
\bar S(\bar z)&=& \lim_{\Delta\rightarrow \frac{1}{2}}\frac{\Delta-\ha}{\pi}\int d^2 z' \frac{1}{(\bar z-\bar z')^3}\mathcal{O}_{\Delta,{+}\frac{3}{2}}(z',\bar{z}')\label{G+3/2def}\ .
\ea
Note that $S(z)$ has conformal weights $(h,\bar h)=(3/2, 0)$. The correlation functions of supercurrents with other operators can be evaluated by using the soft limits (\ref{half1},\ref{half2}). To that end, it is convenient to set the reference point $z_r\to\infty$. This leads to
\begin{eqnarray}\Big\langle S(z)&&\!\!\!\!\!\!\!\!\mathcal{O}_{\Delta_1,\ell_1}(z_1,\bar
 z_1)
\cdots \mathcal{O}_{\Delta_k,\ell_k}(z_k,\bar z_k)
\mathcal{O}_{\Delta_{k+1},\ell_{k+1}^c}(z_{k+1},\bar z_{k+1})\cdots \mathcal{O}_{\Delta_N,\ell_N^c}(z_N,\bar z_N)\Big\rangle\nonumber\\
&=& \frac{1}{\pi}\int d^2 z_s \frac{1}{({z}-{z}_s)^3} \sum_{i=k+1}^N  \frac{{z}_{si}}{\bar z_{si}}(-1)^{\sigma_i}\!\times\label{scurr1}\\ &&\times
\Big\langle \mathcal{O}_{\Delta_1,\ell_1}(z_1,\bar z_1)
\cdots \mathcal{O}_{\Delta_k,\ell_k}(z_k,\bar z_k)\cdots
\mathcal{O}_{\Delta_{i}+\ha,\ell_{i}}(z_{k+1},\bar z_{k+1})\cdots \mathcal{O}_{\Delta_N,\ell_N^c}(z_N,\bar z_N)\Big\rangle \ . \nonumber
\end{eqnarray}
The shadow integral can be performed by using the identities $$\frac{1}{({z}-{z}_s)^3} = \frac{1}{2}\partial_{{z}_s^2}\bigg(\frac{1}{{z}-{z}_s}\bigg)\ ,\qquad \partial_{{z}_s}\bigg(\frac{1}{\bar z_s-\bar z_i}\bigg) =2\pi \delta^{(2)}(z_s-z_i)\ ,$$
so that
\be \frac{1}{\pi}\int d^2 z_s \frac{1}{({z}-{z}_s)^3} \frac{{z}_{si}}{\bar z_{si}}=\frac{1}{z-z_i}\ .\ee
As a result,
\begin{eqnarray}\Big\langle S(z)&&\!\!\!\!\!\!\!\!\mathcal{O}_{\Delta_1,\ell_1}(z_1,\bar
 z_1)
\cdots \mathcal{O}_{\Delta_k,\ell_k}(z_k,\bar z_k)
\mathcal{O}_{\Delta_{k+1},\ell_{k+1}^c}(z_{k+1},\bar z_{k+1})\cdots \mathcal{O}_{\Delta_N,\ell_N^c}(z_N,\bar z_N)\Big\rangle\nonumber\\
&=& \sum_{i=k+1}^N  \frac{(-1)^{\sigma_i}}{z-z_i}\times\label{scurr3}\\
\times&&\!\!\!\!\!\!\!\!
\Big\langle\prod_{j=1}^k \mathcal{O}_{\Delta_j,\ell_j}(z_j,\bar z_j)
\mathcal{O}_{\Delta_{k+1},\ell_{k+1}^c}(z_{k+1},\bar z_{k+1})\cdots
\mathcal{O}_{\Delta_{i}+\ha,\ell_{i}}(z_{i},\bar z_{i})\cdots \mathcal{O}_{\Delta_N,\ell_N^c}(z_N,\bar z_N)\Big\rangle \ . \nonumber
\end{eqnarray}
We recognize the above equation as a Ward identity of local supersymmetry generated by $Q$. Since the gravitino correlators fall off at infinity as $1/z^3$,\footnote{Notice that at $z\rightarrow \infty$, $S(z)$, which has conformal weight $h={3\over 2}$,  decays as $\frac{1}{z^3}$.} we can obtain global Ward identities by integrating $S(z)$  multiplied by any first order polynomial in $z$ (which we can choose as $\langle\eta q\rangle$) over a contour surrounding all points  $z_i$:
\begin{eqnarray}\oint\frac{\langle\eta q\rangle dz}{2\pi i}\Big\langle S(z)&&\!\!\!\!\!\!\!\!\mathcal{O}_{\Delta_1,\ell_1}(z_1,\bar
 z_1)
\cdots \mathcal{O}_{\Delta_k,\ell_k}(z_k,\bar z_k)
\mathcal{O}_{\Delta_{k+1},\ell_{k+1}^c}(z_{k+1},\bar z_{k+1})\cdots \mathcal{O}_{\Delta_N,\ell_N^c}(z_N,\bar z_N)\Big\rangle\nonumber\\
&=& 0= \sum_{i=k+1}^N  (-1)^{\sigma_i}\langle\eta q_i\rangle\times\label{scurr5}\\
\times
\Big\langle\prod_{j=1}^k&&\!\!\!\!\!\!\!\! \mathcal{O}_{\Delta_j,\ell_j}(z_j,\bar z_j)
\mathcal{O}_{\Delta_{k+1},\ell_{k+1}^c}(z_{k+1},\bar z_{k+1})\cdots
\mathcal{O}_{\Delta_{i}+\ha,\ell_{i}}(z_{i},\bar z_{i})\cdots \mathcal{O}_{\Delta_N,\ell_N^c}(z_N,\bar z_N)\Big\rangle\ .\nonumber
\end{eqnarray}
We can also use Eq.(\ref{scurr3}) to read out the OPEs
\be S(z)\mathcal{O}_{\Delta,\ell^c}(w,\bar w) =\frac{1}{z-w}\mathcal{O}_{\Delta+\ha,\ell}(w,\bar w)+\makebox{regular},\label{soprod}\ee
which confirm that $S(z)$  is the supercurrent generating local supersymmetry transformations
associated to the $Q$ generator. Similarly, $\bar S(\bar z)$ generates $\bar Q$ transformations:
\be \bar S(\bz)\mathcal{O}_{\Delta,\ell}(w,\bar w) =\frac{1}{\bz-\bw}\mathcal{O}_{\Delta+\ha,\ell^c}(w,\bar w)+\makebox{regular}.\label{sboprod}\ee

\section{OPEs of super BMS generators}\label{sec:sbms}
\subsection{$TS$}
The calculation is similar to the calculation of $TT$ OPE in \cite{Foto1912} because it involves two shadow transforms. Here, we use the same notation and define
\be
\doublewidetilde{\!\!\cA}(z,w,z_2,\dots)=\int d^2 z_0{1\over (z_0-z)^{4}} \int d^2 z_1{1\over (w-z_1)^{3}}\cA_{\ell_0=-2,\ell_1 = -\frac{3}{2},\ell_2\dots\ell_k,\ell^c_{k+1}\dots \ell_N^c}\ .\ee
Then we have
\ba
\Big\langle  T(z) S(w) &&\!\!\!\!\!\!\!\!\! \prod_{m=2}^k\cO_{\D_m,\ell_m}(z_m,\zbar_m)\!\!\prod_{n=k+1}^N\cO_{\D_n,\ell_n^c}(z_n,\zbar_n)\Big \rangle\\ &=& -{3!\over 4\pi^2}\lim_{\Delta_1\rightarrow\frac{1}{2}}\lim_{\Delta_0\rightarrow 0}{\Delta_0} (\Delta_1-\frac{1}{2})\,\,\doublewidetilde{\!\!\cA}(z,w,z_2,\dots) \ .
\ea
We first take the $\Delta_0\rightarrow 0$ limit:
\begin{equation}
\cA_{\ell_0=-2,\ell_1 = -\frac{3}{2},\ell_2\dots \ell_N^c} \rightarrow \frac{1}{\Delta_0}\sum_{i=1}^{N}\frac{z_{0i}}{\bar{z}_{0i}} \frac{\bar{z}_{ri}}{\bar{z}_{r0}}\Big(  z_{0i}\partial_{z_i} -2h_i \Big) \cA_{\ell_1 = -\frac{3}{2},\ell_2\dots \ell_N^c} \ .
\end{equation}
The integral over $z_0$ can be evaluated in the same way as in \cite{Foto1912} :
\begin{eqnarray}
-\lim_{\Delta_0\rightarrow 0} \frac{3!}{4\pi} \Delta_0\,\,\,\doublewidetilde{\!\!\cA}(z,w,z_2,\dots) =&& \int d^2 z_1{1\over (w-z_1)^{3}} \Big[\frac{h_1}{(z-z_1)^2} +\frac{1}{z-z_1}\partial_{z_1} \Big] \mathcal{A}_{\ell_1 = -\frac{3}{2},\ell_2\dots \ell_N^c} \nonumber\\
&&+\sum_{i=2}^N\Big[  \frac{h_i}{(z-z_i)^2} +\frac{1}{z-z_i}\partial_{z_i}\Big]\widetilde{\mathcal{A}}_{\ell_1 = -\frac{3}{2},\ell_2\dots \ell_N^c} \ ,  \label{Adoublesh}
\end{eqnarray}
where
\begin{equation}
\widetilde{\mathcal{A}}_{\ell_1 = -\frac{3}{2},\ell_2\dots \ell_N^c} = \int d^2 z_1{1\over (w-z_1)^{3}}\mathcal{A}_{\ell_1 = -\frac{3}{2},\ell_2\dots \ell_N^c}\ .
\end{equation}

It is clear from Eq.(\ref{Adoublesh}) that the second term (involving the sum over $i\ge 2$) is non-singular in the limit of $w\rightarrow z$, therefore only the first term needs to be included in the  derivation of OPE:
\begin{eqnarray}
\Big\langle T(z) S(w)&&\!\!\!\!\!\!\!\!
\prod_{m=2}^k\cO_{\D_m,\ell_m}(z_m,\zbar_m)\prod_{n=k+1}^N\cO_{\D_n,\ell_n^c}(z_n,\zbar_n)
\Big \rangle\label{TSstep1}\\
&=&\lim_{\Delta_1\rightarrow \frac{1}{2}}\left(\frac{\Delta_1-\frac{1}{2}}{\pi}\right)\int d^2 z_1{1\over (w-z_1)^{3}} \Big[ \frac{h_1}{(z-z_1)^2} +\frac{1}{z-z_1}\partial_{z_1} \Big] \mathcal{A}_{\ell_1 = -\frac{3}{2},\ell_2\dots \ell_N^c} \ , \nonumber
\end{eqnarray}
where $h_1 = -\frac{1}{2}$ in the $\Delta_1\rightarrow \frac{1}{2}$ limit. To simplify the notation, we define
\begin{eqnarray}\mathcal{G}(z_1,\dots)\equiv
\left(\frac{\Delta_1-\frac{1}{2}}{\pi}\right)\mathcal{A}_{\ell_1 = -\frac{3}{2},\ell_2\dots \ell_N} \ea
and introduce the variables $Z = z-z_1,~ W = w-z_1$.
We rewrite Eq.(\ref{TSstep1}) as
\begin{eqnarray}
&&\lim_{\Delta_1 \rightarrow \frac{1}{2}} \int d^2 z_1 \Big[   \frac{-\frac{1}{2}}{W^3 Z^2} +\frac{1}{W^3 Z}\partial_{z_1}   \Big]\mathcal{G}(z_1,\dots)  \nonumber\\
&=&\lim_{\Delta_1 \rightarrow \frac{1}{2}}\int d^2 z_1\Big[   \frac{-\frac{1}{2}}{W^3 Z^2} + \Big(\frac{1}{W^3}-\frac{1}{W^2 Z} \Big) \frac{1}{z-w}\partial_{z_1} \Big]\mathcal{G}(z_1,\dots)\nonumber\\
&=& -\lim_{\Delta_1 \rightarrow \frac{1}{2}}\int d^2 z_1 \Big( \frac{1}{2W^3 Z^2}+\frac{1}{z-w}\frac{1}{W^2 Z}\partial_{z_1}\Big)\mathcal{G}(z_1,\dots) +\frac{1}{z-w}\langle \partial S(w) \cdots\rangle \label{TSstep2},
\end{eqnarray}
where in the last term, we used
\begin{eqnarray}\lim_{\Delta_1\rightarrow \frac{1}{2}}\frac{1}{z-w} \int \frac{d^2 z_1}{W^3} \partial_{z_1}\mathcal{G}(z_1,\dots)&=&\lim_{\Delta_1\rightarrow \frac{1}{2}} \frac{1}{z-w}\partial_w \int \frac{d^2 z_1}{W^3} \mathcal{G}(z_1,\dots)\\ &=&
\frac{1}{z-w} \langle \partial S(w)\prod_{n=2}^N\cO_{\D_n,\ell_n}(z_n,\zbar_n)\rangle . \nonumber
\end{eqnarray}
After performing integration by parts over $z_1$, the first term in the last line
of Eq.(\ref{TSstep2}) can be rewritten  as
\begin{eqnarray}
&&\lim_{\Delta_1\rightarrow \frac{1}{2}}\Big[ \frac{\frac{3}{2}}{(z-w)^2}\int \frac{d^2 z_1}{W^4} \mathcal{G}(z_1,\cdots) -\frac{3}{2} \frac{1}{(z-w)^2} \int \frac{d^2 z_1}{W Z^2} \mathcal{G}(z_1,\cdots)\Big] \nonumber\\
&&~~~~= \frac{\frac{3}{2}}{(z-w)^2} \langle S(w) \cdots\rangle -\frac{3}{2} \lim_{\Delta_1\rightarrow \frac{1}{2}} \frac{1}{(z-w)^2}\int \frac{d^2 z_1}{W Z^2} \mathcal{G}(z_1,\cdots)  \ .\label{TSstep3}
\end{eqnarray}
In this way, we obtain
\begin{eqnarray}
\langle T(z) S(w) \cdots\rangle
&=& \frac{3}{2}\frac{1}{(z-w)^2}\langle  S(w) \cdots\rangle+\frac{1}{z-w}\langle  \partial S(w) \cdots\rangle\label{TbG+step1}\\
&-& \frac{\frac{3}{2}}{(z-w)^2}\lim_{\Delta_1\rightarrow \frac{1}{2}}\left(\frac{\Delta_1-\frac{1}{2}}{\pi}\right)\int d^2z_1\frac{1}{WZ^2}\langle \mathcal{O}_{\Delta_1,\ell_1 = -\frac{3}{2}}(z_1,\bar{z}_1) \cdots\rangle.
\nonumber\end{eqnarray}

The first two terms are those expected for the OPE of the energy-momentum tensor with a dimension $\D={3\over 2}$ primary with $h = \frac{3}{2}$. The last term will be shown to vanish as a consequence of supersymmetric Ward identities. To that end, we use the soft gravitino limit $\Delta_1 \rightarrow \frac{1}{2}$ of Eq.(\ref{half2}):
\ba &&
\lim_{\Delta_\rightarrow \frac{1}{2}}(\Delta_1-\frac{1}{2})\Big\langle O_{\Delta_1,\ell_1 = -\frac{3}{2}}(z_1,\bar{z}_1) \prod_{m=2}^k\cO_{\D_m,\ell_m}(z_m,\zbar_m)\!\!\prod_{n=k+1}^N\cO_{\D_n,\ell_n^c}(z_n,\zbar_n)
\Big\rangle\\ && =\sum_{i=k+1}^{N}\frac{z_{1i}}{\bar{z}_{1i}}\frac{\bar{z}_{ri}}{\bar{z}_{r1}}(-1)^{\sigma_i}
\Big\langle\prod_{m=2}^k\cO_{\D_m,\ell_m}(z_m,\zbar_m)\cO_{\D_{k+1},\ell_{k+1}^c}(z_{k+1},
\zbar_{k+1}) \cdots \mathcal{O}_{\Delta_i+\frac{1}{2},\ell_i}(z_i)\cdots\Big\rangle \ . \nonumber
\ea
Here, it is convenient to choose the reference point $\bar{z}_r= \bar{w}$. Then the last term of Eq.(\ref{TbG+step1}) becomes
\begin{equation}
-\frac{3}{2\pi (z-w)^2}\sum_{i=k+1}^{N}\int d^2 z_1\frac{z_1-z_i}{(w-z_1)(z-z_1)^2}\frac{(\bar{w}-\bar{z}_i)
(-1)^{\sigma_i}}{(\bar{z}_1-\bar{z}_i)(\bar{w}-\bar{z}_1)}\langle \cdots O_{\Delta_i+\frac{1}{2},\ell_i}(z_i)\cdots\rangle \ .
\end{equation}
We can evaluate the integrals  by using \be
-\frac{1}{2\pi}\int d^2 z_1\frac{z_1-z_i}{(w-z_1)(z-z_1)^2}\frac{1}{(\bar{z}_1-\bar{z}_i)(\bar{w}-\bar{z}_1)} = \Gamma(0)\frac{w-z_i}{(z-w)^2(\bar{w}-\bar{z}_i)} \ .
\end{equation}
In this way, the last term of Eq.(\ref{TbG+step1}) becomes
\begin{equation}
\frac{3\Gamma(0)}{(z-w)^4}\!\!\sum_{i=k+1}^N(w-z_i)(-1)^{\sigma_i}
\Big\langle\!\!\prod_{m=2}^k\cO_{\D_m,\ell_m}(z_m,\zbar_m)\cO_{\D_{k+1},\ell_{k+1}^c}(z_{k+1},
\zbar_{k+1}) \!\cdots\! \mathcal{O}_{\Delta_i+\frac{1}{2},\ell_i}(z_i)\!\cdots\!\!\Big\rangle \ .
\nonumber \end{equation}
This sum is zero due to the supersymmetric Ward identity (\ref{scurr5}) with $\langle\eta q_i\rangle =w-z_i$. The final result is the expected OPE
\begin{equation}
T(z)S(w) = \frac{3}{2}\frac{S(w)}{(z-w)^2}+\frac{\partial S(w)}{z-w}+\makebox{regular} , \label{TG-}
\end{equation}
and similarly,
\begin{equation}
\overline{T}(\bar{z})\bar{S}(\bar{w}) = \frac{3}{2}\frac{\bar{S}(\bar{z})}{(\bar{z}-\bar{w})^2}+\frac{\bar{\partial}\bar{S}(\bar{w})}{\bar{z}-\bar{w}}\label{TbG+}
+\makebox{regular}.\end{equation}

By using the same methods as in \cite{Foto1912}, one can show that the products $T(z)\bar{S}(\bar{w})$ and $\overline{T}(\bar{z})S(w)$ lead to derivatives of $\delta$ functions and can be ignored in  OPEs. Thus the OPEs with $T$ and $\overline{T}$ confirm that $S(w)$ is a primary with conformal weights $h= \frac{3}{2}$, $\bar{h} = 0$ and that $ \bar{S}(\bar{w})$ is a primary with $h = 0$, $\bar{h} = \frac{3}{2}$.
\subsection{$SP$ and $\bar SP$}
The supertranslation current $P(w)$ is defined as a descendant of the $\Delta=1$ graviton operator:
\begin{equation}
P(w) = \lim_{\Delta\rightarrow 1}\Big(\frac{\Delta-1}{4}\Big)\partial_{\bar{z}}\mathcal{O}_{\Delta,\ell=+2}(z,\bar{z}) \ .
\end{equation}
We consider the following correlator:
\begin{eqnarray}
\Big\langle S(z) P(w) &&\!\!\!\!\prod_{m=2}^k\cO_{\D_m,\ell_m}(z_m,\zbar_m)\prod_{n=k+1}^N
\cO_{\D_n,\ell_n^c}(z_n,\zbar_n)\Big\rangle \label{SPstep1}\\
&&= \lim_{\Delta_0\rightarrow \frac{1}{2}}\lim_{\Delta_1\rightarrow 1}\Big( \frac{\Delta_0-\frac{1}{2}}{\pi} \Big)\Big( \frac{\Delta_1-1}{4}\Big)\int \frac{d^2 z_0}{(z-z_0)^3} \partial_{\bar{w}}\mathcal{A}_{\ell_0 = -\frac{3}{2},\ell_1 = +2,\ell_2\dots\ell_N^c}\ . \nonumber
\end{eqnarray}
We first take the $\Delta_0\rightarrow \frac{1}{2}$ limit.
 Since $\ell_0=-\frac{3}{2}$ and $\ell_1 = +2$,  there is no singularity as $z_0= w$, c.f.\ Eq.(\ref{scurr3}). After integrating  over $z_0$, the remaining terms become
\begin{eqnarray}
&&\Big\langle S(z) P(w) \prod_{m=2}^k\cO_{\D_m,\ell_m}(z_m,\zbar_m)\prod_{n=k+1}^N\cO_{\D_n,\ell_n^c}(z_n,\zbar_n)
\Big\rangle \\
&=& \lim_{\Delta_1\rightarrow 1} \Big( \frac{\Delta_1-1}{4}\Big) \sum_{i=k+1}^N (-1)^{\sigma_i}\frac{1}{z-z_i}\partial_{\bar{w}}\langle\mathcal{O}_{\Delta_1,\ell_1=+2}(w,\bar{w})\cdots  \mathcal{O}_{\Delta_i+\frac{1}{2},\ell_i}(z_i,\bar{z}_i) \cdots   \rangle \ . \nonumber
\end{eqnarray}
Here again, there is no singularity at $z= w$.
 Thus we have
\begin{equation}
S(z)P(w)\sim \mbox{regular}\label{G-P}
\end{equation}

Next, we  consider $\bar{S}(\bar{z})P(w)$, starting from the correlator:
\begin{eqnarray}
\Big\langle \bar S(\bar z) P(w) &&\!\!\!\!\prod_{m=2}^k\cO_{\D_m,\ell_m}(z_m,\zbar_m)\prod_{n=k+1}^N
\cO_{\D_n,\ell_n^c}(z_n,\zbar_n) \Big\rangle \label{SbPstep1}
 \\
&&= \lim_{\Delta_0\rightarrow \frac{1}{2}}\lim_{\Delta_1\rightarrow 1}\Big( \frac{\Delta_0-\frac{1}{2}}{\pi} \Big)\Big( \frac{\Delta_1-1}{4}\Big)\int \frac{d^2 z_0}{(\bar{z}-\bar{z}_0)^3}\partial_{\bar{w}}\mathcal{A}_{\ell_0 = +\frac{3}{2},\ell_1 = +2,\ell_2\dots\ell_N^c} \ .  \nonumber
\end{eqnarray}
After taking the $\Delta_0\rightarrow\frac{1}{2}$ gravitino limit, we obtain
\begin{eqnarray}
&& \lim_{\Delta_1\rightarrow 1} \Big( \frac{\Delta_1-1}{4}\Big) \partial_{\bar{w}} \frac{1}{\bz-\bw}\langle\mathcal{O}_{\Delta_{1}+\ha,\ell_1^c=\frac{3}{2}}(w,\bar{w})\cdots   \rangle\label{SbPstep2}\\
&&+ \lim_{\Delta_1\rightarrow 1} \Big( \frac{\Delta_1-1}{4}\Big) \sum_{i=2}^k (-1)^{\sigma_i}\frac{1}{\bz-\bz_i}\partial_{\bar{w}}\langle\mathcal{O}_{\Delta_1,\ell_1=+2}
(w,\bar{w})\cdots  \mathcal{O}_{\Delta_i+\frac{1}{2},\ell_i^c}(z_i,\bar{z}_i) \cdots   \rangle\nonumber .
\end{eqnarray}
The first term disappears in the $\Delta_1=1$ limit while the second term is finite at $z=w$. Hence
\begin{equation}
\bar{S}(\bar{z})P(w)\sim \mbox{regular}\label{G+P}
\end{equation}

\subsection{ $S\bar{S}$ and relation to the supertranslation operator ${\cal P}$}
We consider the correlator
\begin{eqnarray}
\Big\langle S(z)&&\!\!\!\!\bar{S}(\bar{w})\prod_{m=3}^k\cO_{\D_m,\ell_m}(z_m,\zbar_m)\prod_{n=k+1}^N
\cO_{\D_n,\ell_n^c}(z_n,\zbar_n) \Big\rangle= \nonumber \\&&  \label{SSbarstep1} =\lim_{\Delta_1\rightarrow \frac{1}{2}}\lim_{\Delta_2\rightarrow\frac{1}{2}} \frac{(\Delta_1-\frac{1}{2})(\Delta_2-\frac{1}{2})}{\pi^2}\int\frac{d^2z_1}{(z-z_1)^3}\int \frac{d^2z_2}{(\bar{w}-\bar{z}_2)^3} \\ \times\Big\langle &&\!\!\!\! \mathcal{O}_{\Delta_1,\ell_1=-\frac{3}{2}}(z_1,\bar{z}_1)
\mathcal{O}_{\Delta_2,\ell_2^c=\frac{3}{2}}(z_2,\bar{z}_2)
\prod_{m=3}^k\cO_{\D_m,\ell_m}(z_m,\zbar_m)\prod_{n=k+1}^N
\cO_{\D_n,\ell_n^c}(z_n,\zbar_n) \Big\rangle \ .
 \nonumber
\end{eqnarray}
This is a double soft limit with opposite helicities, therefore we expect that the result depends on the order in which the limits are taken.
We take $\Delta_1\rightarrow \frac{1}{2}$ first and perform the first shadow integral, to obtain
\begin{eqnarray}
&&\Big\langle S(z)\bar{S}(\bar{w}) \prod_{m=3}^k\cO_{\D_m,\ell_m}(z_m,\zbar_m)\prod_{n=k+1}^N
\cO_{\D_n,\ell_n^c}(z_n,\zbar_n) \Big\rangle=\lim_{\Delta_2\rightarrow\frac{1}{2}} \frac{(\Delta_2-\frac{1}{2})}{\pi} \int \frac{d^2z_2}{(\bar{w}-\bar{z}_2)^3}\nonumber\\
&&\times \bigg[ \frac{1}{z-z_2}\Big\langle \mathcal{O}_{\Delta_2+\frac{1}{2},\ell_2=2}(z_2,\bar{z}_2)\prod_{m=3}^k
\cO_{\D_m,\ell_m}(z_m,\zbar_m)\prod_{n=k+1}^N
\cO_{\D_n,\ell_n^c}(z_n,\zbar_n) \Big\rangle
\nonumber\\
&&~~~~ +\! \sum_{i=k+1}^N \frac{1}{z-z_i}(-1)^{\sigma_i}\Big\langle \mathcal{O}_{\Delta_2,\ell_2^c=\frac{3}{2}}(z_2,\bar{z}_2) \cdots \mathcal{O}_{\Delta_i+\frac{1}{2},\ell_i}(z_i,\bar{z}_i) \cdots \Big\rangle  \bigg]  \ . \label{SSbarstep2}
\end{eqnarray}
In the limit $\Delta_2\rightarrow \frac{1}{2}$, the dimension of graviton operator present in the first correlator inside the square bracket becomes $\D_2 +\frac{1}{2}\to 1$, therefore we can use
the leading soft graviton limit $(\D\to1)$ and perform the second shadow integral. This yields:
\begin{eqnarray}
\lim_{\Delta_2\rightarrow \frac{1}{2}}\!\! \frac{\Delta_2-\frac{1}{2}}{\pi}&&\!\!\!\!\!\!\!\! \int \!\! \frac{d^2z_2}{(\bar{w}-\bar{z}_2)^3}\frac{1}{z-z_2}\Big\langle \mathcal{O}_{\Delta_2+\frac{1}{2},\ell_2=2}(z_2,\bar{z}_2)
\prod_{m=3}^k
\cO_{\D_m,\ell_m}(z_m,\zbar_m)\!\!\prod_{n=k+1}^N
\cO_{\D_n,\ell_n^c}(z_n,\zbar_n) \Big\rangle\nonumber\\
&=&\frac{1}{\pi}\int \frac{d^2z_2}{(\bar{w}-\bar{z}_2)^3(z-z_2)}\sum_{i=3}^N\frac{\bar{z}_2-\bar{z}_i}{z_2-z_i}
\Big\langle \mathcal{O}_{\Delta_3,\ell_3}\cdots \mathcal{O}_{\Delta_i+1,\ell_i}\cdots \mathcal{O}_{\Delta_N,\ell_N^c}\Big\rangle\\
=\sum_{i=3}^N&&\!\!\!\!\!\!\!\! \bigg[ \frac{1}{(\bar{w}-\bar{z})^2}\frac{\bar{z}-\bar{z}_i}{z-z_i}+\frac{1}{\bar{z}-\bar{w}}
\frac{1}{z-z_i}+\frac{1}{\bar{w}-\bar{z}_i}\frac{1}{z-z_i}\bigg]\Big\langle \mathcal{O}_{\Delta_3,\ell_3}\!\cdots \mathcal{O}_{\Delta_i+1,\ell_i}\!\cdots \mathcal{O}_{\Delta_N,\ell_N^c}\Big\rangle\nonumber . \label{SSbarstep3}
\end{eqnarray}
The  $\Delta_2\rightarrow \frac{1}{2}$ limit of the second correlator inside the square bracket in Eq.(\ref{SSbarstep2}) is the leading soft gravitino limit, which yields
\begin{eqnarray}
&&\lim_{\Delta_2\rightarrow \frac{1}{2}}\frac{\Delta_2-\frac{1}{2}}{\pi} \int \frac{d^2z_2}{(\bar{w}-\bar{z}_2)^3}
\sum_{i=k+1}^N \frac{1}{z-z_i}(-1)^{\sigma_i}
\Big\langle \mathcal{O}_{\Delta_2,\ell_2^c=\frac{3}{2}}(z_2,\bar{z}_2) \cdots \mathcal{O}_{\Delta_i+\frac{1}{2},\ell_i}(z_i,\bar{z}_i) \cdots \!\Big\rangle \nonumber\\
&&~~~~=-\frac{1}{\pi}\int \frac{d^2z_2}{(\bar{w}-\bar{z}_2)^3} \bigg[\sum_{i=k+1}^N \frac{1}{z-z_i}\frac{\bar{z}_{2i}}{z_{2i}}\Big\langle
\mathcal{O}_{\Delta_3,\ell_3} (z_3,\bar{z}_3)\cdots \mathcal{O}_{\Delta_i+1,\ell_i^c}(z_i,\bar{z}_i)\cdots \!\Big\rangle \label{SSbarstep4}\\
&&+\! \sum_{i=k+1}^N \,\,\sum_{j = 3}^{k}\frac{1}{z-z_i}(-1)^{\sigma_i+\sigma_j}
\frac{\bar{z}_{2j}}{z_{2j}}\Big\langle  \mathcal{O}_{\Delta_3,\ell_3}\cdots \mathcal{O}_{\Delta_j+\frac{1}{2},\ell_j^c}\cdots
\mathcal{O}_{\Delta_i+\frac{1}{2},\ell_i}\cdots
\mathcal{O}_{\Delta_N\ell_N^c}\Big\rangle\bigg] \ . \nonumber
\end{eqnarray}
After performing the shadow integral, the r.h.s.\ becomes
\begin{eqnarray}
&&~~~~-\sum_{i=k+1}^N \frac{1}{z-z_i}\frac{1}{\bar{w}-\bar{z}_i}\Big\langle
\mathcal{O}_{\Delta_3,\ell_3} (z_3,\bar{z}_3)\cdots \mathcal{O}_{\Delta_i+1,\ell_i^c}(z_i,\bar{z}_i)\cdots \!\Big\rangle \label{SSbarstep5}\\
&&+\! \sum_{i=k+1}^N \,\,\sum_{j = 3}^{k}(-1)^{\sigma_i+\sigma_j}\frac{1}{z-z_i}\frac{1}{\bar{w}-\bar{z}_j}\Big\langle  \mathcal{O}_{\Delta_3,\ell_3} \cdots \mathcal{O}_{\Delta_j+\frac{1}{2},\ell_j^c}\cdots
\mathcal{O}_{\Delta_i+\frac{1}{2},\ell_i}\cdots
\mathcal{O}_{\Delta_N\ell_N^c}\Big\rangle \ . \nonumber
\end{eqnarray}
Combining Eqs.(\ref{SSbarstep2})-(\ref{SSbarstep5}), we obtain
\begin{eqnarray}
&&\Big\langle S(z)\bar{S}(\bar{w}) \prod_{m=3}^k\cO_{\D_m,\ell_m}(z_m,\zbar_m)\prod_{n=k+1}^N
\cO_{\D_n,\ell_n^c}(z_n,\zbar_n) \Big\rangle=\nonumber\\
=~\sum_{i=3}^N&&\!\!\! \bigg[ \frac{1}{(\bar{w}-\bar{z})^2}\frac{\bar{z}-\bar{z}_i}{z-z_i}+\frac{1}{\bar{z}-\bar{w}}
\frac{1}{z-z_i}\bigg]\Big\langle \mathcal{O}_{\Delta_3,\ell_3}\!\cdots \mathcal{O}_{\Delta_i+1,\ell_i}\!\cdots \mathcal{O}_{\Delta_N,\ell_N^c}\Big\rangle\label{SSbarresult}\\
&&+ \,\sum_{i=3}^k\frac{1}{\bar{w}-\bar{z}_i}\frac{1}{z-z_i}\Big\langle \mathcal{O}_{\Delta_3,\ell_3}\!\cdots \mathcal{O}_{\Delta_i+1,\ell_i}\!\cdots \mathcal{O}_{\Delta_N,\ell_N^c}\Big\rangle\nonumber\\
&&+ \sum_{i=k+1}^N \,\,\sum_{j = 3}^{k}(-1)^{\sigma_i+\sigma_j}\frac{1}{z-z_i}\frac{1}{\bar{w}-\bar{z}_j}\Big\langle  \mathcal{O}_{\Delta_3,\ell_3} \cdots \mathcal{O}_{\Delta_j+\frac{1}{2},\ell_j^c}\cdots
\mathcal{O}_{\Delta_i+\frac{1}{2},\ell_i}\cdots
\mathcal{O}_{\Delta_N\ell_N^c}\Big\rangle  \ . \nonumber
\end{eqnarray}
Notice that the above expression involves anti-holomorphic poles at $\bz=\bw$, but it does not contain the holomorphic ones. This reflects the order of soft limits: $\D_1\to \ha$ first, followed by $\D_2\to \ha$. In order to construct an order-independent quantity, we consider
\begin{eqnarray}
\Big\langle \bar S(\bar z)&&\!\!\!\!{S}({w})\prod_{m=3}^k\cO_{\D_m,\ell_m}(z_m,\zbar_m)\prod_{n=k+1}^N
\cO_{\D_n,\ell_n^c}(z_n,\zbar_n) \Big\rangle= \nonumber \\&&  \label{SSbarstep11} =\lim_{\Delta_1\rightarrow \frac{1}{2}}\lim_{\Delta_2\rightarrow\frac{1}{2}} \frac{(\Delta_1-\frac{1}{2})(\Delta_2-\frac{1}{2})}{\pi^2}\int\frac{d^2z_1}{(z-z_1)^3}\int \frac{d^2z_2}{(\bar{w}-\bar{z}_2)^3}\\ \times\Big\langle &&\!\!\!\!
\mathcal{O}_{\Delta_1,\ell_1^c=\frac{3}{2}}(z_1,\bar{z}_1)
\mathcal{O}_{\Delta_2,\ell_2=-\frac{3}{2}}(z_2,\bar{z}_2)
\prod_{m=3}^k\cO_{\D_m,\ell_m}(z_m,\zbar_m)\prod_{n=k+1}^N
\cO_{\D_n,\ell_n^c}(z_n,\zbar_n) \Big\rangle \ .
 \nonumber
\end{eqnarray}
After repeating the same steps as before, we find
\begin{eqnarray}
&&\Big\langle \bar S(\bz){S}({w}) \prod_{m=3}^k\cO_{\D_m,\ell_m}(z_m,\zbar_m)\prod_{n=k+1}^N
\cO_{\D_n,\ell_n^c}(z_n,\zbar_n) \Big\rangle=\nonumber\\
=~\sum_{i=3}^N&&\!\!\! \bigg[ \frac{1}{(w-z)^2}\frac{{z}-{z}_i}{\bz-\bz_i}+\frac{1}{z-w}
\frac{1}{\bz-\bz_i}\bigg]\Big\langle \mathcal{O}_{\Delta_3,\ell_3}\!\cdots \mathcal{O}_{\Delta_i+1,\ell_i}\!\cdots \mathcal{O}_{\Delta_N,\ell_N^c}\Big\rangle\label{SSbarresult2}\\
&&+ \,\sum_{i=k+1}^N\frac{1}{w-{z}_i}\frac{1}{\bz-\bz_i}\Big\langle \mathcal{O}_{\Delta_3,\ell_3}\!\cdots \mathcal{O}_{\Delta_i+1,\ell_i}\!\cdots \mathcal{O}_{\Delta_N,\ell_N^c}\Big\rangle\nonumber\\
&&- \sum_{i=k+1}^N \,\,\sum_{j = 3}^{k}(-1)^{\sigma_i+\sigma_j}\frac{1}{\bz-\bz_j}\frac{1}{w-{z}_i}\Big\langle  \mathcal{O}_{\Delta_3,\ell_3} \cdots \mathcal{O}_{\Delta_j+\frac{1}{2},\ell_j^c}\cdots
\mathcal{O}_{\Delta_i+\frac{1}{2},\ell_i}\cdots
\mathcal{O}_{\Delta_N\ell_N^c}\Big\rangle  \ . \nonumber
\end{eqnarray}
Note opposite signs of the last terms in Eqs.(\ref{SSbarresult}) and (\ref{SSbarresult2}) which are due to the ordering of operators. $S$ acts on the right cluster of $\{\ell^c\}$'s
while $\bar S$  on the left cluster of $\{\ell\}$'s. Once we add Eqs.(\ref{SSbarresult}) and (\ref{SSbarresult2}), we find a combination that does not depends on the order of limits:
\begin{eqnarray}
&&\Big\langle [S(z)\bar{S}(\bar{w}) + \bar S(\bz){S}({w})] \prod_{m=3}^k\cO_{\D_m,\ell_m}(z_m,\zbar_m)\prod_{n=k+1}^N
\cO_{\D_n,\ell_n^c}(z_n,\zbar_n) \Big\rangle=\nonumber\\
=\bigg\{\sum_{i=3}^N&&\!\!\! \bigg[ \frac{1}{(\bar{w}-\bar{z})^2}\frac{\bar{z}-\bar{z}_i}{z-z_i}
 +\frac{1}{(w-z)^2}\frac{{z}-{z}_i}{\bz-\bz_i}
+\frac{1}{\bar{z}-\bar{w}}\frac{1}{z-z_i}+\frac{1}{z-w}
\frac{1}{\bz-\bz_i}
\bigg]\label{SSbarresult3}\\
&+&\sum_{i=3}^k\frac{1}{\bar{w}-\bar{z}_i}\frac{1}{z-z_i}
+ \,\sum_{i=k+1}^N\frac{1}{w-{z}_i}\frac{1}{\bz-\bz_i}\bigg\}\Big\langle \mathcal{O}_{\Delta_3,\ell_3}\!\cdots \mathcal{O}_{\Delta_i+1,\ell_i}\!\cdots \mathcal{O}_{\Delta_N,\ell_N^c}\Big\rangle\nonumber\\
&+& \sum_{i=k+1}^N \,\,\sum_{j = 3}^{k}(-1)^{\sigma_i+\sigma_j}\bigg[\frac{1}{z-z_i}\frac{1}{\bar{w}-\bar{z}_j}
-\frac{1}{\bz-\bz_j}\frac{1}{w-{z}_i}\bigg]\Big\langle
\cdots \mathcal{O}_{\Delta_j+\frac{1}{2},\ell_j^c}\cdots
\mathcal{O}_{\Delta_i+\frac{1}{2},\ell_i}\cdots
\Big\rangle \ . \nonumber
\end{eqnarray}
The singular terms involve insertions of the graviton operators. We can compare them with the known graviton correlation functions. As a result, we obtain the following OPE:
\ba
S(z)\bar{S}(\bar{w}) + \bar S(\bz){S}({w})&=&\lim_{\D\to 1}(\D-1)\bigg[\frac{1}{(\bar{z}-\bar{w})^2}\mathcal{O}_{\Delta,\ell=+2}(z,\bar{z})
+\frac{1}{(z-w)^2}\mathcal{O}_{\Delta,\ell^c=-2}(z,\bar{z})\bigg]\nonumber\\
&&+\,\frac{1}{\bar{z}-\bar{w}}4P(z)+\frac{1}{z-w}4\bar P(\bz)+\makebox{regular}.\label{sbarsope}
\ea
From Eq.(\ref{SSbarresult3}) we can also extract the finite piece remaining at $z=w$:
\begin{eqnarray}
&&\Big\langle \!:\![S(z)\bar{S}(\bar{z}) + \bar S(\bz){S}({z})]\!: \prod_{m=3}^k\cO_{\D_m,\ell_m}(z_m,\zbar_m)\prod_{n=k+1}^N
\cO_{\D_n,\ell_n^c}(z_n,\zbar_n) \Big\rangle=\nonumber\\
&&~~~~~~~~~=
\sum_{i=3}^N\frac{1}{z-{z}_i}\frac{1}{\bz-\bz_i}\Big\langle \mathcal{O}_{\Delta_3,\ell_3}\!\cdots \mathcal{O}_{\Delta_i+1,\ell_i}\!\cdots \mathcal{O}_{\Delta_N,\ell_N^c}\Big\rangle\label{spresult}
\end{eqnarray}
We recognize this as the correlator of the BMS supertranslation operator ${\cal P}(z, \bz)$ \cite{Barnich1703}, a primary field  ($h=\frac{3}{2}$, $\bar{h} = \frac{3}{2}$) defined by the Laurent expansion:
\be\label{pfield}
{\cal P}(z,\bz) \equiv \sum\limits_{n,m \in \mathbb{Z}} P_{n-\ha, m-\ha}z^{-n-1} \bz^{-m-1}\ ,
\ee
where $P_{n-\ha, m-\ha}$ generate supertranslations. Its OPEs with other primaries read
\be {\cal P}(w,\bw)\mathcal{O}_{h, \bh}(z, \bz)=\frac{1}{w-z}\frac{1}{\bw-\bz}\mathcal{O}_{h+\ha, \bh+\ha}(z, \bz)+\makebox{regular}\ .\ee
We conclude that
\begin{equation}
: \! S(z)\bar{S}(\bar{z}) +\bar{S}(\bar{z})S(z) \!  : \;=~ \mathcal{P}(z,\bar{z})\ .
\end{equation}

\subsection{$SS$ and $\bar{S}\bar{S}$}
We consider the correlator
\begin{eqnarray}
\Big\langle S(z)&&\!\!\!\!S(w)\prod_{m=3}^k\cO_{\D_m,\ell_m}(z_m,\zbar_m)\prod_{n=k+1}^N
\cO_{\D_n,\ell_n^c}(z_n,\zbar_n) \Big\rangle= \nonumber \\&&  \label{SSstep1} =\lim_{\Delta_1\rightarrow \frac{1}{2}}\lim_{\Delta_2\rightarrow\frac{1}{2}} \frac{(\Delta_1-\frac{1}{2})(\Delta_2-\frac{1}{2})}{\pi^2}\int\frac{d^2z_1}{(z-z_1)^3}\int \frac{d^2z_2}{(w-z_2)^3} \\ \times\Big\langle &&\!\!\!\! \mathcal{O}_{\Delta_1,\ell_1=-\frac{3}{2}}(z_1,\bar{z}_1)
\mathcal{O}_{\Delta_2,\ell_2=-\frac{3}{2}}(z_2,\bar{z}_2)
\prod_{m=3}^k\cO_{\D_m,\ell_m}(z_m,\zbar_m)\prod_{n=k+1}^N
\cO_{\D_n,\ell_n^c}(z_n,\zbar_n) \Big\rangle \ .
 \nonumber
\end{eqnarray}
This is a double soft limit with identical helicities, therefore the result does not depend on the order in which the limits are taken.
We take $\Delta_1\rightarrow \frac{1}{2}$ first and perform the first shadow integral, to obtain
\begin{eqnarray}
&&\Big\langle S(z)S(w) \prod_{m=3}^k\cO_{\D_m,\ell_m}(z_m,\zbar_m)\prod_{n=k+1}^N
\cO_{\D_n,\ell_n^c}(z_n,\zbar_n) \Big\rangle=\lim_{\Delta_2\rightarrow\frac{1}{2}} \frac{(\Delta_2-\frac{1}{2})}{\pi} \int \frac{d^2z_2}{(w-z_2)^3}\nonumber\\
&&~~~~  \sum_{i=k+1}^N \frac{1}{z-z_i}(-1)^{\sigma_i}\Big\langle \mathcal{O}_{\Delta_2,\ell_2=-\frac{3}{2}}(z_2,\bar{z}_2) \cdots \mathcal{O}_{\Delta_i+\frac{1}{2},\ell_i}(z_i,\bar{z}_i) \cdots \Big\rangle    \ . \label{SSstep2}
\end{eqnarray}
Next, we take $\Delta_2\rightarrow \frac{1}{2}$ and perform the second shadow integral. This yields
\begin{eqnarray}
&&\lim_{\Delta_2\rightarrow\frac{1}{2}} \frac{(\Delta_2-\frac{1}{2})}{\pi} \int \frac{d^2z_2}{(w-z_2)^3}
\sum_{i=k+1}^N \frac{1}{z-z_i}(-1)^{\sigma_i}\Big\langle \mathcal{O}_{\Delta_2,\ell_2=-\frac{3}{2}}(z_2,\bar{z}_2) \cdots \mathcal{O}_{\Delta_i+\frac{1}{2},\ell_i}(z_i,\bar{z}_i) \cdots \Big\rangle   \nonumber\\
&&=\sum_{i=k+1}^N \, \,\sum_{j\neq i,j=k+1}^N\frac{1}{z-z_i}\frac{1}{w-z_j}(-1)^{\sigma_i+\sigma_j}\Big\langle \mathcal{O}_{\Delta_3,\ell_3} \cdots \mathcal{O}_{\Delta_i+\frac{1}{2},\ell_i} \cdots\mathcal{O}_{\Delta_j+\frac{1}{2},\ell_j}\cdots \Big\rangle \ . \label{SSresult}
\end{eqnarray}
We see that there are no singularities when
 $z\rightarrow w$, therefore
\begin{equation}
S(z)S(w)\sim \mbox{regular} \ .\label {ssreg}
\end{equation}
Similarly,
\begin{equation}
\bar{S}(\bar{z})\bar{S}(\bar{w})\sim \mbox{regular} \ .\label{ssreg2}
\end{equation}
\section{The algebra $\mathfrak{sbms}_4$ of super $BMS_4$ generators}\label{sec:modexp}
As shown in the previous section, the supercurrent $S(z)$ is a primary field
with chiral weights $h = \frac{3}{2}$, $\bar{h} = 0$. We can write it in the form of a Laurent expansion:
\begin{eqnarray}\label{eq:tGlaurent-}
S(z) = \sum_{n\in \mathbb{Z}+\frac{1}{2}}\frac{G_n}{z^{n+\frac{3}{2}}}\ ,\qquad
\mbox{with}\quad G_n = \oint dz \, z^{n+1/2}\, S(z)\ .
\end{eqnarray}
Similarly,
\begin{eqnarray}\label{eq:tGlaurent+}
\bar S(\bz) = \sum_{n\in \mathbb{Z}+\frac{1}{2}}\frac{\bar G_n}{z^{n+\frac{3}{2}}}\ ,\qquad
\mbox{with}\quad\bar G_n = \oint d\bz \, \bz^{n+1/2}\, \bar S(\bz)\ .
\end{eqnarray}
The OPEs of Eqs.(\ref{soprod}) and (\ref{sboprod}) imply
\begin{eqnarray}
\,[G_n,\mathcal{O}_{\Delta,\ell^c }(w,\bar{w})] &=& w^{n+1/2}\,\mathcal{O}_{\Delta+\frac{1}{2},\,\ell}(w,\bar{w})\label{G-mode1},\\
\label{G+mode1}
\,[\bar{G}_n,\mathcal{O}_{\Delta,\ell}(w,\bar{w})]&=& \bar{w}^{n+1/2}\,\mathcal{O}_{\Delta+\frac{1}{2},\,\ell^c}(w,\bar{w}).
\end{eqnarray}
By comparing with Eqs.(\ref{qsusy}-\ref{qbarsusy}), we can make the following identification between SUSY generators and supercurrent modes:
\begin{eqnarray}
Q_{1}&\rightarrow& G_{+1/2} \ ,\qquad Q_{2}\rightarrow G_{-1/2}  \ , \nonumber\\[1mm]
\bar{Q}_{\dot{1}}&\rightarrow& \bar{G}_{+1/2}\ ,\qquad \bar{Q}_{\dot{2}}\rightarrow \bar{G}_{-1/2}\ .  \nonumber
\end{eqnarray}
These are the four-dimensional SUSY generators  realization in CCFT. In addition to superrotations and superstranslations, $\mathfrak{sbms}_4$ contains ``super'' supersymmetries generated by the infinite set of generators  $G_n$ and $\bar{G}_m$.

If we apply (\ref{G-mode1}) and (\ref{G+mode1}) consecutively to a primary operator, we find
\begin{eqnarray}
\,[\{ G_n, \bar{G}_m\},\,\mathcal{O}_{\Delta}(z,\bar{z})] =z^{n+\frac{1}{2}}\,\bar{z}^{m+\frac{1}{2}}\mathcal{O}_{\Delta+1}(z,\bar{z})=[P_{n,m},\,
\mathcal{O}_{\Delta}(z,\bar{z})]\label{G-G+mode} \ ,
\end{eqnarray}
where $P_{n,m}$ are modes of the supertranslation operator \cite{Foto1912}
${\cal P}(z,\bz)$, c.f.\ Eq.(\ref{pfield}). We conclude that
\be \label{eq:tGtGanticom}
\{G_n,\bar{G}_m\} = P_{n,m} \ .
\ee
Furthermore, from Eqs.(\ref{ssreg}) and (\ref{ssreg2}) it follows that
\be
\{G_n,G_m\} = \{\bar G_n,\bar{G}_m\}=0 \ .
\ee

The holomorphic and anti-holomorphic supertranslation currents were previously expanded
in \cite{Foto1912} as
\begin{equation}
P(z) = \sum_{n\in \mathbb{Z}}\hat{P}_{n-\frac{1}{2}}z^{-n-1}\ ,\qquad
\bar{P}(\bar{z}) = \sum_{n\in \mathbb{Z}}\hat{\bar{P}}_{n-\frac{1}{2}}\bar{z}^{-n-1}
\end{equation}
These modes form a subset of the mode expansion (\ref{pfield}) of $\mathcal{P}(z,\bz)$:
$\hat{P}_{n-\frac{1}{2}} = \frac{1}{4}P_{n-\frac{1}{2},-\frac{1}{2}}$ and
$\hat{\bar{P}}_{n-\frac{1}{2}} = \frac{1}{4}P_{-\frac{1}{2},n-\frac{1}{2}}$. Then from Eqs.(\ref{G-P})  and (\ref{G+P}) it follows that
\begin{eqnarray}\label{eq:comtGP}
\big[G_{m},P_{n-\frac{1}{2},-\frac{1}{2}}] =[\bar{G}_{m},P_{n-\frac{1}{2},-\frac{1}{2}}]= [G_{m},P_{-\frac{1}{2},n-\frac{1}{2}}] =
[\bar{G}_{m},P_{-\frac{1}{2},n-\frac{1}{2}}] =0\ .
\end{eqnarray}
The commutators of $G$'s with the remaining supertranslation generators can be obtained by successive applications of the operators to generic primary operators. In this way, we find
\begin{eqnarray}
\lbrack G_n, P_{k,l}\rbrack ~=~
\lbrack \bar{G}_m, P_{k,l}\rbrack ~=~ 0\label{G+Pmode}\ .
\end{eqnarray}

The well-known mode expansions of the energy momentum tensor are
\begin{eqnarray}
T(z)= \sum_m \frac{L_m}{z^{m+2}}\ ,\qquad
\overline{T}(\bar{z}) &=& \sum_m\frac{\bar{L}_m}{\bar{z}^{m+2}}\ ,
\end{eqnarray}
where $L_m$ and $\bar L_n$ are the Virasoro operators.
Then from Eqs.(\ref{TG-}) and (\ref{TbG+}), we find the following algebra:
\begin{eqnarray}\label{eq:comtGT}
\left[L_m,G_n\right] &=&(\frac{1}{2}m-n)G_{m+n} \ , \\ \nonumber
\left[\bar{L}_m,\bar{G}_n\right] &=& (\frac{1}{2}m-n)\bar{G}_{m+n}\ .
\end{eqnarray}
On the other hand, the OPEs of
$T\bar{S}$ and $\overline{T}S$ are regular, therefore
\begin{eqnarray}
[L_m, \bar{G}_n] = [\bar{L}_m,G_n] = 0\ .
\end{eqnarray}

After collecting all of the above we obtain the following $\mathfrak{sbms_4}$ algebra:
\begin{eqnarray}
&& \{G_m,\bar{G}_n\} = P_{m,n}\nonumber\\[1mm]
&& \{G_m,{G}_n\} = \{\bar G_m,\bar{G}_n\} ~=~ 0\nonumber\\[1mm]
&& \lbrack P_{k,l},G_n\rbrack=
\lbrack P_{k,l},\bar{G}_m\rbrack ~=~ 0\label{smbs}\\[1mm]
&& \left[L_m,G_k\right] =\Big(\frac{1}{2}m -k\Big)G_{m+k}\nonumber \\ &&
\left[\bar{L}_m, \bar{G}_l\right] =\Big(\frac{1}{2}m -l\Big)\bar{G}_{m+l}\nonumber\\[1.5mm]
&& [L_m, \bar{G}_n] = [\bar{L}_m,G_n] ~=~ 0 \ ,
\nonumber
\end{eqnarray}
together with the remaining commutators of $\mathfrak{bms_4}$ \cite{Foto1912}.
This is an infinite-dimensional  symmetry algebra of ${\cal N}=1$ supersymmetric theory on ${\cal CS}_2$. It can be compared with the supersymmetric BMS algebra in three dimensions \cite{Barnich1407, Lodato1610,Fuentealba1706} and in four dimensions \cite{Awada1985}.
In this context, infinite-dimensional extension of supersymmetry appears as a ``square root'' of supertranslations. Unlike in superstring theory, there is no ``world-sheet'' two-dimensional supersymmetry on ${\cal CS}_2$; it is possible though that it can appear in some limits of celestial amplitudes \cite{Stieberger1806}.

\section{Conclusions}\label{sec:conclusions}
In the present paper, we extended our earlier work \cite{Foto1912} to include supersymmetry. We used on-shell supersymmetry transformations to construct chiral and gauge supermultiplets of conformal primary wave functions. The oscillator mode expansion of the fields reveals that in order to have a consistent description of the theory one needs to extend the theory to include states with conformal dimensions beyond the normalizable  Re$(\D) = 1$. In the bosonic case this was the case due to the action of supertranslations on the primary states of the theory. Here we see that supersymmetry leads to the same conclusion. This is even more evident when one considers supersymmetric Ward identities. Their Mellin transforms  lead to relations between fermionic and bosonic correlators on ${\cal CS}_2$ with Re$(\D) = \ha$. The nature of these states is not yet clear \cite{Donnay2005} and requires further investigation.

We also discussed in detail fermionic conformal soft theorems, both leading and subleading and the associated CCFT Ward identities. We exhibited an intricate pattern of supersymmetric Ward identities that relates fermionic and bosonic soft theorems, at both leading and subleading levels. It would be very interesting to understand if this chain of relations is sufficient to prove that all soft theorems, leading and subleading are not renormalized in supersymmetric theories \cite{He1405,Cachazo1405,Bern1406}.

By using soft theorems, we constructed  $\mathfrak{sbms_4}$ --  ${\cal N}{=}1$ supersymmetric version of the extended BMS algebra. Most of the previous studies of CCFT have concentrated on the bosonic sector of the theory. One of the main questions regarding the bosonic CCFT is the status of the symmetries under quantum corrections and in particular the value of the central charge. Given the usual non-renormalization theorems for supersymmetric theories in 4d, we hope that our work will be a useful step towards addressing the fate of extended BMS symmetry beyond the tree level.

Another interesting aspect of our work is the relation between the supertranslation field ${\cal P}(z,\zbar)$ and two-dimensional supercurrents $S(z)$ and $\bar{S}(\bar{z})$. This field appears in the product of holomorphic and anti-holomorphic supercurrents. Previously,  in \cite{Foto1912}, we were able to identify the supertranslation currents $P(z)$ and $ \bar{P}(\bz)$ as the descendants of conformally soft graviton modes. No such state was found for ${\cal P}(z,\bz)$. Now we learn that in supersymmetric theory,  ${\cal P}(z,\bz)$, which generates all superstranslations, can be identified as a composite operator.

There were several puzzles encountered in the course of this work. One, already mentioned before, is the role of states with Re$(\D)\neq 1$. Are they physical, independent states? If yes, what do they represent?
Another striking point is that the subleading gravitino and sub-subleading graviton limits do not seem to play any role in the construction of  $\mathfrak{sbms_4}$. Do they generate further extension of the symmetry algebra? These are just two examples of many questions that need to be addressed in order to develop CCFT into a viable candidate for a holographic description of four-dimensional physics.
\\[3mm]
\leftline{\noindent{\bf Acknowledgments}}
\vskip 1mm
\noindent
We are grateful to Nima Arkani-Hamed, Glenn Barnich, Monica Pate, Andrea Puhm, Ana-Maria Raclariu and Andy Strominger for stimulating conversations.
This material is based in part upon work supported by the National Science Foundation
under Grant Number PHY--1913328.
Any opinions, findings, and conclusions or recommendations
expressed in this material are those of the authors and do not necessarily
reflect the views of the National Science Foundation.

\newpage

\renewcommand{\thesection}{A}
\setcounter{equation}{0}
\renewcommand{\theequation}{A.\arabic{equation}}
\renewcommand{\thesection}{A}
\setcounter{equation}{0}
\renewcommand{\theequation}{A.\arabic{equation}}
\vskip 3mm
\section{Soft and collinear limits of gauginos and  gravitinos}\label{App:col}
\subsection{Soft gaugino limit}
We start from the soft limit of gaugino in ${\cal N}{=}1$ supersymmetric Einstein-Yang-Mills theory (SEYM). The singular contributions to a soft gaugino emission arise from Feynman diagrams in which internal gluon goes on-shell, shown in Fig.(\ref{soft2gaugino}) or an internal gaugino goes on-shell, shown in Fig.(\ref{softgginoboson}). To be specific, we assume that the soft gaugino, which carries momentum $p_1\to 0$, has positive helicity.

\begin{figure}[h]
\begin{center}
\begin{picture}(300,100)(0,-10)
\SetColor{Black}
\Text(100,90)[l]{$a$}
\Text(200,90)[l]{$b$}
\Text(130,30)[l]{$c$}
\Text(130,50)[l]{$c$}
\Photon(150,20)(150,60){5}{4}\Vertex(150,60){2}
\ArrowLine(170,30)(170,48)
\Text(175,38)[l]{$p_1+p_2$}
\ArrowLine(150,60)(120,90)
\ArrowLine(120,65)(104,80)
\Text(95,70)[l]{$p_1$}
\ArrowLine(180,90)(150,60)
\ArrowLine(175,65)(190,80)
\Text(200,65)[l]{$p_2$}
\GCirc(150,0){22}{0.5}
\Vertex(125,18){1}
\Vertex(122.5,10.5){1}
\Vertex(120,3){1}
\Vertex(175,18){1}
\Vertex(178.5,10.5){1}
\Vertex(180,3){1}
\Vertex(135,-26){1}
\Vertex(150,-26){1}
\Vertex(165,-26){1}
\end{picture}
\end{center}
\caption{Feynman diagrams leading to soft gaugino singularities.}\label{soft2gaugino}
\end{figure}
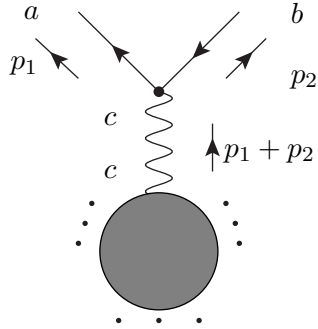

 The interactions of gauginos and gauge boson is given by the vertex:
\be
V_{\tilde{g}\tilde{g}g}^{\mu}= if^{abc}\sigma^{\mu} \ ,
\ee
where $\sigma^{\mu}=(\mathbb{1},\vec{\sigma})$. To evaluate the contribution from Fig.(\ref{soft2gaugino}),  we define the splitting vector:
\begin{equation}
S^{\mu} = \tilde{\lambda}_1\lambda_2 D^{\mu\alpha}(p_1+p_2)V_{\tilde{g}\tilde{g}g}(p_1,p_2)_{\alpha} \ ,
\end{equation}
where
\begin{eqnarray}
\tilde{\lambda}_1 = |1], \qquad   \lambda_2 = \langle 2| \ .   \qquad
\end{eqnarray}
$D^{\mu\alpha}$ is the gauge boson propagator
\be
D^{\mu\alpha}(p_1+p_2) = \frac{-i}{(p_1+p_2)^2}g^{\mu\alpha} \ .
\ee
When $p_1$ becomes soft, the gauge boson propagator goes on-shell. In the soft limit,
the splitting vector becomes
\begin{equation}
\lim_{\omega_1\rightarrow 0} S^{\mu}(+\frac{1}{2}, -\frac{1}{2}) = f^{abc}\frac{1}{z_{12}\sqrt{\omega_1\omega_2}}\epsilon_2^{-\mu} \ , 
\label{spl1}
\end{equation}
where $\epsilon_2^{-\mu}$ denotes $\epsilon^{\mu}_{2,\ell_2=-1}$. Notice that the the leading order of the soft limit is ${\cal O}(\frac{1}{\sqrt{\omega_1}})$.

\begin{figure}[h]
\begin{center}
\begin{picture}(300,100)(0,-10)
\SetColor{Black}
\Text(100,90)[l]{$a$}
\Text(200,90)[l]{$b$}
\Text(130,30)[l]{$c$}
\Text(130,50)[l]{$c$}
\ArrowLine(150,20)(150,60)\Vertex(150,60){2}
\ArrowLine(170,30)(170,48)
\Text(175,38)[l]{$p_1+p_2$}
\ArrowLine(150,60)(120,90)
\ArrowLine(120,65)(104,80)
\Text(95,70)[l]{$p_1$}
\Photon(150,60)(180,90){5}{4}
\ArrowLine(175,70)(190,85)
\Text(190,70)[l]{$p_2$}
\GCirc(150,0){22}{0.5}
\Vertex(125,18){1}
\Vertex(122.5,10.5){1}
\Vertex(120,3){1}
\Vertex(175,18){1}
\Vertex(178.5,10.5){1}
\Vertex(180,3){1}
\Vertex(135,-26){1}
\Vertex(150,-26){1}
\Vertex(165,-26){1}
\end{picture}
\end{center}
\caption{Feynman diagrams leading to soft gaugino singularities.}\label{softgginoboson}
\end{figure}
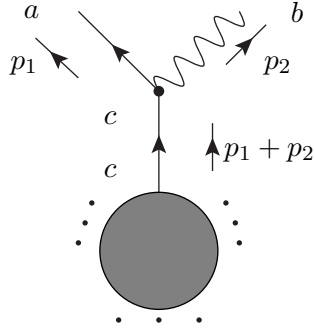

Next to evaluate the contribution from Fig.(\ref{softgginoboson}), we define the splitting spinor:
\be
S = \tilde{\lambda}_1\epsilon_2^{+\mu}V_{\tilde{g}g\tilde{g}}(p_1,p_2)_{\mu}D(p_1+p_2) \ ,
\ee
where $\epsilon_2^{+\mu}$ denotes $\epsilon^{\mu}_{2,\ell_2 = +1}$. $D(p_1+p_2)$ is the gaugino propagator
\ba
D(p_1+p_2) &=& i\frac{\lambda_P\tilde{\lambda}_P}{(p_1+p_2)^2}\ ,\label{fpro}
\ea
with the numerator factorized assuming that $P=p_1+p_2$ is also on-shell, $P^2=0$.
In the soft limit,
\be
\lim_{\omega_1\rightarrow 0}S(+\frac{1}{2},+1)= f^{abc}\frac{1}{z_{12}\sqrt{\omega_1\omega_2}}\tilde{\lambda}_2 \ .
\label{spl2}
\ee

Other possible gaugino production channels involve gravitational interactions. They are shown in Figs.(\ref{softgginograviton})-(\ref{softgginogvitino}). It is easy to check that these channels do not contribute at the   ${\cal O}(\frac{1}{\sqrt{\omega_1}})$ order.\par
\begin{figure}[h]
\begin{center}
\begin{picture}(300,100)(0,-10)
\SetColor{Black}
\ArrowLine(150,20)(150,60)\Vertex(150,60){2}
\ArrowLine(170,30)(170,48)
\Text(175,38)[l]{$p_1+p_2$}
\ArrowLine(150,60)(120,90)
\ArrowLine(120,65)(104,80)
\Text(95,70)[l]{$p_1$}
\Photon(150,60)(180,90){5}{4}
\Photon(152,60)(182,90){5}{4}
\ArrowLine(175,70)(190,85)
\Text(190,70)[l]{$p_2$}
\GCirc(150,0){22}{0.5}
\Vertex(125,18){1}
\Vertex(122.5,10.5){1}
\Vertex(120,3){1}
\Vertex(175,18){1}
\Vertex(178.5,10.5){1}
\Vertex(180,3){1}
\Vertex(135,-26){1}
\Vertex(150,-26){1}
\Vertex(165,-26){1}
\end{picture}
\end{center}
\caption{Double wavy line represents the graviton}\label{softgginograviton}
\end{figure}
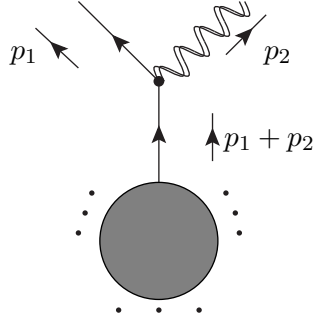
\begin{figure}[h]
\begin{center}
\begin{picture}(300,100)(0,-10)
\SetColor{Black}
\Photon(150,20)(150,60){5}{4}
\Photon(152,20)(152,60){5}{4}
\Vertex(150,60){2}
\ArrowLine(170,30)(170,48)
\Text(175,38)[l]{$p_1+p_2$}
\ArrowLine(150,60)(120,90)
\ArrowLine(120,65)(104,80)
\Text(95,70)[l]{$p_1$}
\ArrowLine(180,90)(150,60)
\ArrowLine(175,70)(190,85)
\Text(190,70)[l]{$p_2$}
\GCirc(150,0){22}{0.5}
\Vertex(125,18){1}
\Vertex(122.5,10.5){1}
\Vertex(120,3){1}
\Vertex(175,18){1}
\Vertex(178.5,10.5){1}
\Vertex(180,3){1}
\Vertex(135,-26){1}
\Vertex(150,-26){1}
\Vertex(165,-26){1}
\end{picture}
\end{center}
\caption{Virtual graviton decaying into a gaugino pair.}
\end{figure}
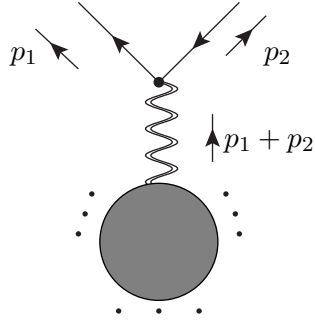
\begin{figure}[h]
\begin{center}
\begin{picture}(300,100)(0,-10)
\SetColor{Black}
\Photon(150,20)(150,60){5}{4}
\ArrowLine(150,20)(150,60)
\Vertex(150,60){2}
\ArrowLine(170,30)(170,48)
\Text(175,38)[l]{$p_1+p_2$}
\ArrowLine(150,60)(120,90)
\ArrowLine(120,65)(104,80)
\Text(95,70)[l]{$p_1$}
\Photon(180,90)(150,60){5}{4}
\ArrowLine(175,70)(190,85)
\Text(190,70)[l]{$p_2$}
\GCirc(150,0){22}{0.5}
\Vertex(125,18){1}
\Vertex(122.5,10.5){1}
\Vertex(120,3){1}
\Vertex(175,18){1}
\Vertex(178.5,10.5){1}
\Vertex(180,3){1}
\Vertex(135,-26){1}
\Vertex(150,-26){1}
\Vertex(165,-26){1}
\end{picture}
\end{center}
\caption{The double wavy-solid line represents the gravitino decaying into a gauge boson and gaugino.}
\end{figure}
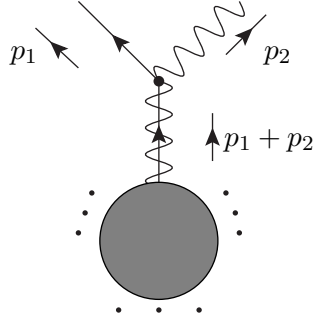
\begin{figure}[h]
\begin{center}
\begin{picture}(300,100)(0,-10)
\SetColor{Black}
\Photon(150,20)(150,60){5}{4}
\Vertex(150,60){2}
\ArrowLine(170,30)(170,48)
\Text(175,38)[l]{$p_1+p_2$}
\ArrowLine(150,60)(120,90)
\ArrowLine(120,65)(104,80)
\Text(95,70)[l]{$p_1$}
\Photon(180,90)(150,60){5}{4}
\ArrowLine(180,90)(150,60)
\ArrowLine(175,70)(190,85)
\Text(190,70)[l]{$p_2$}
\GCirc(150,0){22}{0.5}
\Vertex(125,18){1}
\Vertex(122.5,10.5){1}
\Vertex(120,3){1}
\Vertex(175,18){1}
\Vertex(178.5,10.5){1}
\Vertex(180,3){1}
\Vertex(135,-26){1}
\Vertex(150,-26){1}
\Vertex(165,-26){1}
\end{picture}
\end{center}
\caption{Virtual gauge boson decaying into gaugino and gravitino.}\label{softgginogvitino}
\end{figure}
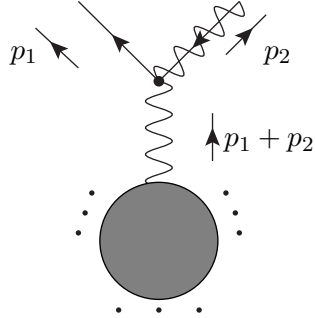\noindent
Note that in all singular contributions, the helicity of the non-soft particle is decreased by $\frac{1}{2}$. After summing all contributions of partial amplitudes with color factors, we find the leading soft gaugino limit:
\begin{equation}
\lim_{\omega_s\rightarrow0} \mathcal{M}(s^{a,+\frac{1}{2}},1^{a_1,\ell_2},\cdots,N^{a_N,\ell_N}) =\frac{1}{\sqrt{\omega_s}}\sum_{i=1}^Nf^{aa_ic}\frac{1}{(z_s-z_i)
\sqrt{\omega_i}}(-1)^{\sigma_i}\mathcal{M}(\cdots i^{c,\ell_i-\frac{1}{2}},\cdots) \ ,
\end{equation}
where the non-zero terms on the right hand side require $\ell_i-\frac{1}{2} = +\frac{1}{2}$ or $\ell_i-\frac{1}{2} = -1$. Here,
$\sigma_i$ is the number of fermions {\em preceding} particle $i$, i.e.\ the number of fermion creation/annihilation operators inserted to the left of the operator creating or annihilating soft gaugino.
After performing Mellin transforms, we find the $\Delta_s\rightarrow \frac{1}{2}$ limit of the celestial amplitude:
\begin{eqnarray}
&&\mathcal{A}_{\ell_s = +\frac{1}{2}, \ell_1,\dots,\ell_N}(\Delta_s,\Delta_1\dots\Delta_N)\nonumber\\
&&\rightarrow \frac{1}{\Delta_s-\frac{1}{2}}\sum_{i=1}^N\frac{f^{aa_ic}}{z_s-z_i}(-1)^{\sigma_i}\mathcal{A}_{\ell_1,\dots \ell_i-\frac{1}{2},\dots\ell_N}(\Delta_1,\dots,\Delta_i-\frac{1}{2},\dots,\Delta_N) \ .
\end{eqnarray}
%

\subsection{Soft gravitino limits}

Feynman diagrams contributing to  soft singularities are shown in Figs.(\ref{softgravitino&gauge}) and (\ref{softgravitino&gaugino}).
We assume that the soft gravitino, which carries momentum $p_1\to 0$, has helicity $+\frac{3}{2}$.
\par
To evaluate Fig.(\ref{softgravitino&gauge}),  we use the gauge boson-gaugino-gravitino vertex and the following polarization vectors for gravitino and gauge boson: 
\begin{eqnarray}
&&\text{Gravitino:} \qquad\epsilon^{\mu}_{\ell = +\frac{3}{2}}(p_1,r) = \epsilon^{\mu}_{\ell=+1}(p_1,r) |1] = \epsilon^{+\mu}_1|1] \ , 
\end{eqnarray}
with the polarization vectors
\be \epsilon^\mu_{\ell=+1}(p,r)=\frac{\langle r|\sigma^\mu| p]}{\sqrt{2}\langle rp\rangle} \ .\qquad
\ee
where $\langle r|$ is a reference spinor.\par
The expression for the vertex is given by
\begin{equation}
V_{\mu\nu}(p_1,p_2) = \kappa\left[ \frac{i}{2}(-(p_2)_{\mu}\sigma_{\nu}+\eta_{\mu\nu}p_2\cdot \sigma)+\frac{1}{2}p_2^{\rho}\epsilon_{\mu\nu\rho\kappa}\sigma^{\kappa} \right]  \ , \label{gravitinogaugevert}
\end{equation}
where $\sigma^{\mu}=(\mathbb{1},\vec{\sigma})$. In our conventions, $\kappa = 2$. 
\begin{figure}[h]
\begin{center}
\begin{picture}(300,100)(0,-10)
\SetColor{Black}
\ArrowLine(150,20)(150,60)\Vertex(150,60){2}
\ArrowLine(170,30)(170,48)
\Text(175,38)[l]{$p_1+p_2$}
\ArrowLine(150,60)(120,90)
\Photon(150,60)(120,90){5}{4}
\ArrowLine(120,65)(104,80)
\Text(95,70)[l]{$p_1$}
\Text(95,90)[l]{$\mu$}
\Photon(150,60)(180,90){5}{4}
\ArrowLine(175,70)(190,85)
\Text(190,70)[l]{$p_2$}
\Text(190,90)[l]{$\nu$}
\GCirc(150,0){22}{0.5}
\Vertex(125,18){1}
\Vertex(122.5,10.5){1}
\Vertex(120,3){1}
\Vertex(175,18){1}
\Vertex(178.5,10.5){1}
\Vertex(180,3){1}
\Vertex(135,-26){1}
\Vertex(150,-26){1}
\Vertex(165,-26){1}
\end{picture}
\end{center}
\caption{Feynman diagrams leading to soft gravitino singularities.}\label{softgravitino&gauge}
\end{figure}
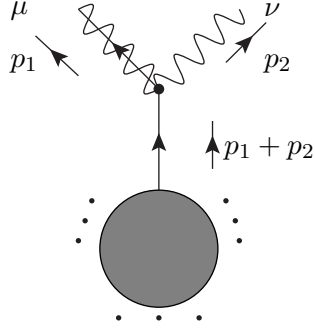
The splitting spinor of this diagram is defined as
\begin{equation}
S(+\frac{3}{2},+1) = \epsilon_{1,\ell_1=+\frac{3}{2}}^{\mu} \epsilon_{2}^{+\nu} V_{\mu\nu}(p_1,p_2) D(p_1+p_2) \ ,
\end{equation}
with the fermion propagator given in Eq.(\ref{fpro}).
After contracting with the polarization vectors, we obtain
\begin{eqnarray}
S(+\frac{3}{2},+1)
&=&\frac{[21]}{\langle 12\rangle}\frac{\langle P r\rangle}{\langle 1 r\rangle}|P]\nonumber\\
&=& \frac{\bar{z}_{12}}{z_{12}}\frac{z_{Pr}}{z_{1r}}\Big(\frac{\omega_P}{\omega_1}\Big)^{\frac{1}{2}}|P] \ .
\end{eqnarray}
When $\omega_1\rightarrow 0$, the leading term is of order ${\cal O}(\frac{1}{\sqrt{\omega_1}})$:

\begin{equation}
\lim_{\omega_1\rightarrow 0}S(+\frac{3}{2},+1) = \frac{\bar{z}_{12}}{z_{12}}\frac{z_{2r}}{z_{1r}}\Big(\frac{\omega_2}{\omega_1}\Big)^{\frac{1}{2}}|2] \ .
\end{equation}
It is also easy to see that if
the helicity of the gauge boson is $-1$, there is no ${\cal O}(\frac{1}{\sqrt{\omega_1}})$ term.

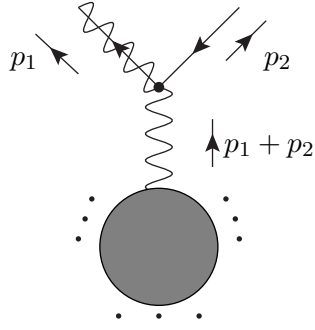
\begin{figure}[h]
\begin{center}
\begin{picture}(300,100)(0,-10)
\SetColor{Black}
\Photon(150,20)(150,60){5}{4}
\Vertex(150,60){2}
\ArrowLine(170,30)(170,48)
\Text(175,38)[l]{$p_1+p_2$}
\ArrowLine(150,60)(120,90)
\Photon(150,60)(120,90){5}{4}
\ArrowLine(120,65)(104,80)
\Text(95,70)[l]{$p_1$}
\ArrowLine(180,90)(150,60)
\ArrowLine(175,70)(190,85)
\Text(190,70)[l]{$p_2$}
\GCirc(150,0){22}{0.5}
\Vertex(125,18){1}
\Vertex(122.5,10.5){1}
\Vertex(120,3){1}
\Vertex(175,18){1}
\Vertex(178.5,10.5){1}
\Vertex(180,3){1}
\Vertex(135,-26){1}
\Vertex(150,-26){1}
\Vertex(165,-26){1}
\end{picture}
\end{center}
\caption{Feynman diagrams leading to soft gravitino singularities.}\label{softgravitino&gaugino}
\end{figure}

Next to evaluate the contribution from Fig.(\ref{softgravitino&gaugino}), consider the case when the gravitino has positive helicity and the gaugino has negative helicity. The corresponding splitting vector is
\begin{equation}
S^{\gamma}(+\frac{3}{2},-\frac{1}{2}) = \epsilon_{1,\ell_1=+\frac{3}{2}}^{\mu}\,\lambda_2\,V(p_1,p_2)_{\mu\nu}D^{\nu\gamma}(p_1+p_2) \ .
\end{equation}
We find
\begin{equation}
S^{\gamma}\Big(+\frac{3}{2},-\frac{1}{2}\Big) = \frac{[21]}{\langle 12\rangle}\frac{\langle 2r\rangle}{\langle 1r\rangle} \epsilon_2^{-\gamma} \ ,
\end{equation}
hence
\begin{equation}
\lim_{\omega_1\rightarrow0}S^{\gamma}\Big(+\frac{3}{2},-\frac{1}{2}\Big) = \frac{\bar{z}_{12}}{z_{12}}\frac{z_{2r}}{z_{1r}}\Big(\frac{\omega_2}{\omega_1}\Big)^{\frac{1}{2}}\epsilon_2^{-\gamma} \ .
\end{equation}
If the helicity of the gaugino is $+\frac{1}{2}$, then there is no ${\cal O}(\frac{1}{\sqrt{\omega_1}})$ term.
Now we can write down the full expression of soft gravitino limit of amplitudes containing one soft gravitino with other particles being gauginos and gauge bosons:
\begin{eqnarray}
&&\lim_{\omega_s\rightarrow 0}\mathcal{M}(p_s\, \ell_s = +\frac{3}{2}, p_1 \, \ell_1, \cdots ,p_N \, \ell_N)\nonumber\\
 &=&  \sum_{i=1}^N  \frac{\bar{z}_{si}}{z_{si}}\frac{z_{ri}}{z_{rs}}\Big(\frac{\omega_i}{\omega_s}\Big)^{\frac{1}{2}}(-1)^{\sigma_i}\mathcal{M}(p_1\,\ell_1,\cdots p_i\,\ell_i-\frac{1}{2},\cdots,p_N,\,\ell_N) +\mathcal{O}(\omega_s^{\frac{1}{2}})\ , \label{softgravi1}
\end{eqnarray}
where $z_r$ is the reference point. Note that on the right hand side $\ell_i-\frac{1}{2}$ can be equal to either $\frac{1}{2}$ or $-1$.
As shown in \cite{Avery1512}, the soft gravitino limit expression above is also valid for amplitudes that contain multiple gravitini and gravitons. Here we rederive this result by using Feynman diagrams. The additional Feynman diagrams are shown in Figs.(\ref{softgravitinosugra1}) and (\ref{softgravitinosugra2}).
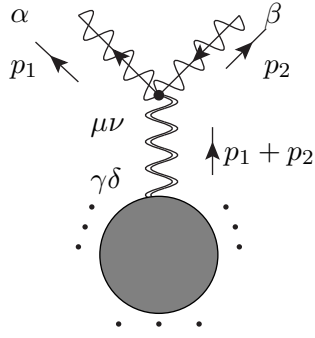
\begin{figure}[h]
\begin{center}
\begin{picture}(300,100)(0,-10)
\SetColor{Black}
\Photon(150,20)(150,60){5}{4}
\Photon(152,20)(152,60){5}{4}
\Vertex(150,60){2}
\ArrowLine(170,30)(170,48)
\Text(175,38)[l]{$p_1+p_2$}
\Text(125,48)[l]{$\mu\nu$}
\Text(125,28)[l]{$\gamma\delta$}
\ArrowLine(150,60)(120,90)
\Photon(150,60)(120,90){5}{4}
\ArrowLine(120,65)(104,80)
\Text(95,70)[l]{$p_1$}
\Text(95,90)[l]{$\alpha$}
\ArrowLine(180,90)(150,60)
\Photon(180,90)(150,60){5}{4}
\ArrowLine(175,70)(190,85)
\Text(190,70)[l]{$p_2$}
\Text(190,90)[l]{$\beta$}
\GCirc(150,0){22}{0.5}
\Vertex(125,18){1}
\Vertex(122.5,10.5){1}
\Vertex(120,3){1}
\Vertex(175,18){1}
\Vertex(178.5,10.5){1}
\Vertex(180,3){1}
\Vertex(135,-26){1}
\Vertex(150,-26){1}
\Vertex(165,-26){1}
\end{picture}
\end{center}
\caption{Feynman diagrams leading to soft gravitino singularities interacting with non-soft gravitino.}\label{softgravitinosugra1}
\end{figure}

The contribution of Fig.(\ref{softgravitinosugra1}) involves the splitting tensor $S^{\gamma\delta}$:
\begin{equation}
S^{\gamma\delta} = \epsilon_{1\alpha, \ell_1 = \pm\frac{3}{2}}\,\epsilon_{2\beta, \ell_2 = \mp\frac{3}{2}}V^{\alpha\beta,\mu\nu}(p_1,p_2)D_{\mu\nu}^{\gamma\delta}(p_1+p_2) \ ,
\end{equation}
where the three point vertex $V^{\alpha\beta,\mu\nu}(p_1,p_2)$ is given by \cite{Bjerrum1002}:
\begin{equation}
V^{\alpha\beta,\mu\nu}(p_1,p_2) = i\frac{\kappa}{2}\sigma^{\mu}((p_1-p_2)^{\nu}g^{\alpha\beta}+ (2p_2+p_1)^{\alpha}g^{\beta\nu}+(-2p_1-p_2)^{\beta}g^{\nu\alpha})\ . \label{gravitivert}
\end{equation}
The graviton propagator is
\begin{equation}
D_{\mu\nu}^{\gamma\delta}(p_1+p_2) = \frac{i}{2}(\delta^{\mu}_{\gamma}\delta^{\nu}_{\delta}+\delta^{\mu}_{\delta}\delta^{\nu}_{\gamma}-g^{\mu\nu}g_{\gamma\delta})\frac{1}{(p_1+p_2)^2}\ . \label{graprop}
\end{equation}
The second (hard) gravitino must carry helicity $-\frac{3}{2}$, therefore
\begin{equation}
\epsilon^{\alpha}_{1,\ell_1 =+ \frac{3}{2}} = \epsilon_1^{+\alpha}|1], \qquad \epsilon^{\beta}_{2,\ell_2 = -\frac{3}{2}} = \epsilon_2^{-\beta}|2\rangle \ .
\end{equation}
We find that when $\omega_1\rightarrow 0$, the splitting tensor
\begin{eqnarray}
\lim_{\omega_1\rightarrow0}S^{\gamma\delta}(+\frac{3}{2},-\frac{3}{2}) &=&\frac{\bar{z}_{12}}{z_{12}}\frac{z_{r2}}{z_{r1}}\Big(\frac{\omega_2}{\omega_1}\Big)^{\frac{1}{2}}  \epsilon_2^{-\gamma}\epsilon_2^{-\delta}\nonumber\\
&=&\frac{\bar{z}_{12}}{z_{12}}\frac{z_{r2}}{z_{r1}}\Big(\frac{\omega_2}{\omega_1}\Big)^{\frac{1}{2}} \epsilon^{\gamma\delta}_{2,\ell_2 = -2} \ .
\end{eqnarray}
\begin{figure}[h]
\begin{center}
\begin{picture}(300,100)(0,-10)
\SetColor{Black}
\Photon(150,20)(150,60){5}{4}
\ArrowLine(150,20)(150,60)
\Vertex(150,60){2}
\ArrowLine(170,30)(170,48)
\Text(175,38)[l]{$p_1+p_2$}
\Text(135,50)[l]{$\beta$}
\Text(135,25)[l]{$\gamma$}
\ArrowLine(150,60)(120,90)
\Photon(150,60)(120,90){5}{4}
\ArrowLine(120,65)(104,80)
\Text(95,70)[l]{$p_1$}
\Text(95,90)[l]{$\alpha$}
\Photon(152,60)(182,90){5}{4}
\Photon(180,90)(150,60){5}{4}
\ArrowLine(175,70)(190,85)
\Text(190,70)[l]{$p_2$}
\Text(190,90)[l]{$\mu\nu$}
\GCirc(150,0){22}{0.5}
\Vertex(125,18){1}
\Vertex(122.5,10.5){1}
\Vertex(120,3){1}
\Vertex(175,18){1}
\Vertex(178.5,10.5){1}
\Vertex(180,3){1}
\Vertex(135,-26){1}
\Vertex(150,-26){1}
\Vertex(165,-26){1}
\end{picture}
\end{center}
\caption{Feynman diagrams leading to soft gravitino singularities interacting with non-soft graviton.}\label{softgravitinosugra2}
\end{figure}

Next, to evaluate the contribution from Fig.(\ref{softgravitinosugra2}), we define the splitting spinors
\begin{equation}
S^\gamma = \epsilon_{1\alpha}\epsilon_{2\mu\nu}V^{\alpha,\mu\nu,\beta}(p_1,p_2)D_{\beta}^{\gamma}(p_1+p_2) \ ,
\end{equation}
where the vertex is related to eq.(\ref{gravitivert}) by crossing and the gravitino propagator is
\begin{eqnarray}
D_{\beta}^{\gamma}(p_1+p_2) &=& D(p_1+p_2)\delta_{\beta}^{\gamma}\ .
\end{eqnarray}
Let us assume that the external graviton carries helicity $+2$. Then
\begin{equation}
\epsilon^{\alpha}_{1,\ell_1 =+ \frac{3}{2}} = \epsilon_1^{+\alpha}|1], \qquad \epsilon^{\mu\nu}_{2,\ell_2 = +2} = \epsilon_2^{+\mu}\epsilon_2^{+\nu} \ .
\end{equation}
We find that when $\omega_1\rightarrow 0$, the splitting spinor
\begin{eqnarray}
\lim_{\omega_1\rightarrow0}S^\gamma(+\frac{3}{2},+2) &=& \frac{\bar{z}_{12}}{z_{12}}\frac{z_{r2}}{z_{r1}}\Big(\frac{\omega_2}{\omega_1}\Big)^{\frac{1}{2}} \epsilon_2^{+\gamma}|2]\nonumber\\
&=& \frac{\bar{z}_{12}}{z_{12}}\frac{z_{r2}}{z_{r1}}\Big(\frac{\omega_2}{\omega_1}\Big)^{\frac{1}{2}} \epsilon_2^{\gamma}(\ell = +\frac{3}{2}) \ .
\end{eqnarray}
In the case of $S^{\gamma}(+\frac{3}{2},-2)$, it is easy to check there is no $O(\frac{1}{\sqrt{\omega_1}})$term  in the soft gravitino limit.\par
Now we are ready to generalize Eq.(\ref{softgravi1}) to the amplitudes involving also gravitinos and gravitons:
\begin{eqnarray}
&&\lim_{\omega_s\rightarrow 0}\mathcal{M}(p_s\, \ell_s = +\frac{3}{2}, p_1 \, \ell_1, \cdots ,p_N \, \ell_N)\nonumber\\
 &=&  \sum_{i=1}^N  \frac{\bar{z}_{si}}{z_{si}}\frac{z_{ri}}{z_{rs}}\Big(\frac{\omega_i}{\omega_s}\Big)^{\frac{1}{2}}(-1)^{\sigma_i}\mathcal{M}(p_1\,\ell_1,\cdots p_i\,\ell_i-\frac{1}{2},\cdots,p_N,\,\ell_N)+\mathcal{O}(\omega_s^{\frac{1}{2}})\ , \label{softgravi2}
\end{eqnarray}
where on the right hand side, $\ell_i-\frac{1}{2}$ must be equal to $\frac{1}{2}$  for gaugino, $-1$ for gauge boson, $-2$ for graviton or $+\frac{3}{2}$ for gravitino, therefore belong to the set of $\{\ell^c\}$ defined in (\ref{lsusy}). Other helicities of these particles do not appear in the soft limit of gravitino with helicity $+\frac{3}{2}$. The complement set of $\{\ell\}$ appears in the soft limit of helicity $-\frac{3}{2}$.

\subsection{Collinear limit: Kinematics}
In this section the kinematics of collinear limits will be reviewed. As in Ref.\cite{Foto1912}, we use the conventions of \cite{tt}.  We will parametrize two light-like momenta $p_1$ and $p_2$ and introduce two light-like vectors $P$ and $r$ in the following way \cite{Stieberger:2015kia}:
\ba\lambda_{1}=\lambda_P\cos\theta-\epsilon\lambda_r\sin\theta\ ,
&&\tilde\lambda_{1}=\tilde\lambda_P\cos\theta
-\tilde\epsilon\tilde\lambda_r\sin\theta\ ,\label{l11}
\\[1mm]
\lambda_{2}=\lambda_P\sin\theta+\epsilon\lambda_r\cos\theta\ ,
&&\tilde\lambda_{2}=\tilde\lambda_P\sin\theta+
\tilde\epsilon\tilde\lambda_r\cos\theta\ ,\label{l22}
\ea
thus we have
\ba p_{1}& ={\bf c}^2P-{\bf s c} (\epsilon
\lambda_r\tilde\lambda_P+\tilde\epsilon\lambda_P\tilde\lambda_r)+ \epsilon\tilde\epsilon{\bf s}^2 r\ ,\label{pon}\\[1mm]
p_{2}& ={\bf s}^2P+{\bf sc} (
\epsilon\lambda_r\tilde\lambda_P+\tilde\epsilon\lambda_P\tilde\lambda_r)+
\epsilon\tilde\epsilon{\bf c}^2 r\ ,\label{ptw}
\ea
where
\be {\bf{c}}\equiv\cos\theta=\sqrt{x}~,\qquad\qquad {\bf{s}}\equiv\sin\theta=\sqrt{1-x}\ .\ee
Then
\begin{equation}
\langle 12\rangle=\epsilon\,\langle Pr\rangle~, \qquad [12]=\tilde\epsilon\,[Pr]~.
\end{equation}
The collinear limit of $p_1$ and $p_2$ occurs when $\epsilon = \tilde{\epsilon} = 0$. In this limit,
\be \omega_P = \omega_1+\omega_2 \ee
\be
{\bf c}^2 = \frac{\omega_1}{\omega_P}, \qquad {\bf s}^2 = \frac{\omega_2}{\omega_P}~.
\ee

\subsection{Collinear limits involving gaugino}
We first we consider the collinear limit of two gauginos. The Feymann diagram contributing to the collinear singularities is shown in Fig.(\ref{2gauginofig}).
\begin{figure}[h]
\begin{center}
\begin{picture}(300,100)(0,-10)
\SetColor{Black}
\Text(100,90)[l]{$a$}
\Text(200,90)[l]{$b$}
\Text(130,30)[l]{$c$}
\Text(130,50)[l]{$c$}
\Photon(150,20)(150,60){5}{4}\Vertex(150,60){2}
\ArrowLine(170,30)(170,48)
\Text(175,38)[l]{$p_1+p_2$}
\ArrowLine(120,90)(150,60)
\ArrowLine(120,65)(104,80)
\Text(95,70)[l]{$p_1$}
\ArrowLine(150,60)(180,90)
\ArrowLine(175,65)(190,80)
\Text(200,65)[l]{$p_2$}
\GCirc(150,0){22}{0.5}
\Vertex(125,18){1}
\Vertex(122.5,10.5){1}
\Vertex(120,3){1}
\Vertex(175,18){1}
\Vertex(178.5,10.5){1}
\Vertex(180,3){1}
\Vertex(135,-26){1}
\Vertex(150,-26){1}
\Vertex(165,-26){1}
\end{picture}
\end{center}
\caption{Feynman diagrams leading to collinear gaugino singularities.}\label{2gauginofig}
\end{figure}
Assume that particle 1 is a gaugino with negative helicity and particle 2 is a gaugino with positive helicity. We define the splitting vector:
\begin{equation}
S^{\mu} = \lambda_1\tilde{\lambda}_2 D^{\mu\alpha}(p_1+p_2)V_{\tilde{g}\tilde{g}g}(p_1,p_2)_{\alpha} \ ,
\end{equation}
where $D^{\mu,\alpha}$ is the gauge boson propagator
\be
D^{\mu\alpha}(p_1+p_2) = \frac{-i}{(p_1+p_2)^2}g^{\mu\alpha}
\label{gaugeprop}
\ee
and the vertex is
\be
V_{\tilde{g}\tilde{g}g}^{\mu}= if^{abc}\sigma^{\mu} \ .
\ee
After inserting $\lambda_1$ and $\tilde{\lambda}_2$ of Eqs.(\ref{l11}) and (\ref{l22}) we find that at the leading order,
\ba
S^{\mu}(-\frac{1}{2},+\frac{1}{2}) &=& f^{abc}\Big( \frac{1}{\bar{z}_{12}\sqrt{\omega_1\omega_2}}{\bf s^2}\epsilon_P^{+\mu} + \frac{1}{z_{12}\sqrt{\omega_1\omega_2} }{\bf c}^2\epsilon_P^{-\mu}  \Big)\nonumber\\
&=&f^{abc}\Big( \frac{1}{\bar{z}_{12}}\frac{\omega_2^{\frac{1}{2}}}{\omega_1^{\frac{1}{2}}\omega_P}\epsilon_P^{+\mu} + \frac{1}{z_{12} }\frac{\omega_1^{\frac{1}{2}}}{\omega_2^{\frac{1}{2}}\omega_P}\epsilon_P^{-\mu}  \Big) \ .
\ea
Next, we include the gravitational production channel shown in Fig.(\ref{2ggino1gra}).
\begin{figure}[h]
\begin{center}
\begin{picture}(300,100)(0,-10)
\SetColor{Black}
\Photon(150,20)(150,60){5}{4}
\Photon(152,20)(152,60){5}{4}
\Vertex(150,60){2}
\ArrowLine(170,30)(170,48)
\Text(175,38)[l]{$p_1+p_2$}
\ArrowLine(120,90)(150,60)
\ArrowLine(120,65)(104,80)
\Text(95,70)[l]{$p_1$}
\ArrowLine(150,60)(180,90)
\ArrowLine(175,70)(190,85)
\Text(190,70)[l]{$p_2$}
\GCirc(150,0){22}{0.5}
\Vertex(125,18){1}
\Vertex(122.5,10.5){1}
\Vertex(120,3){1}
\Vertex(175,18){1}
\Vertex(178.5,10.5){1}
\Vertex(180,3){1}
\Vertex(135,-26){1}
\Vertex(150,-26){1}
\Vertex(165,-26){1}
\end{picture}
\end{center}
\caption{Feynman diagrams leads to collinear gaugino singularities.}\label{2ggino1gra}
\end{figure}
Here, the splitting tensor is
\begin{equation}
S^{\mu\nu} = \lambda_1\tilde{\lambda}_2 D^{\mu\nu}_{\gamma\delta}(p_1+p_2)V^{\gamma\delta}_{\tilde{g}\tilde{g}h}(p_1,p_2) \ ,
\end{equation}
where the graviton propagator is given by eq.(\ref{graprop}) and the gaugino-graviton vertex
\begin{equation}
V^{\gamma\delta}_{\tilde{g}\tilde{g}h}(p_1,p_2) = i\delta^{ab}\sigma^{\gamma}(p_2-p_1)^{\delta} \ .
\label{gginograviton}
\end{equation}
We obtain
\begin{eqnarray}
S^{\mu\nu}\Big(-\frac{1}{2},+\frac{1}{2}\Big) &=& \delta^{ab}\Big( - \frac{z_{12}}{\bar{z}_{12}}{\bf s^3 c}\epsilon_P^{+\mu}\epsilon_P^{+\nu} - \frac{\bar{z}_{12}}{z_{12}}{\bf s c^3}\epsilon_P^{-\mu}\epsilon_P^{-\nu}\Big) \nonumber\\
&=&\delta^{ab} \Big(  - \frac{z_{12}}{\bar{z}_{12}}\frac{\omega_2^{\frac{3}{2}}\omega_1^{\frac{1}{2}}}{\omega_P^2}\epsilon_P^{+\mu}\epsilon_P^{+\nu} - \frac{\bar{z}_{12}}{z_{12}}\frac{\omega_2^{\frac{1}{2}}\omega_1^{\frac{3}{2}}}{\omega_P^2}\epsilon_P^{-\mu}\epsilon_P^{-\nu}\Big) \ .
\end{eqnarray}

Thus the collinear limit of the amplitude is given by
\ba
\mathcal{M}(1^{a,-\frac{1}{2}},2^{b,+\frac{1}{2}},\cdots) &=& f^{abc}\Big(\frac{1}{\bar{z}_{12}}\frac{\omega_2^{\frac{1}{2}}}{\omega_1^{\frac{1}{2}}\omega_P}\mathcal{M}(P^{c,+1},\cdots) + \frac{1}{z_{12}}\frac{\omega_1^{\frac{1}{2}}}{\omega_2^{\frac{1}{2}}\omega_P}\mathcal{M}(P^{c,-1},\cdots) \Big)\nonumber\\
+ \delta^{ab}\!\!\!\!\!\!\!&& \Big(- \frac{z_{12}}{\bar{z}_{12}}\frac{\omega_2^{\frac{3}{2}}\omega_1^{\frac{1}{2}}}{\omega_P^2}\mathcal{M}(P^{+2},\cdots) - \frac{\bar{z}_{12}}{z_{12}}\frac{\omega_2^{\frac{1}{2}}\omega_1^{\frac{3}{2}}}{\omega_P^2}\mathcal{M}(P^{-2},\cdots)\Big) \ .
\ea
In terms of partial amplitudes, this corresponds to
\ba
M(1^{-\frac{1}{2}},2^{+\frac{1}{2}},\cdots) &=& \frac{1}{\bar{z}_{12}}\frac{\omega_2^{\frac{1}{2}}}{\omega_1^{\frac{1}{2}}\omega_P}M(P^{+1},\cdots) + \frac{1}{ z_{12}}\frac{\omega_1^{\frac{1}{2}}}{\omega_2^{\frac{1}{2}}\omega_P}M(P^{-1},\cdots)\nonumber\\
&-& \frac{z_{12}}{\bar{z}_{12}}\frac{\omega_2^{\frac{3}{2}}\omega_1^{\frac{1}{2}}}{\omega_P^2}M(P^{+2},\cdots) - \frac{\bar{z}_{12}}{z_{12}}\frac{\omega_2^{\frac{1}{2}}\omega_1^{\frac{3}{2}}}{\omega_P^2}M(P^{-2},\cdots) \ .
\ea
In a similar way, we obtain the collinear limit of $1^{+\frac{1}{2}}$ and $2^{-\frac{1}{2}}$:
\ba
M(1^{+\frac{1}{2}},2^{-\frac{1}{2}},\cdots) &=&  \frac{1}{{z}_{12}}\frac{\omega_2^{\frac{1}{2}}}{\omega_1^{\frac{1}{2}}\omega_P}M(P^{-1},\cdots) + \frac{1}{ \bar{z}_{12}}\frac{\omega_1^{\frac{1}{2}}}{\omega_2^{\frac{1}{2}}\omega_P}M(P^{+1},\cdots)\nonumber\\
&-& \frac{\bar{z}_{12}}{{z}_{12}}\frac{\omega_2^{\frac{3}{2}}\omega_1^{\frac{1}{2}}}{\omega_P^2}M(P^{-2},\cdots) - \frac{{z}_{12}}{\bar{z}_{12}}\frac{\omega_2^{\frac{1}{2}}\omega_1^{\frac{3}{2}}}{\omega_P^2}M(P^{+2},\cdots) \ .
\label{f+f-}
\ea

Next we consider the case of collinear gaugino and gauge boson. The Feynman diagram is shown in Fig.(\ref{1gaugino1bosonfig}).
\begin{figure}[h]
\begin{center}
\begin{picture}(300,100)(0,-10)
\SetColor{Black}
\Text(100,90)[l]{$a$}
\Text(200,90)[l]{$b$}
\Text(130,30)[l]{$c$}
\Text(130,50)[l]{$c$}
\ArrowLine(150,20)(150,60)\Vertex(150,60){2}
\ArrowLine(170,30)(170,48)
\Text(175,38)[l]{$p_1+p_2$}
\ArrowLine(150,60)(120,90)
\ArrowLine(120,65)(104,80)
\Text(95,70)[l]{$p_1$}
\Photon(150,60)(180,90){5}{4}
\ArrowLine(175,70)(190,85)
\Text(190,70)[l]{$p_2$}
\GCirc(150,0){22}{0.5}
\Vertex(125,18){1}
\Vertex(122.5,10.5){1}
\Vertex(120,3){1}
\Vertex(175,18){1}
\Vertex(178.5,10.5){1}
\Vertex(180,3){1}
\Vertex(135,-26){1}
\Vertex(150,-26){1}
\Vertex(165,-26){1}
\end{picture}
\end{center}
\caption{Feynman diagrams leading to collinear gaugino and gauge boson singularities.}\label{1gaugino1bosonfig}
\end{figure}
The corresponding splitting spinor is
\be
S = \tilde{\lambda}_1\epsilon_2^{\mu}V_{\tilde{g}g\tilde{g}}(p_1,p_2)_{\mu}D(p_1+p_2) \ .
\ee
Let us assume that the external gauge boson has positive helicity. After inserting
$\tilde{\lambda}_1$ and $\epsilon_2^{+\mu}$, we find
\be
S(+\frac{1}{2},+1) = f^{abc}\frac{1}{z_{12}\sqrt{\omega_1\omega_2}} \frac{1}{{\bf s}}\tilde{\lambda}_P  = f^{abc}\frac{1}{z_{12}}\frac{\omega_P^{1/2}}{\omega_1^{1/2}\omega_2} \tilde{\lambda}_P \ .
\ee
{}For a gauge boson with negative helicity, we find
\begin{equation}
S(+\frac{1}{2},-1) = f^{abc}\frac{1}{\bar{z}_{12}\sqrt{\omega_1\omega_2}} \frac{{\bf c^2}}{{\bf s}}\tilde{\lambda}_P = f^{abc}\frac{1}{\bar{z}_{12}}\frac{\omega_1^{1/2}}{\omega_2\omega_P^{1/2}}\tilde{\lambda}_P \ .
\end{equation}
The diagram describing gaugino-gauge boson gravitational production channel is shown in Fig.(\ref{gginogbsongviti}).
\begin{figure}[h]
\begin{center}
\begin{picture}(300,100)(0,-10)
\SetColor{Black}
\Text(100,90)[l]{$a$}
\Text(200,90)[l]{$b$}
\ArrowLine(150,20)(150,60)\Vertex(150,60){2}
\Photon(150,20)(150,60){5}{4}
\ArrowLine(170,30)(170,48)
\Text(175,38)[l]{$p_1+p_2$}
\ArrowLine(150,60)(120,90)
\ArrowLine(120,65)(104,80)
\Text(95,70)[l]{$p_1$}
\Photon(150,60)(180,90){5}{4}
\ArrowLine(175,70)(190,85)
\Text(190,70)[l]{$p_2$}
\GCirc(150,0){22}{0.5}
\Vertex(125,18){1}
\Vertex(122.5,10.5){1}
\Vertex(120,3){1}
\Vertex(175,18){1}
\Vertex(178.5,10.5){1}
\Vertex(180,3){1}
\Vertex(135,-26){1}
\Vertex(150,-26){1}
\Vertex(165,-26){1}
\end{picture}
\end{center}
\caption{Feynman diagrams leading to collinear gaugino and gauge boson singularities.}\label{gginogbsongviti}
\end{figure}
It yields
\begin{equation}
S^{\mu} = \tilde{\lambda}_1 \epsilon_{2\nu}V_{\tilde{g}g\tilde{h}}^{\nu\gamma}(p_1,p_2)D_{\gamma}^{\mu}(p_1+p_2) \ ,
\end{equation}
which leads to
\begin{eqnarray}
S^{\mu}(+\frac{1}{2},+1) &=& \delta^{ab} {\bf c}\,\epsilon_P^{+\mu}\tilde{\lambda}_P +\cdots\nonumber\\
&=&\delta^{ab} \frac{\omega_1^{1/2}}{\omega_P^{1/2}}\epsilon_P^{+\mu}\tilde{\lambda}_P +\cdots
\end{eqnarray}
\begin{eqnarray}
S^{\mu}(+\frac{1}{2},-1) &=&\delta^{ab} \frac{z_{12}}{\bar{z}_{12}}{\bf c^3} \epsilon_P^{+\mu}\tilde{\lambda}_P +\cdots \nonumber\\
&=&\delta^{ab}\frac{z_{12}}{\bar{z}_{12}}\frac{\omega_1^{3/2}}{\omega_P^{3/2}}\epsilon_P^{+\mu}\tilde{\lambda}_P +\cdots
\end{eqnarray}

We conclude that in the collinear limit of $1^{\frac{1}{2}}$ and $2^{+1}$, the amplitude becomes
\ba
\mathcal{M}(1^{a,+\frac{1}{2}},2^{b,+1},\cdots) &=& f^{abc}\frac{1}{z_{ 12}\sqrt{\omega_1\omega_2}}\Big(\frac{\omega_P}{\omega_2}\Big)^{\frac{1}{2}}\mathcal{M}(P^{c,+\frac{1}{2}},\cdots)\nonumber\\
&+& \delta^{ab}\frac{\omega_1^{1/2}}{\omega_P^{1/2}}\mathcal{M}(P^{+\frac{3}{2}},\cdots) \ .
\ea
In terms of partial amplitudes this corresponds to
\ba
M(1^{+\frac{1}{2}},2^{+1},\cdots) &=& \frac{1}{z_{12}\sqrt{\omega_1\omega_2}}\Big(\frac{\omega_P}{\omega_2}\Big)^{\frac{1}{2}}{M}(P^{+\frac{1}{2}},\cdots)\nonumber\\
&+&\frac{\omega_1^{1/2}}{\omega_P^{1/2}}{M}(P^{+\frac{3}{2}},\cdots) \  .
\label{f+g+}
\ea
In the collinear limit of $1^{\frac{1}{2}}$ and $2^{-1}$,
\begin{eqnarray}
\mathcal{M}(1^{a,+\frac{1}{2}},2^{b,-1},\cdots) &=&  f^{abc}\frac{1}{\bar{z}_{12}}\frac{\omega_1^{1/2}}{\omega_2\omega_P^{1/2}}\mathcal{M}(P^{c,+\frac{1}{2}},\cdots) \nonumber\\
&+&\delta^{ab}\frac{z_{12}}{\bar{z}_{12}}\frac{\omega_1^{3/2}}{\omega_P^{3/2}}\mathcal{M}(P^{+\frac{3}{2}},\cdots) \ .
\end{eqnarray}
In terms of partial amplitudes, this corresponds to
\begin{eqnarray}
{M}(1^{+\frac{1}{2}},2^{-1},\cdots) &=&  \frac{1}{\bar{z}_{12}}\frac{\omega_1^{1/2}}{\omega_2\omega_P^{1/2}}{M}(P^{+\frac{1}{2}},\cdots) \nonumber\\
&+&\frac{z_{12}}{\bar{z}_{12}}\frac{\omega_1^{3/2}}{\omega_P^{3/2}}{M}(P^{+\frac{3}{2}},\cdots) \ .
\end{eqnarray}

Finally, we consider collinear gaugino and graviton, shown in Fig.(\ref{collgginoh}).
\begin{figure}[h]
\begin{center}
\begin{picture}(300,100)(0,-10)
\SetColor{Black}
\ArrowLine(150,20)(150,60)\Vertex(150,60){2}
\ArrowLine(170,30)(170,48)
\Text(175,38)[l]{$p_1+p_2$}
\Photon(150,60)(120,90){5}{4}
\Photon(148,60)(118,90){5}{4}
\ArrowLine(120,65)(104,80)
\Text(95,70)[l]{$p_1$}
\ArrowLine(150,60)(180,90)
\ArrowLine(175,70)(190,85)
\Text(190,70)[l]{$p_2$}
\GCirc(150,0){22}{0.5}
\Vertex(125,18){1}
\Vertex(122.5,10.5){1}
\Vertex(120,3){1}
\Vertex(175,18){1}
\Vertex(178.5,10.5){1}
\Vertex(180,3){1}
\Vertex(135,-26){1}
\Vertex(150,-26){1}
\Vertex(165,-26){1}
\end{picture}
\end{center}
\caption{Feynman diagrams leading to collinear gaugino and graviton singularities.}\label{collgginoh}
\end{figure}
We find the splitting spinors:
\begin{equation}
S(-2,+\frac{1}{2}) = \frac{z_{12}}{\bar{z}_{12}}\frac{\omega_2^{3/2}}{\omega_1\omega_P^{1/2}}|P]
\end{equation}
\begin{equation}
S(+2,+\frac{1}{2}) =  \frac{\bar{z}_{12}}{{z}_{12}} \frac{\omega_2^{1/2}\omega_P^{1/2}}{\omega_1}|P\rangle \ .
\end{equation}
Thus the collinear gaugino-graviton limits  are
\begin{equation}
M(1^{-2},2^{+\frac{1}{2}},\cdots) =  \frac{z_{12}}{\bar{z}_{12}}\frac{\omega_2^{3/2}}{\omega_1\omega_P^{1/2}}M(P^{+\frac{1}{2}},\cdots) \ ,
\end{equation}
\begin{equation}
M(1^{+2},2^{+\frac{1}{2}},\cdots) = \frac{\bar{z}_{12}}{{z}_{12}} \frac{\omega_2^{1/2}\omega_P^{1/2}}{\omega_1}M(P^{+\frac{1}{2}},\cdots) \ .
\end{equation}
\subsection{Collinear limits involving gravitino}
The Feynman diagrams contributing to the limit of a gravitino collinear with a gauge boson and with a gaugino are shown in Figs.(\ref{collgravitino&gauge}) and (\ref{collgravitino&gaugino}), respectively.
\begin{figure}[h]
\begin{center}
\begin{picture}(300,100)(0,-10)
\SetColor{Black}
\ArrowLine(150,20)(150,60)\Vertex(150,60){2}
\ArrowLine(170,30)(170,48)
\Text(175,38)[l]{$p_1+p_2$}
\ArrowLine(150,60)(120,90)
\Photon(150,60)(120,90){5}{4}
\ArrowLine(120,65)(104,80)
\Text(95,70)[l]{$p_1$}
\Photon(150,60)(180,90){5}{4}
\ArrowLine(175,70)(190,85)
\Text(190,70)[l]{$p_2$}
\GCirc(150,0){22}{0.5}
\Vertex(125,18){1}
\Vertex(122.5,10.5){1}
\Vertex(120,3){1}
\Vertex(175,18){1}
\Vertex(178.5,10.5){1}
\Vertex(180,3){1}
\Vertex(135,-26){1}
\Vertex(150,-26){1}
\Vertex(165,-26){1}
\end{picture}
\end{center}
\caption{Feynman diagrams leading to collinear gravitino and gauge boson singularities.}\label{collgravitino&gauge}
\end{figure}
We can use the same definition for splitting vector or spinor as  in the soft limit. In the collinear limit, Fig.(\ref{collgravitino&gauge}) leads to the splitting spinor
\begin{equation}
S\Big(+\frac{3}{2},+1\Big) = \frac{\bar{z}_{12}}{z_{12}}\Big( \frac{\omega_P}{\omega_1}\Big)^{\frac{1}{2}}|P] \ ,
\end{equation}
which leads to
\begin{equation}
M(1^{+\frac{3}{2}},2^{+1}\cdots) = \frac{\bar{z}_{12}}{z_{12}}\Big( \frac{\omega_P}{\omega_1}\Big)^{\frac{1}{2}}M(P^{+\frac{1}{2}},\cdots) \ .
\end{equation}
\begin{figure}[h]
\begin{center}
\begin{picture}(300,100)(0,-10)
\SetColor{Black}
\Photon(150,20)(150,60){5}{4}
\Vertex(150,60){2}
\ArrowLine(170,30)(170,48)
\Text(175,38)[l]{$p_1+p_2$}
\ArrowLine(150,60)(120,90)
\Photon(150,60)(120,90){5}{4}
\ArrowLine(120,65)(104,80)
\Text(95,70)[l]{$p_1$}
\ArrowLine(180,90)(150,60)
\ArrowLine(175,70)(190,85)
\Text(190,70)[l]{$p_2$}
\GCirc(150,0){22}{0.5}
\Vertex(125,18){1}
\Vertex(122.5,10.5){1}
\Vertex(120,3){1}
\Vertex(175,18){1}
\Vertex(178.5,10.5){1}
\Vertex(180,3){1}
\Vertex(135,-26){1}
\Vertex(150,-26){1}
\Vertex(165,-26){1}
\end{picture}
\end{center}
\caption{Feynman diagrams leading to collinear gravitino and gaugino singularities.}\label{collgravitino&gaugino}
\end{figure}
Fig.(\ref{collgravitino&gaugino}) leads to the splitting vector
\begin{equation}
S^{\gamma}\Big(+\frac{3}{2},-\frac{1}{2}\Big) = \frac{\bar{z}_{12}}{z_{12}}\frac{\omega_2^{3/2}}{\sqrt{\omega_1}\omega_P}\epsilon_P^{-\gamma} \ .
\end{equation}
In this case,
\begin{equation}
M(1^{+\frac{3}{2}},2^{-\frac{1}{2}}\cdots) =\frac{\bar{z}_{12}}{z_{12}}\frac{\omega_2^{3/2}}{\sqrt{\omega_1}\omega_P}M(P^{-1},\cdots) \ .
\end{equation}

The Feynman diagram is contributing to the collinear limit of two gravitini is shown in Fig.(\ref{coll2gravitino}).
\begin{figure}[h]
\begin{center}
\begin{picture}(300,100)(0,-10)
\SetColor{Black}
\Photon(150,20)(150,60){5}{4}
\Photon(152,20)(152,60){5}{4}
\Vertex(150,60){2}
\ArrowLine(170,30)(170,48)
\Text(175,38)[l]{$p_1+p_2$}
\ArrowLine(120,90)(150,60)
\Photon(150,60)(120,90){5}{4}
\ArrowLine(120,65)(104,80)
\Text(95,70)[l]{$p_1$}
\ArrowLine(150,60)(180,90)
\Photon(180,90)(150,60){5}{4}
\ArrowLine(175,70)(190,85)
\Text(190,70)[l]{$p_2$}
\GCirc(150,0){22}{0.5}
\Vertex(125,18){1}
\Vertex(122.5,10.5){1}
\Vertex(120,3){1}
\Vertex(175,18){1}
\Vertex(178.5,10.5){1}
\Vertex(180,3){1}
\Vertex(135,-26){1}
\Vertex(150,-26){1}
\Vertex(165,-26){1}
\end{picture}
\end{center}
\caption{Feynman diagrams leading to collinear gravitini singularities.}\label{coll2gravitino}
\end{figure}
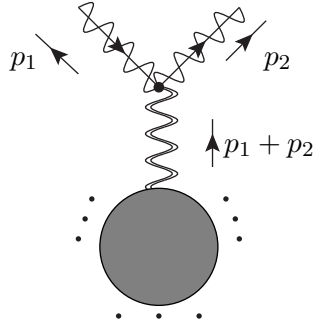
It yields
\begin{eqnarray}
M(1^{-\frac{3}{2}},2^{+\frac{3}{2}},\cdots) &=& \frac{{\bf s^5}}{{\bf c}} \frac{\langle 21\rangle}{[12]}M(P^{+2},\cdots)+\frac{{\bf c^5}}{{\bf s}} \frac{[21]}{\langle 12\rangle}M(P^{-2},\cdots) \nonumber\\
&=&\frac{z_{12}}{\bar{z}_{12}}\frac{\omega_2^{5/2}}{\omega_1^{1/2}\omega_P^2}M(P^{+2},\cdots) +\frac{\bar{z}_{12}}{z_{12}}\frac{\omega_1^{5/2}}{\omega_2^{1/2}\omega_P^2}M(P^{-2},\cdots) \ .
\end{eqnarray}
The Feynman diagram contributing to the gravitino-graviton colliner limit is shown in Fig.(\ref{collgravi&gra}).
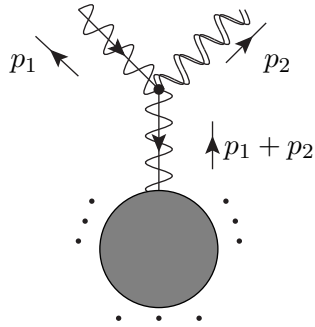
\begin{figure}[h]
\begin{center}
\begin{picture}(300,100)(0,-10)
\SetColor{Black}
\Photon(150,20)(150,60){5}{4}
\ArrowLine(150,60)(150,20)
\Vertex(150,60){2}
\ArrowLine(170,30)(170,48)
\Text(175,38)[l]{$p_1+p_2$}
\ArrowLine(120,90)(150,60)
\Photon(150,60)(120,90){5}{4}
\ArrowLine(120,65)(104,80)
\Text(95,70)[l]{$p_1$}
\Photon(152,60)(182,90){5}{4}
\Photon(180,90)(150,60){5}{4}
\ArrowLine(175,70)(190,85)
\Text(190,70)[l]{$p_2$}
\GCirc(150,0){22}{0.5}
\Vertex(125,18){1}
\Vertex(122.5,10.5){1}
\Vertex(120,3){1}
\Vertex(175,18){1}
\Vertex(178.5,10.5){1}
\Vertex(180,3){1}
\Vertex(135,-26){1}
\Vertex(150,-26){1}
\Vertex(165,-26){1}
\end{picture}
\end{center}
\caption{Feynman diagrams leading to collinear gravitino and graviton singularities.}\label{collgravi&gra}
\end{figure}
It leads to
\begin{equation}
M(1^{-\frac{3}{2}},2^{-2},\cdots)  = \frac{\bar{z}_{12}}{z_{12} }\frac{\omega_1^{5/2}}{\omega_2\omega_P^{3/2}}M(P^{-3/2},\cdots) \ ,
\end{equation}
\begin{equation}
M(1^{-\frac{3}{2}},2^{+2},\cdots)  = \frac{z_{12}}{\bar{z}_{12}}\frac{\omega_P^{3/2}}{\omega_2\omega_1^{1/2}}M(P^{-3/2},\cdots) \ .
\end{equation}

\renewcommand{\thesection}{B}
\setcounter{equation}{0}
\renewcommand{\theequation}{B.\arabic{equation}}
\renewcommand{\thesection}{B}
\setcounter{equation}{0}
\renewcommand{\theequation}{B.\arabic{equation}}
\vskip 3mm

\section{Gluon soft theorems in Mellin space}\label{App:GluonSoft}

We start with the BCFW representation for the $n+1$--point gluon amplitude $A_{n{+}1}$
\begin{align}\label{CSsoft}
A_{n{+}1} \Big(\{  \lambda_1, \tilde{\lambda}_1 \}\, , &\ldots \,,
\{  \lambda_n, \tilde{\lambda}_n \}\, , \{  \lambda_s,  \tilde{\lambda}_s \}^+\Big)={ \langle n 1 \rangle  \over \langle n s\rangle \langle s 1\rangle  }\nonumber\\
&\times A_{n}\lf(
\Big\{  \lambda_1,
\tilde{\lambda}_1
 + { \langle s \, n \rangle \over \langle 1 \, n\rangle } \tilde{\lambda}_s \Big\}   ,
\ldots,
\Big\{  \lambda_n, \tilde{\lambda}_{n}
 + { \langle s \, 1 \rangle \over \langle n \, 1\rangle } \tilde{\lambda}_s \Big\} \ri)\
 +\ldots \ ,
 \end{align}
with the shifts: $\lambda_s(z)=\lambda_s+z\lambda_n$ and $\tilde\lambda_n(z)=\tilde\lambda_n-z\tilde\lambda_s$ \cite{Casali:2014xpa}. The dots denote regular terms in $z=-\tfrac{\braket{s1}}{\braket{n1}}$.

A specific Lorentz generator which only acts on antichiral spinors can be introduced as in Ref.\cite{Guevara1906}:
\be\label{lorenzJ}
J=\fc{1}{4}\ (p_s^\mu\epsilon^\nu-p_s^\mu\epsilon^\mu)\ \tilde\sigma_\mu\sigma_\nu=\h\ |s][s|
\ee
in terms of spinors $\lambda_s,\tilde\lambda_s$ referring to particle $s$.
With this Lorentz generator \req{lorenzJ} in \req{CSsoft} the shifts  in the antichiral spinors $\tilde\lambda_1$ and $\tilde\lambda_n$ can be furnished by acting with $J_1$ and $J_n$  on gluon $1$ and $n$, respectively.
Since
\begin{align}
\exp\lf\{\frac{J_1}{(\epsilon_s p_1)}\ri\}\tilde\lambda_1&=\tilde\lambda_1+\frac{\braket{ns}}{\braket{n1}}\tilde\lambda_s\ ,\nonumber\\
\exp\lf\{-\frac{p_sp_1}{p_sp_n}\frac{J_n}{(\epsilon_s p_1)}\ri\}\tilde\lambda_n&=\tilde\lambda_n+\frac{\braket{1s}}{\braket{1n}}\tilde\lambda_s\ ,\label{Jops}
\end{align}
we have:
\begin{align}\label{cssoft}
A_{n{+}1} \Big(\{  \lambda_1, \tilde{\lambda}_1 \}\, , &\ldots \,,
\{  \lambda_n, \tilde{\lambda}_n \}\, , \{  \lambda_s,  \tilde{\lambda}_s \}^+\Big)={ \langle n 1 \rangle  \over \langle n s\rangle \langle s 1\rangle  }
\nonumber\\
&\times \exp\left\{\frac{1}{(\epsilon_s p_1)}\lf(J_1-\fc{p_sp_1}{p_sp_n}J_n\ri)\ri\}\
A_{n} \Big(\{  \lambda_1, \tilde{\lambda}_1 \}\, , &\ldots \,,
\{  \lambda_n, \tilde{\lambda}_n \}\Big)+\ldots\, .
 \end{align}
To translate this formula into Mellin space we introduce celestial coordinates. In this basis
the operators  \req{Jops} act as conformal transformations on the gluon operators  number $1$ and $n$. Concretely, with $\epsilon_sp_1=\h\fc{\braket{n1}[s1]}{\braket{ns}}$  we have \cite{Guevara1906}
\be
\frac{J}{(\epsilon_s p_1)}=\fc{\alpha_1}{\bar z_{s1}}(\bar z_s^2 l_{-1}-2\bar z_sl_0+l_1)\ \ \ ,\ \ \ \alpha_1=\fc{\epsilon_s\omega_sz_{ns}}{\epsilon_1\omega_1 z_{n1}}\ .
\ee
In terms of the $SL(2,C)$ generators $l_{-1}=\p_{\bar z},\, l_0=\bar z\p_{\bar z}+\bar h$ and $l_1=\bar z^2\p_{\bar z}+2\bar z\bar h$ \cite{Stieberger1812}
acting on some function $f(\bar z)$ depending on the celestial coordinate $\bar z$  we get:
\be
\lf.\exp\lf\{\frac{J_1}{(\epsilon_s p_1)}\ri\}f(\bar z)=\exp\lf\{\frac{\alpha_1}{\bar z_{s1}}[(\bar z_s-\bar z)^2\p_{\bar z}-2\bar h_1(\bar z_s-\bar z)]\ri\}f(\bar z)\ri|_{\bar z=\bar z_1}\ .
\ee
Furthermore, we obtain
\begin{align}
\exp\lf\{-\frac{p_sp_1}{p_sp_n}\frac{J_n}{(\epsilon_s p_1)}\ri\}f(\bar z)&\lf.=\exp\lf\{\frac{\alpha_n}{\bar z_{sn}}[(\bar z_s-\bar z)^2\p_{\bar z}-2\bar h_n(\bar z_s-\bar z)]\ri\}f(\bar z)\ri|_{\bar z=\bar z_n}\ ,\label{JOps}
\end{align}
with $\alpha_n=\fc{\epsilon_s\omega_sz_{1s}}{\epsilon_1\omega_n z_{1n}}$.
Eventually, with this information for \req{cssoft} we obtain the following expression in celestial coordinates:
\begin{align}
\tilde A_{n+1}&=\fc{z_{n1}}{z_{ns}z_{s1}}\fc{1}{\omega_s}\nonumber\\
&\times
\exp\lf\{\fc{\alpha_1}{\bar z_{s1}}[(\bar z_s-\bar z)^2\partial_{\bar z}-2\bar h_1(\bar z_s-\bar z)]+\fc{\alpha_n}{\bar z_{sn}}[(\bar z_s-\bar z')^2\partial_{\bar z'}-2\bar h_n(\bar z_s-\bar z')]\ri\}\nonumber\\
&\times \lf.\tilde A_n(\{z_1,\bar z,\Delta_1,J_1\},\ldots,\{z_n,\bar z',\Delta_n,J_n\})\ri|_{\bar z=\bar z_1\atop \bar z'=\bar z_n}\ .\label{Gluonsoft}
\end{align}

The lowest order  $\omega_s^{-1}$ in $\omega_s$ is related to the limit $\Delta_s\ra1$
\be\label{soft1}
\Delta_s=1:\ \ \ \fc{z_{n1}}{z_{ns}z_{s1}} \tilde A_n(\{z_1,\bar z_1,\Delta_1,J_1\},\ldots,\{z_n,\bar z_n,\Delta_n,J_n\})\ ,
\ee
which corresponds to the soft--limit of the amplitude $A_{n+1}$.
On the other hand, the next order $\omega_s^0$ corresponds to the limit $\Delta_s\ra0$
\begin{align}
\Delta_s=0:\ \ \ &\Big\{\fc{1}{\omega_1}\fc{1}{z_{s1}}\ (\bar z_{s1}\p_{\bar z_1}-2\bar h_1)+\fc{1}{\omega_n}\fc{1}{z_{ns}}\ (\bar z_{sn}\p_{\bar z_n}-2\bar h_n)\Big\}\nonumber\\
& \tilde A_n(\{z_1,\bar z_1,\Delta_1,J_1\},\ldots,\{z_n,\bar z_n,\Delta_n,J_n\})\ ,\label{soft0}
\end{align}
which gives rise  to the subleading soft--limit of the amplitude $A_{n+1}$. The result  \req{soft0} agrees with the expression given in \cite{Mason1905} after translating the latter into celestial coordinates.

\end{document}